\documentclass[traditabstract]{aa} 
\usepackage{color}
\usepackage{graphicx}
\usepackage{natbib}
\bibpunct{(}{)}{;}{a}{}{,}
\usepackage{amssymb}
\usepackage{amsmath}
\usepackage{graphicx}
\usepackage{multirow}
\usepackage{txfonts}
\newcommand\be{\begin{equation}}
\newcommand\en{\end{equation}}
\newcommand{\msun}{\mbox{\rm $M_{\odot}$}}

\newcommand{\lsun}{\mbox{\rm $L_{\odot}$}~}

\newcommand{\arcs}{\hbox{$^{\prime\prime}$}}

\newcommand{\arcm}{\mbox{$^{\prime}$}}

\newcommand{\degree}{\mbox{$^{\circ}$}}
\newcommand{\av}{\mbox{$A_{\rm V}$~}}

\newcommand{\hii}{\mbox{H~{\sc ii}~}}

\newcommand\kms{km~s$^{-1}$}

\newcommand\nht{N(H$_{2})$}

\defcitealias{maserpap}{C09}

\begin{document}
   \title{Star formation in the filament of S254-S258 OB complex: a cluster in the process of making}

   \author{M. R. Samal\inst{1}
\and D.K. Ojha\inst{2}
\and J. Jose\inst{3}
\and A. Zavagno\inst{1}
\and S. Takahashi\inst{4}
\and B. Neichel\inst{1}
\and J. S. Kim\inst{5}
\and N. Chauhan\inst{6}
\and A. K. Pandey\inst{7}
\and I. Zinchenko\inst{8}
\and M. Tamura\inst{9}
\and S. K. Ghosh\inst{10}
          }

   \institute{Aix Marseille Universit\'e, CNRS, LAM (Laboratoire d'Astrophysique de
Marseille) UMR 7326, 13388 Marseille, France
              \email{manash.samal@lam.fr}
              \and Tata Institute of Fundamental Research, Mumbai 400 005, India 
              \and Kavli Institute for Astronomy and Astrophysics, Peking University, Beijing 100871, China
               \and Joint ALMA Observatory, Alonso de Córdova 3107, Vitacura, Santiago, Chile  
               \and Steward Observatory, University of Arizona, 933 North Cherry Avenue, Tucson, AZ 85721-0065, USA
               \and Institute of Astronomy, National Central University, Chung-Li 32054, Taiwan
              \and Aryabhatta Research Institute of Observational Sciences (ARIES), Manora Peak, Nainital 263129, India  
               \and Institute of Applied Physics of the Russian Academy of Sciences, 46 Ulyanov St., Nizhny Novgorod 603950, Russia
              \and National Astronomical Observatory of Japan, Mitaka, Tokyo 181-8588, Japan
              \and National Centre for Radio Astrophysics, Tata Institute of Fundamental Research, Pune 411007, India
             }

 
  \abstract
   {Infrared Dark Clouds (IRDCs) are ideal laboratories to study the initial processes of high-mass star and star cluster formation. We investigated  star formation activity 
   of an unexplored filamentary dark cloud (size $\sim$ 5.7 pc $\times$ 1.9 pc), which itself is part of a large
filament ($\sim$ 20 pc)  located in the S254-S258 OB complex at a distance of 2.5 kpc. Using  Multi-band Imaging Photometer (MIPS)  {\it Spitzer} 24 $\mu$m data, 
we uncover  49 sources with signal-to-noise ratio greater than 5. We identified 45 sources as candidate young stellar objects (YSOs) of Class I, Flat-spectrum, and Class II nature. 
Additional 17 candidate YSOs (9 Class I and 8 Class II) are also identified using JHK and Wide-field Infrared Survey Explorer (WISE) photometry. 
We find that the protostar to Class II sources ratio ($\sim$ 2) and the protostar fraction ($\sim$ 70 \%) of the region are high.
When the protostar fraction compared to other young clusters, it suggests that the  star formation in the dark cloud was possibly 
started  only  1 Myr ago. 
Combining the near-infrared photometry of the YSO candidates with the theoretical evolutionary models, we infer that most  of the candidate 
YSOs formed in the dark cloud are low-mass ($<$ 2 \msun) in nature. 
 We examine the spatial
distribution of the YSOs and find that majority of them are linearly aligned  along the
highest column density line (\nht $\sim$ 1 $\times$ 10$^{22}$ cm$^{-2}$) of the dark cloud along its long axis at mean nearest 
neighbor separation of $\sim$ 0.2 pc. 
Using  observed properties  of the  YSOs, physical conditions of the cloud and a simple cylindrical model, 
 we explore the  possible star formation process of this filamentary dark cloud and suggest that gravitational fragmentation within the filament 
should have played a dominant role in the formation of the YSOs. From  the total mass of the YSOs, gaseous mass associated with the dark cloud, and surrounding environment, we infer that the region is presently 
forming stars at an efficiency $\sim$ 3\% 
and a rate $\sim$ 30 $\msun$~ Myr$^{-1}$, and may emerge to a richer cluster.
}   
 \titlerunning{A cluster in the process of formation}

 \maketitle
%

\section{Introduction}\label{intro}
Recently, filaments and infrared dark clouds (IRDCs) have received special attention because they are the potential 
progenitors of cluster formation  \citep[e.g.,][]{rath06, beuth07, pere09, myers09, henn10, bata11, ragan12,long12, russeil13}. 
Results from the {\it Herschel} Space Telescope, particularly the
Gould Belt  \citep{andre10} and Hi-GAL  \citep{molinari10} Legacy Surveys, have also emphasized the crucial role of
filamentary structures on star and cluster formation. However, due to the complex nature of the interstellar medium (ISM), 
the modes of fragmentation and the physical processes that govern star formation in  filaments are still under active 
discussion \citep[e.g.,][]{hei09,semadeni09,pon11,andre13,smith14}.   The identification and characterization of young 
protostars represent a  key point to partly answer some of these questions, as they still carry the imprint of the 
fragmentation of the primordial cloud.   However, filaments and IRDCs are characterized by high column densities and, therefore, 
the identification of such sources is often difficult.
In the last decade, observations at longer infrared wavelengths with Mid-infrared Imaging Photometer
\citep[MIPS;][]{carey09} and 
Wide-field Infrared Survey Explorer \citep[WISE;][]{wright10} 
allowed us to overcome the high extinction in  such regions and  explore the properties and
 distribution of associated young sources.
 In this context,
 we present an investigation of a filamentary dark cloud located in the S254-S258 
OB complex.

The S254-S258 complex, situated at a distance of 2.5 kpc,  is part of the 
Gemini OB association \citep[][]{chava08,beig09,ojha11}. 
The complex contains six  \hii regions (S254, S255, S256, S257, S258 and S255B) 
projected on or in the close vicinity of a long ($\sim$ 20 pc) filamentary  cloud (see Fig. 1). 
The search for young stellar objects (YSOs) in the complex has been  carried out by \citet{chava08}, \citet{ojha11}, and \citet{mucci11}, 
but these studies mainly focus on the central part of the complex, i.e., roughly over  $\leq$ 13 square arcmin area around the massive  young cluster 
S255-IR \citep{wang11,ojha11,zinch12} located between the two evolved \hii regions S255 and S257.  
These studies have shown that the S254-S258 complex contains a rich population of 
Class I, Class II, Class III, and near-infrared (NIR)-excess sources  formed in  groups,  
and also in distributed mode  across the complex. 

 The present study focuses on the star formation activity of a part of the long filament (centered at $\alpha_{2000}=06^{h}13^{m}47^{s}$, $\delta_{2000}=17^{\circ}54^{\prime}37^{\prime\prime}$ or 
$l$=192\degree.76, $b$=00\degree.10).  
The filamentary area under investigation is of size $\sim$ 5.7 pc in 
length and 1.9 pc in width (marked with a rectangular box in Fig. 1). The region  
is highly obscured at optical wavelength regime, and corresponds to a dense part of the long filament.   
The region is dense as revealed by the presence of CS (J=2-1) emission, 
which is a tracer of dense ($>$ 10$^{4}$ cm$^{-3}$) gas.  
The distribution of CS emission \citep [see Fig. 4 of][]{carpen95} is also seen elongated 
in shape (size $\sim$ 5.6 pc $\times$ 1.7 pc), with its long axis along the long axis of the large filament. 
IRDCs are dense molecular clouds seen as extinction features against the bright  background.
Molecular line and dust continuum studies of  IRDCs have shown that they are cold (T $<$ 25 K), dense  ($N({\rm H_2})\ga 10^{22}$ cm$^{-2}$), and 
massive ($\sim $ $10^2{-}10^5~M_{\odot}$) structures with sizes 1-15 pc \citep{carey98,simon06}. 
Like other IRDCs \citep[e.g.,][]{simon06}, the area under investigation is dark in optical (e.g., in DSS2 Survey images) as well as in mid-infrared  (e.g., in
MSX Survey images; \citet{price01}), and also cold (T $\leq$ 15 K) and dense  ($N({\rm H_2})\sim 10^{22}$ cm$^{-2}$) 
in nature (discussed in Sects. 3.5 and 4.2).  Hence, following the general nomenclature of the IRDCs \citep[][]{rath06,pere09}, hereafter, we designate 
the dense cloud under investigation as ``IRDC G192.76+00.10''. 

Massive OB2 stars ionize the surrounding ISM and create \hii region around them.  During expansion, H II region drives the shock front 
preceding ionization front. The shock front sweeps up the ISM to form a cold dense shell and star formation can be induced by the instability of the shell \citep[
e.g., ``collect and collapse`` scenario;] []{elm77}.
In the S254-S258 complex, \citet{ojha11} and \citet{mucci11} suggested the presence of induced second generation star formation  at the peripheries of
S255 and S257. Similar to here, evidences of second generation triggered star formations have been  observed at the peripheries of several \hii regions
\citep[e.g.,][]{zava07, inde07,chu08, chen11, brand11,dehar12,getman12,jose13, samal14}, suggesting that the presence of an \hii region can influence star formation
processes of a complex. However, in the S254-S258 complex, the IRDC G192.76+00.10 region is 
relatively isolated from the feedback effects, as it is situated farther from the 
known evolved  \hii regions (e.g., S254, S255, and S257). 
Moreover,  shell-like  filamentary structures resulted due to sweeping and compression  of expanding \hii regions  are generally located  
parallel to the \hii region's ionization front \citep[e.g.,][]{deha10,jose13}, whereas  the filament axis of the IRDC G192.76+00.10 
region is perpendicular  
to the ionization front of the evolved \hii regions of the complex.
Also the IRDC G192.76+00.10 area  does not seem to be contaminated by polycyclic 
aromatic hydrocarbon (PAH) emissions at 11.3 $\mu$m and 12.7 $\mu$m of WISE 12 $\mu$m band (not shown); PAH emissions are generally found 
at the interaction zone of \hii region and molecular cloud \citep[e.g.,][]{samal07,deha10}.  
These evidences suggest that the star formation in the IRDC G192.76+00.10 region is unlikely to be influenced by the evolved \hii regions of the complex. 

Star formation in the IRDC G192.76+00.10 region is  
unexplored so far; only partly covered by \citet{chava08} at {\it Spitzer}-IRAC wavelengths (i.e., 3.6-8.0 $\mu$m) in their YSOs search.
In the present work, along with $^{13}\mathrm{CO}$ data, we make use of MIPS, WISE, and NIR-JHK band data sets   to study
the star formation activity of  IRDC G192.76+00.10. We present our work with the following layout.
We describe the observations and data reduction techniques in Sect. 2.
In Sect. 3, the observational results are presented, which include
morphology of the region and identification, distribution, and 
characterization of  YSOs.  
We  discuss the  properties and star formation processes of the IRDC G192.76+00.10 cloud and its overall relation 
with the main filament in Sect. 4. 
\begin{figure*}
\centering
\resizebox{16.0cm}{14.0cm}{\includegraphics{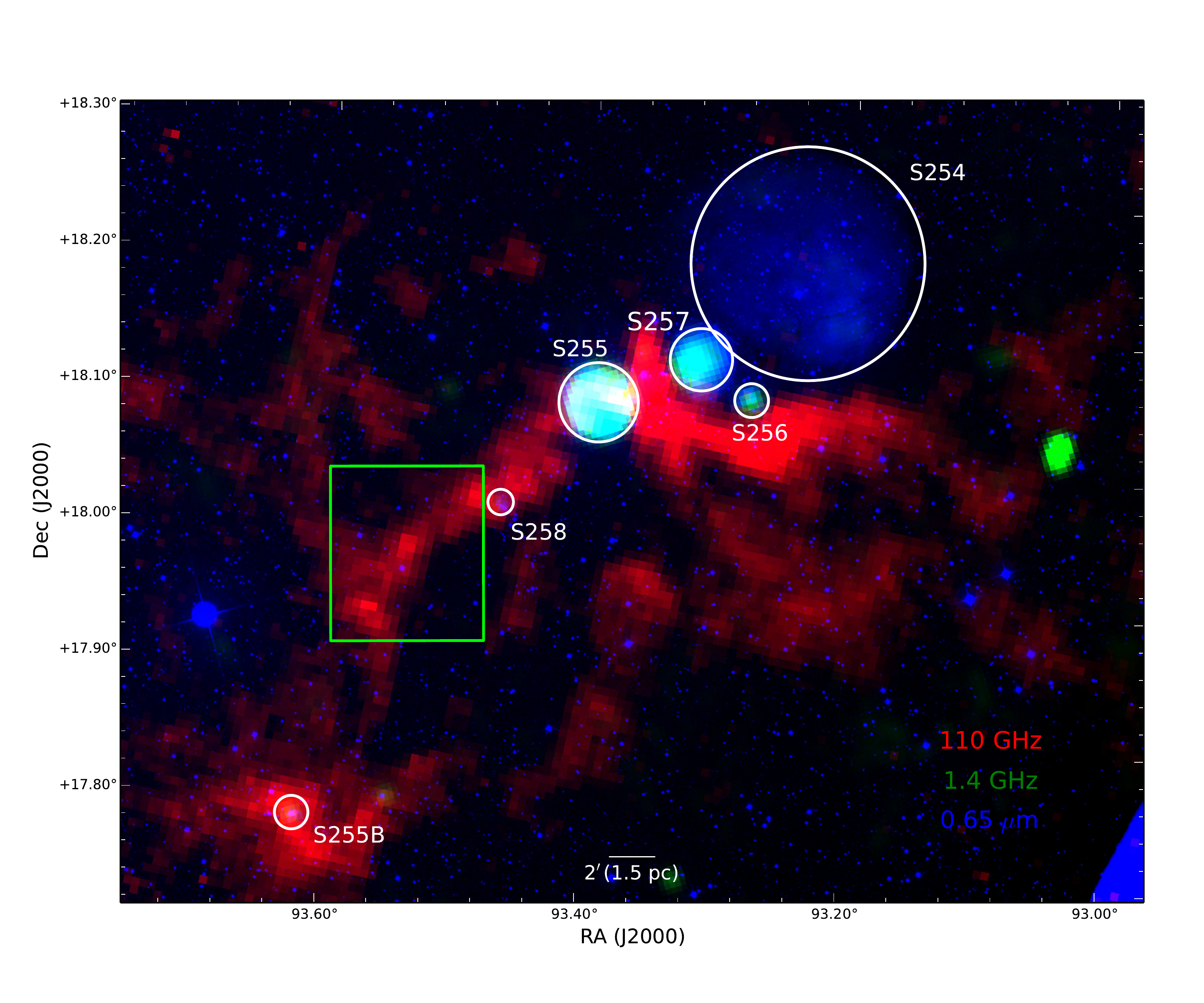}}
\caption{ Colour composite image of the complex obtained using the  $^{13}\mathrm{CO}$ (110 GHz) column density map in red (from \citet {chava08}), radio 
emission in green (from NVSS survey  at 1.4 
GHz; \citet {condon98}) and optical emission in blue (from DSS2 survey). The abscissa (RA) and ordinate (Dec) are in J2000 epoch. North is up and 
east is left. The \hii regions discussed in the text are marked in white circles.
 The rectangular box represents the area studied in this work (the corresponding image at 24 $\mu$m is shown in Fig. 2). }
\label{NGC6823_field.ps}
\end{figure*} 
\section{Observations and Data Reduction}
 The 24 $\mu$m observations of the IRDC G192.76+00.10 region were downloaded from the {\it Spitzer} archive
(Program ID 20635),
and cover an area $\sim$ 7\farcm0 $\times$ 8\arcm0, centered  at $\alpha_{2000}=06^{h}13^{m}47^{s}$, $\delta_{2000}=17^{\circ}53^{\prime}40^{\prime\prime}$. 
We downloaded the corrected basic-calibrated data (CBCD) images  and the corresponding
uncertainty files and performed source detection and photometry using the MOPEX/APEX software.
To extract the flux,  we applied the APEX 
point-response function (PRF)  
fitting method to the detected sources. 
We used the zero-point value of 7.17 Jy from the MIPS
Data Handbook to convert  flux densities to magnitudes. We considered only those 49 
sources whose signal-to-noise (S/N) ratio was found to be greater than 5. 
The magnitudes of these 49 sources along with their corresponding uncertainties are given in Table 1. 
 We point out that the reported uncertainties are lower limit  to the actual values, as the  uncertainty in the 
absolute flux densities at 24 $\mu$m  is $\sim$ 4\% \citep{enge07}.

We also downloaded MIPS 70 $\mu$m CBCD image from the {\it Spitzer} archive (Program ID 20635), and performed
photometry using APEX-PRF fitting method as discussed above. The area covered by the MIPS  70 $\mu$m image is
$\sim$  5\farcm7 $\times$  2\farcm0, centered  at $\alpha_{2000}=06^{h}13^{m}47^{s}$, $\delta_{2000}=17^{\circ}54^{\prime}45^{\prime\prime}$.
We detected only four point sources with S/N $>$ 5. Zero-point value of 0.778 Jy, adopted from the MIPS
Data Handbook, was used  to convert the PRF fitted flux densities of these sources to magnitudes. 

Deep NIR observations of the IRDC G192.76+00.10 region 
in  $J$ ($\lambda$ = 1.25  $\mu$m), $H$ ($\lambda$ = 1.63
$\mu$m), and  $K$ ($\lambda$ = 2.14  $\mu$m) bands were obtained
from the WFCAM Science Archive \citep{hambly08}, taken with the WFCAM instrument \citep{casali07} at 
the United Kingdom Infrared Telescope (UKIRT). 
We performed photometry  on the retrieved  stacked images produced by WFCAM pipeline 
at Cambridge Astronomical Survey Unit (CASU) for an area of $\sim$ 15$\arcm$ $\times$ 12$\arcm$ around
the IRDC region centered at  $\alpha_{2000}=06^{h}13^{m}48^{s}$ \& $\delta_{2000}=17^{\circ}52^{\prime}50^{\prime\prime}$.
Photometry on the images  was done using the  PSF algorithm of DAOPHOT package (Stetson 1987) in   IRAF.
The PSF was determined from the bright and isolated stars of the field.
For photometric calibration, we used isolated Two Micron
All Sky Survey (2MASS) point sources \citep[][]{cutri03} having error $<$ 0.1 mag and rd-flag ``123".
Rd-flag values of '1', '2' or '3' generally indicate the best quality detections, photometry and astrometry, respectively.
A mean calibration dispersion of $\leq$ 0.07 mag is observed in each band,
indicating that our photometry is reliable within $\sim$ 0.07 mag. Saturated sources in our catalog were replaced by 2MASS sources.

Only 9 out of the 49 MIPS detected sources have a counterpart in the NIR and {\it Spitzer}-IRAC bands from the catalog of \citet[][]{chava08}.
We thus mostly used WISE and our NIR point source catalogs to classify YSOs 
and for the construction of their spectral energy distributions (SEDs).
The WISE survey \citep{cutri12}  provides  photometry at four wavelengths:  3.4, 4.6, 12 and 22 $\mu$m, 
with an angular resolution of 6\farcs1, 6\farcs4, 6\farcs5 and 12\farcs0, respectively. 
We merged the MIPS catalog with the WISE and NIR catalogs  
using a matching radius of 3\farcs0 (following \citet[][]{koenig12}). 

 \citet{chava08} conducted observations in the J=1{\mbox{--}} 0 spectral 
lines of $^{12}\mathrm{CO}\,$ and $^{13}\mathrm{CO}\,$  using the Five College Radio Astronomy Observatory (FCRAO) 14 m telescope. 
The FCRAO beam size is 45$^{\prime \prime}$ in $^{12}\mathrm{CO}\,$ and $46^{\prime \prime }$ in $^{13}\mathrm{CO}\,$. 
 We used the $^{13}\mathrm{CO}\,$ column density
map of \citet{chava08}  to study the gas content of the region.

\begin{table}
\centering
\scriptsize
\caption{ MIPS 24 $\mu$m photometry for the 49 sources detected in the IRDC G192.76+00.10 region with a S/N  
ratio greater than 5}
\begin{tabular}{ccccccc}
\hline\hline
\multicolumn {1}{c} {ID} & \multicolumn {1}{c} {RA (deg)} & \multicolumn{1}{c}  {DEC (deg)} &\multicolumn{1}{c}{[24] $\mu$m } &\multicolumn{1}{c}{$\alpha$} \\
\multicolumn {1}{c} {} & \multicolumn{1}{c} {J2000}  & \multicolumn{1}{c} {J2000} & \multicolumn{1}{c} {mag} &\multicolumn{1}{c}{} \\
\hline
1 & 93.424041 & 17.944051 & 8.568	$\pm$	0.051 & -0.43 \\
2 & 93.416244 & 17.943596 & 7.645	$\pm$	0.032 &-0.29\\
3 & 93.386224 & 17.934850 & 5.809	$\pm$	0.017 &0.43\\
4 & 93.483044 & 17.933043 & 8.339	$\pm$	0.044 &-3.06\\
5 & 93.433941 & 17.926279 & 8.407	$\pm$	0.034 &-0.82\\
6 & 93.446990 & 17.921782 & 8.060	$\pm$	0.024 &-0.02\\
7 & 93.442216 & 17.921563 & 6.423	$\pm$	0.016 &-0.18\\
8 & 93.436925 & 17.918236 & 4.256	$\pm$	0.011 &0.93\\
9 & 93.424662 & 17.917668 & 6.554	$\pm$	0.017 &-0.31\\
10 & 93.447529 & 17.914416 & 4.802	$\pm$	0.011 &1.83\\
11 & 93.432413 & 17.912164 & 8.354	$\pm$	0.032 &-1.17\\
12 & 93.460228 & 17.911043 & 3.570	$\pm$	0.011 &1.92\\
13 & 93.457060 & 17.906121 & 8.200	$\pm$	0.029 &---\\
14 & 93.466088 & 17.904649 & 7.684	$\pm$	0.018 &-0.39\\
15 & 93.453099 & 17.903396 & 5.898	$\pm$	0.014 &-0.34\\
16 & 93.461708 & 17.903433 & 5.564	$\pm$	0.013 &1.27\\
17 & 93.441803 & 17.901178 & 8.424	$\pm$	0.033 &-0.85\\
18 & 93.457880 & 17.899842 & 9.098	$\pm$	0.065 &-2.51\\
19 & 93.468669 & 17.895977 & 6.052	$\pm$	0.014 &-2.61\\
20 & 93.476514 & 17.889089 & 6.750	$\pm$	0.010 &-0.64\\
21 & 93.503164 & 17.888787 & 4.377	$\pm$	0.012 &1.37\\
22 & 93.494384 & 17.886028 & 7.768	$\pm$	0.028 &0.34\\
23 & 93.483423 & 17.882087 & 8.463	$\pm$	0.057 &-0.55\\
24 & 93.498257 & 17.880100 & 3.893	$\pm$	0.001 &1.10\\
25 & 93.500148 & 17.878929 & 5.609	$\pm$	0.006 &0.52\\
26 & 93.496984 & 17.877505 & 5.850	$\pm$	0.008 &0.26\\
27 & 93.459530 & 17.877001 & 8.918	$\pm$	0.073 &---\\
28 & 93.482464 & 17.875690 & 6.792	$\pm$	0.017 &-0.13\\
29 & 93.392728 & 17.875741 & 7.388	$\pm$	0.032 &-2.62\\
30 & 93.482421 & 17.872994 & 7.248	$\pm$	0.026 &----\\
31 & 93.478138 & 17.863109 & 8.028	$\pm$	0.050 &-0.29\\
32 & 93.490105 & 17.861847 & 6.729	$\pm$	0.016 &-0.46\\
33 & 93.491665 & 17.856101 & 5.204	$\pm$	0.004 &2.08\\
34 & 93.481174 & 17.852611 & 8.152	$\pm$	0.061 &-0.32\\
35 & 93.497793 & 17.852376 & 5.718	$\pm$	0.017 &---\\
36 & 93.496107 & 17.850057 & 5.564	$\pm$	0.016 &0.26\\
37 & 93.493245 & 17.848029 & 6.739	$\pm$	0.019 &---\\
38 & 93.479544 & 17.894587 & 9.160	$\pm$	0.091 &---\\
39 & 93.474166 & 17.930534 & 9.010	$\pm$	0.075 &---\\
40 & 93.467083 & 17.908493 & 9.617	$\pm$	0.131 &---\\
41 & 93.477897 & 17.937458 & 9.523	$\pm$	0.140 &---\\
44 & 93.492631 & 17.906890 & 4.921	$\pm$	0.012 &0.17\\
45 & 93.437399 & 17.921868 & 6.806	$\pm$	0.018 &-0.56\\
46 & 93.450619 & 17.906530 & 7.407	$\pm$	0.035 &---\\
47 & 93.471334 & 17.903989 & 6.356	$\pm$	0.176 &1.83\\
48 & 93.444723 & 17.913772 & 7.178	$\pm$	0.250 &0.36 \\ 
49 & 93.450026 & 17.905202 & 6.620	$\pm$	0.180 &1.10\\
\hline
\end{tabular}
 \end{table}

\begin{figure}
\resizebox{9.0cm}{9.5cm}{\includegraphics{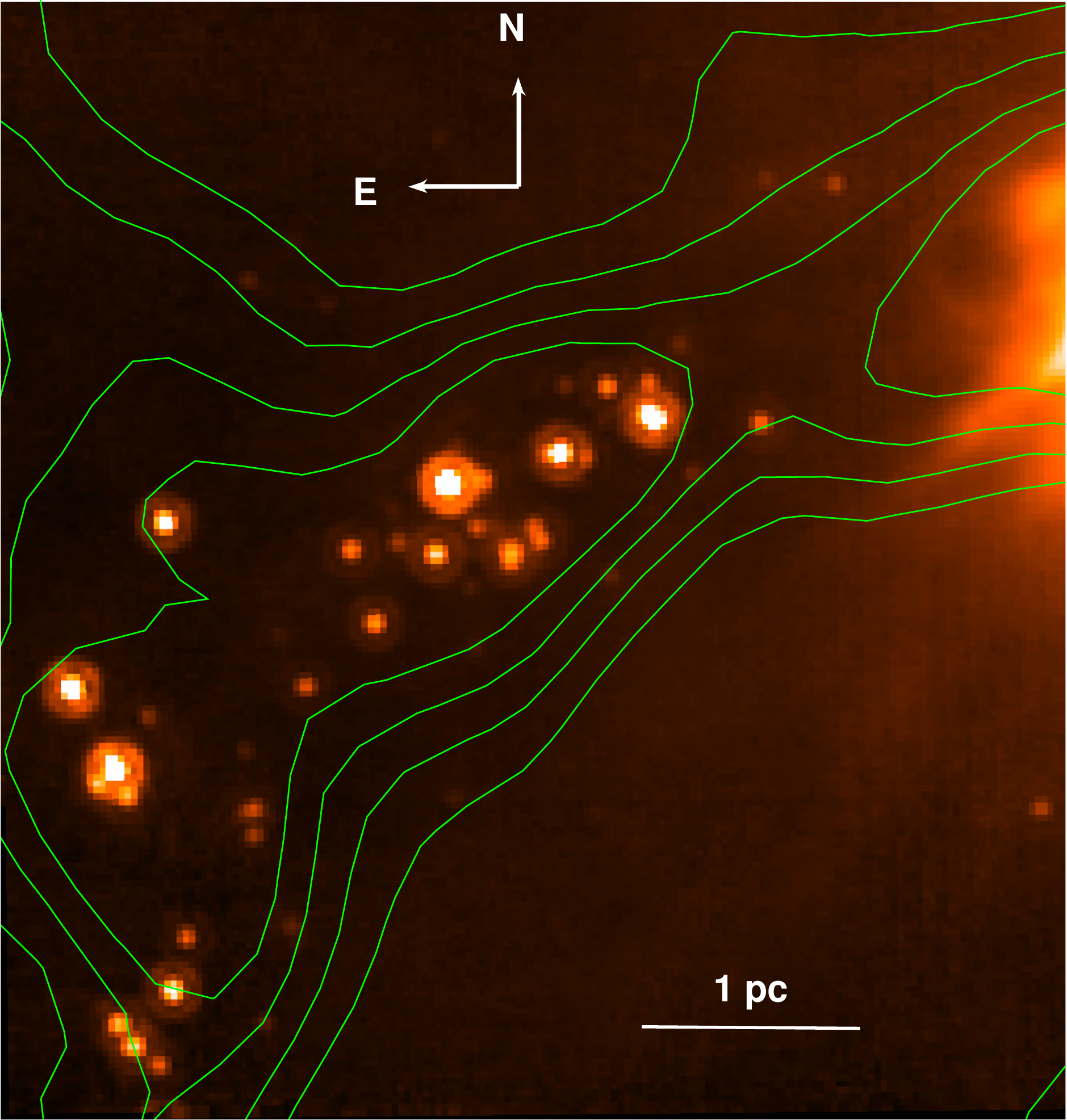}}
\caption{       {\it Spitzer} MIPS 24 $\mu$m  image
  of IRDC G192.76+00.10.
   The image has a field of
view $\sim$7\farcm0 $\times$ 8\farcm0 centered at
$\alpha_{2000}=06^{h}13^{m}47^{s}$, $\delta_{2000}=17^{\circ}54^{\prime}37^{\prime\prime}$ or $l$=192\degree.76, $b$=00\degree.10.  The $^{13}$CO column density contours are also shown (contour levels are at 0.7, 0.9, 1.1, 1.4 $\times$ 10$^{16}$ 
cm$^{-2}$).}
\label{NGC6823_field.ps}
\end{figure}

\section{Results}
\subsection{Morphology}
The MIPS 24 $\mu$m image of the IRDC G192.76+00.10 region is  shown in Fig. 2. 
 The image displays a significant number of point-like sources  aligned  roughly in a linear sequence from south-east to north-west,  and 
 most of the point sources seem  to be bounded by an elongated structure of $^{13}\mathrm{CO}$ gas.
Since dark cloud and dense molecular gas are  the sites of new star formation, these sources
are possibly young protostars in their early evolutionary stages. 
In order to study the star formation activity of the region, it is necessary to 
characterize and discuss the nature  of these sources. 

\subsection{Identification of Young Stellar Objects}
\begin{figure}
\centering
\resizebox{8.5cm}{8.5cm}{\includegraphics{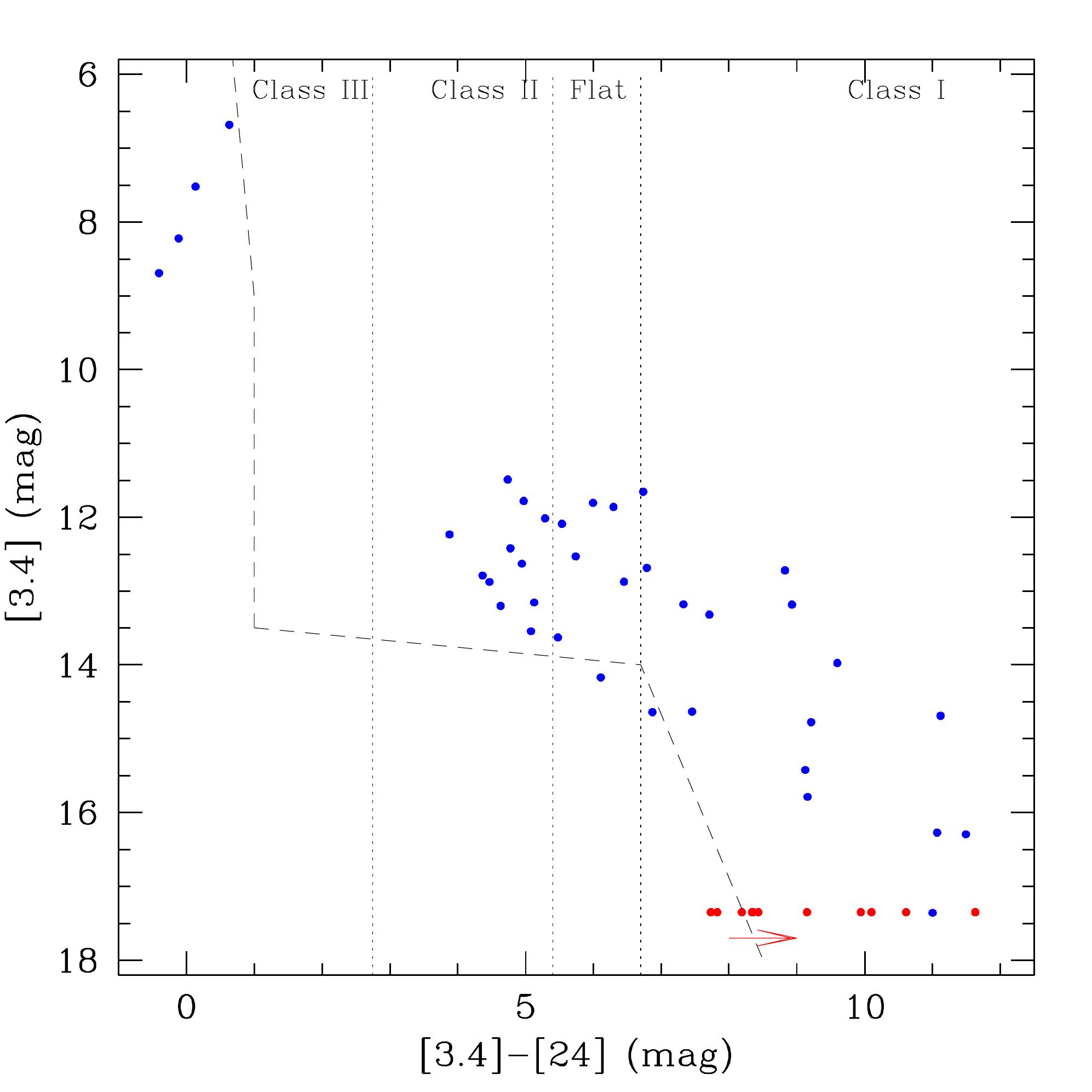}}
\caption{The [3.4] - [24] vs. [3.4] colour-magnitude diagram  for the detected MIPS 24 $\mu$m sources. 
The dotted lines separate  the regions of Class I, Flat-spectrum, Class II, and Class III objects.
The dashed lines denote the dividing line between the region occupied by contaminated sources (galaxies and disk-less stars) and YSOs \citep[see also][]{rebul11}. 
The sources for which we do not have WISE 3.4 $\mu$m detection are represented in red. For these sources, we 
consider the magnitude of the  faintest  3.4 $\mu$m  counterpart of   our 24 $\mu$m detections as an upper-limit. 
The arrow represents the direction of their colours, thus they are likely to be Class I YSOs.}
\label{NGC6823_field.ps}
\end{figure}
The  circumstellar dust emission from the disk and infalling envelope  of young stars  gradually
disappears with time as a function of their evolutionary phases.  
 A number of
classification methods are employed in the literature to classify evolutionary phases of YSOs; often used
are the near- to mid-infrared spectral index $\alpha$ \citep{lada87}, 
the ratio of submillimeter to bolometric luminosity $L_{submm}/L_{bol}$ \citep[e.g.,][]{and00}, and the
bolometric temperature $T_{bol}$  \citep[e.g.,][]{myers98}.
Generally, submillimeter 
luminosity is taken to be the integrated luminosity at wavelengths $\lambda$ $\ge$ 350~ $\mu$m.
The bolometric luminosity is
calculated by integrating the SED   over extensive wavelength coverage, covering far-infrared and millimeter  domain. Similarly, the bolometric temperature is defined as the 
temperature of a black body  with the same mean frequency as the source SED over wide wavelength coverage. Since, the identified point sources typically have flux
measurements  at $\lambda < 24~\mu$m,  accurate determinations
of $L_{bol}$, $T_{bol}$, and $L_{submm}$ is not feasible with the data; thus, in this work, we  classified  the sources based 
on their $\alpha$ values.

Out of the 49 MIPS detections (see Sect. 2), 38  have WISE counterparts in all the four WISE bands. 
 We estimated spectral index ($\alpha = d\,\log(\lambda F_{\lambda})/d\,\log(\lambda)$, where  F$_{\lambda}$ is the flux as a 
function of wavelength, $\lambda$) for these 38  sources  from a linear fit to the fluxes in the range 3.4  to 24 $\mu$m 
(four WISE bands and MIPS 24 $\mu$m band), and then classified sources 
as Class I  ($\alpha$ $\geq$ 0.3), Flat-spectrum (0.3 $>$ $\alpha$  $\geq$ -0.3), Class II (-0.3 $>$ $\alpha$ $\geq$ -1.6), and  Class III 
($\alpha$  $<$ -1.6) YSOs following \citet{evans09}. Using the above approach, we find that 
out of the 38 sources, 16, 7, 11, and 4 sources are  Class I, Flat-spectrum, Class II, and  
Class III YSOs, respectively. The $\alpha$ values of the 38 sources are also tabulated in Table 1.
 Here we would like to mention that, though $\alpha$ is one of the most commonly used methods for the 
classification of YSOs, it is  highly susceptible to disk geometry and source inclination \citep[e.g.,][]{robi06,crapsi08}. 
Also in this approach the distinction of the Class 0 YSOs from the Class I is not possible as 
no well-defined $\alpha$ criteria exists for Class 0 sources. This is 
because, Class~0 spectrum has lowest flux densities from $1.25 - 24~ \mu$m as expected for deeply embedded sources 
with  massive, extincting envelopes, thus were largely not identified in the mid-IR prior to {\it Spitzer}.
However, it is possible to distinguish them from the Class I sources based on their $L_{submm}/L_{bol}$ and $T_{bol}$  
estimations as discussed in \citet[][]{and00} and \citet[][]{myers98}.

It has been found that  a strong correlation between the  colour and spectral index exists, which provides an 
acceptable proxy to classify YSOs \citep[see, e.g.,][]{rebul07,rebul10}. We used
the [3.4] vs. [3.4] - [24] colour-magnitude  diagram (see Fig. 3) to classify  YSOs \citep[][]{guieu10}. 
In Fig. 3, the regions occupied by Class III/main-sequence (MS), Class II, Flat-spectrum, and Class I objects are indicated. 
 In this diagram, 34 sources  are located in the regions occupied by  Class I, Flat-spectrum, and Class II YSOs, whereas 
4 sources are found  in the zone of disk-less stars. We confirmed the nature of these four sources as 
disk-less stars with SED modeling (see Sect. 3.6); these sources will not be considered as YSOs in the followings.
For 11 MIPS detections,  3.4 $\mu$m counterparts are not available. For these sources we then considered the magnitude of the  faintest 3.4 $\mu$m  counterpart of  
the 24 $\mu$m detections  as an upper limit, in order to determine their positions 
on the [3.4] vs. [3.4] - [24] diagram.
The approximate positions of these 11 sources are shown (red dots) in Fig. 3; they appear to be Class I YSOs.
 Out of these 11 sources, 4 sources  have K-band detection. We find that 
they also fall in the Class I zone in the K vs. K-[24] diagram \citep[e.g.,][]{rebul07}, suggesting that these sources 
are likely to be Class I YSOs. For 4 Class I YSOs, we have 70 $\mu$m flux, we thus used the [24] vs. [24]-[70] 
diagram \citep[e.g.,][]{rebul07} to be more precise about their nature; their positions on the [24] vs. [24]-[70] diagram suggest
that they are indeed Class I YSOs.

In Fig. 3, most of the sources are having [3.4]-[24] colour $>$ 4. A  normal MS star has [3.4]-[24] colour around 0.0.  Thus, a normal MS star would require 
a foreground visual extinction (\av) $\sim$  210 mag to redden up to the location  [3.4]-[24] $\sim$ 4, whereas the mean \av
of the region is $\sim$ 10 mag (see Sect. 3.3), suggesting most of the MIPS sources are most-likely  YSOs with disks and/or envelopes.
However,  the YSO sample can be contaminated by
background dusty active galactic nuclei (AGN) and star-forming galaxies, as
they  have  colours similar to that of YSOs.  Hence, in Fig. 3, we  marked the approximate zones of such sources based
on the {\it Spitzer} Wide-area Infrared Extra-galactic Survey catalog \citep{londse03} observations of the ELAIS N1 extra-galactic field. 
 Two  sources of our sample (i.e., sources with [3.4]-[24] colour between 6 and 7 mag) fall in the extra-galactic zone.  However, these sources 
 are found close to other YSO candidates, and are likely members.
Similar to  galaxies, the colours  of the  
asymptotic giant branch (AGB) stars can also mimic with the colour of YSOs.
 AGB stars show a steep spectral index at long
wavelengths. Based on their SEDs, \citet{robi08a} used a 
criterion  [8.0] - [24] $<$ 2.5 to identify them.
In the absence of IRAC 8.0 $\mu$m data, we are unable to apply \citet{robi08a} criterion
to eliminate such sources. However, based on AGB stars distribution, \citet{robi08a} 
established an empirical relationship that roughly reflects the expected number of AGB stars per square 
degree area on the sky for a given Galactic longitude and latitude. 
Using \citet{robi08a} relationship, we calculated 
that the  AGB star contamination to our YSO sample is likely to be less than  
one.

Above discussions suggest that the contamination of 
non-YSO sources to our MIPS identified YSO sample should be
negligible. We therefore considered all the 45 MIPS identified YSO candidates of different 
classes (27 Class I, 7 Flat-spectrum, and 11 Class II) for further analyses. 

\begin{figure}[ht]
\resizebox{8.5cm}{8.5cm}{\includegraphics[clip=true,angle=0]{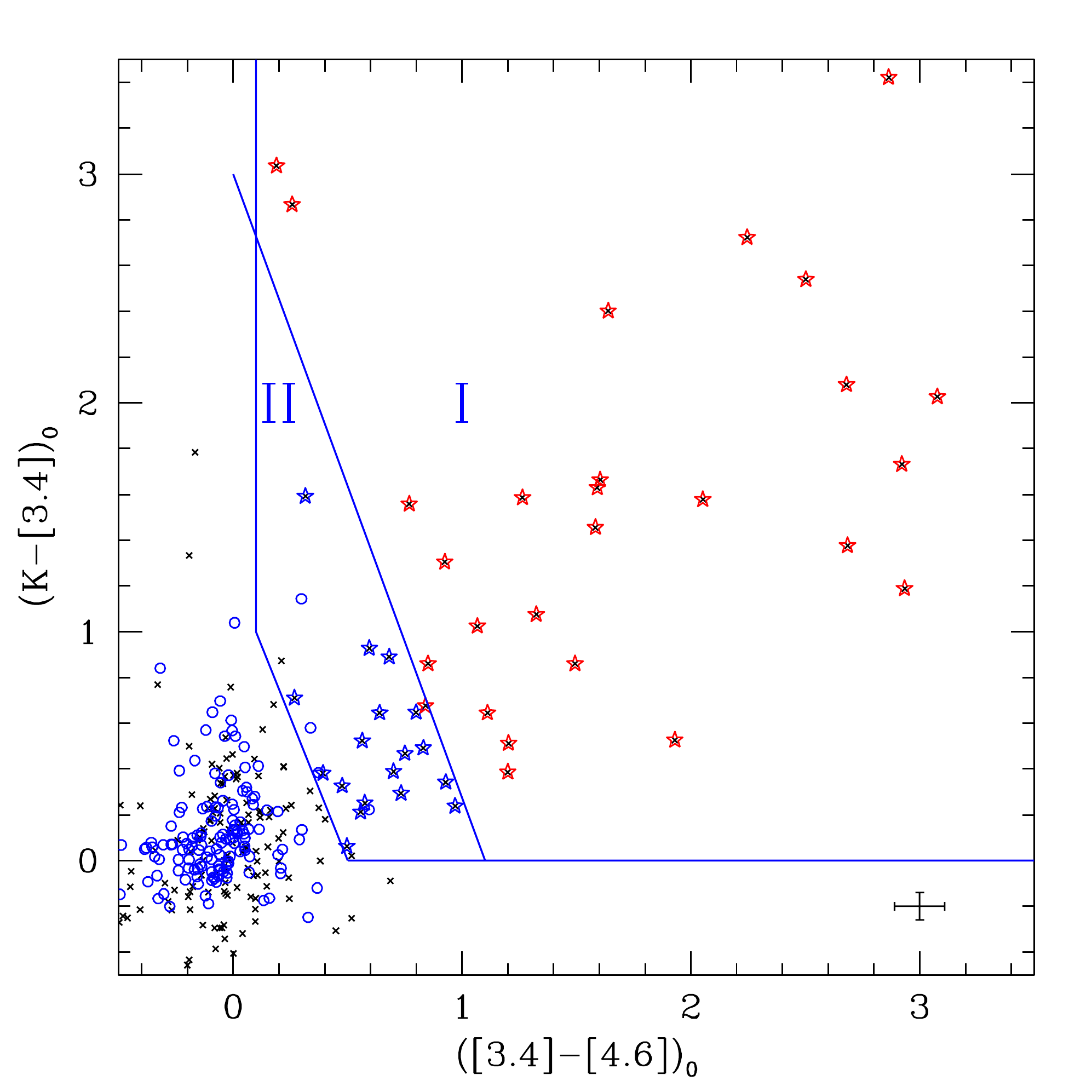}}
\caption{  Intrinsic K - [3.4] vs. [3.4] - [4.6] colour-colour diagram showing the distribution of field stars/MS sources (cross symbols), Class II YSOs (blue stars)
and Class I  YSOs (red stars).
Slanted line indicates the boundary between the Class I and Class II YSOs. The control field sources  are marked with open circles. 
The error bars in the bottom right corner show average errors  in 
the colours.}
\label{NGC6823_field.ps}
\end{figure}
\subsection{Additional YSO candidates}
 The YSO detection using  the above method is primarily limited by 
the 24 $\mu$m detection limit. Thus, we may be missing 
faint YSOs of the  region. 
To overcome this problem, we matched the WISE catalog to
 our  $K$-band point-source catalog, and then selected YSO candidates 
using the intrinsic $K$ - [3.4] vs. [3.4] - [4.6] 
colour–colour diagram as suggested by \citet[][]{koenig12}. 
The  intrinsic $K$ - [3.4] vs. [3.4] - [4.6] diagram is shown in Fig. 4, and  the zones of Class II and Class I YSOs are also marked. 
To construct intrinsic colour–colour diagram, 
we measured the visual extinction   towards individual YSOs  from the H$_2$ column density map (see Sect. 3.5), and 
then used extinction laws of \citet{bohlin78} and  \citet{fla07} to compute their dereddened $K$ - [3.4] and [3.4]-[4.6] colours. 
We then used \citet[][]{koenig12} criterion  to select the Class I and Class II YSOs from the intrinsic $K$ - [3.4] vs. [3.4] - [4.6] diagram. 
This helped us to identify 17 additional YSO candidates,  which include 8 Class II and 9 Class I YSOs.
To quantify  the contamination of other sources to this selection, we selected a control field devoid of CO gas and located
outside the filamentary area, but within the field of view of our NIR observations. We then looked for 
the distribution  of control field sources on $K$ - [3.4] vs. [3.4] - [4.6] diagram. These sources are also shown in Fig. 4 (marked with open circles). 
Their distributions suggest that the majority of them are field stars, indicating that the contamination 
of other sources to the YSOs sample, selected based on $K$ -[3.4] vs. [3.4]-[4.6] diagram, should only be a few.

In summary, we have identified 62 YSO candidates in the IRDC G192.76+00.10 region with excess IR emission. Of these, 43 have excess consistent with Class I plus Flat-spectrum 
YSOs, and 19 have excess consistent 
with Class II  YSOs. We note that even though the near to mid-infrared colours are very useful to identify YSOs, the YSO selection and classification
can be biased due to disk geometry and/or source inclination along the line of sight. 
Infrared spectroscopic observations  provide many useful indicators  that are helpful for  the YSO confirmation and  disentanglement of deeply embedded protostars
from the evolved YSOs  \citep[e.g.,][]{reach07,spe08, carspi08,conn10}, however  foreground 
absorption and edge-on disks can still confuse  the YSOs classification \citep[e.g.,][]{ponto05}. 
Nevertheless, it has been found that the  YSO selection 
scheme based on photometric colours provides a good representation of YSOs 
identification in star-forming complexes. As an example, \citet{spe08} with spectroscopic 
study of Chameleon II cloud found that 96\% of the YSOs identified in {\it Spitzer} c2d Legacy survey \citep{eva03} using 
photometric colours were true members.

\begin{figure}[ht]
\resizebox{8.5cm}{6.5cm}{\includegraphics[trim=0cm 2cm 1cm 2cm, clip=true,angle=0]{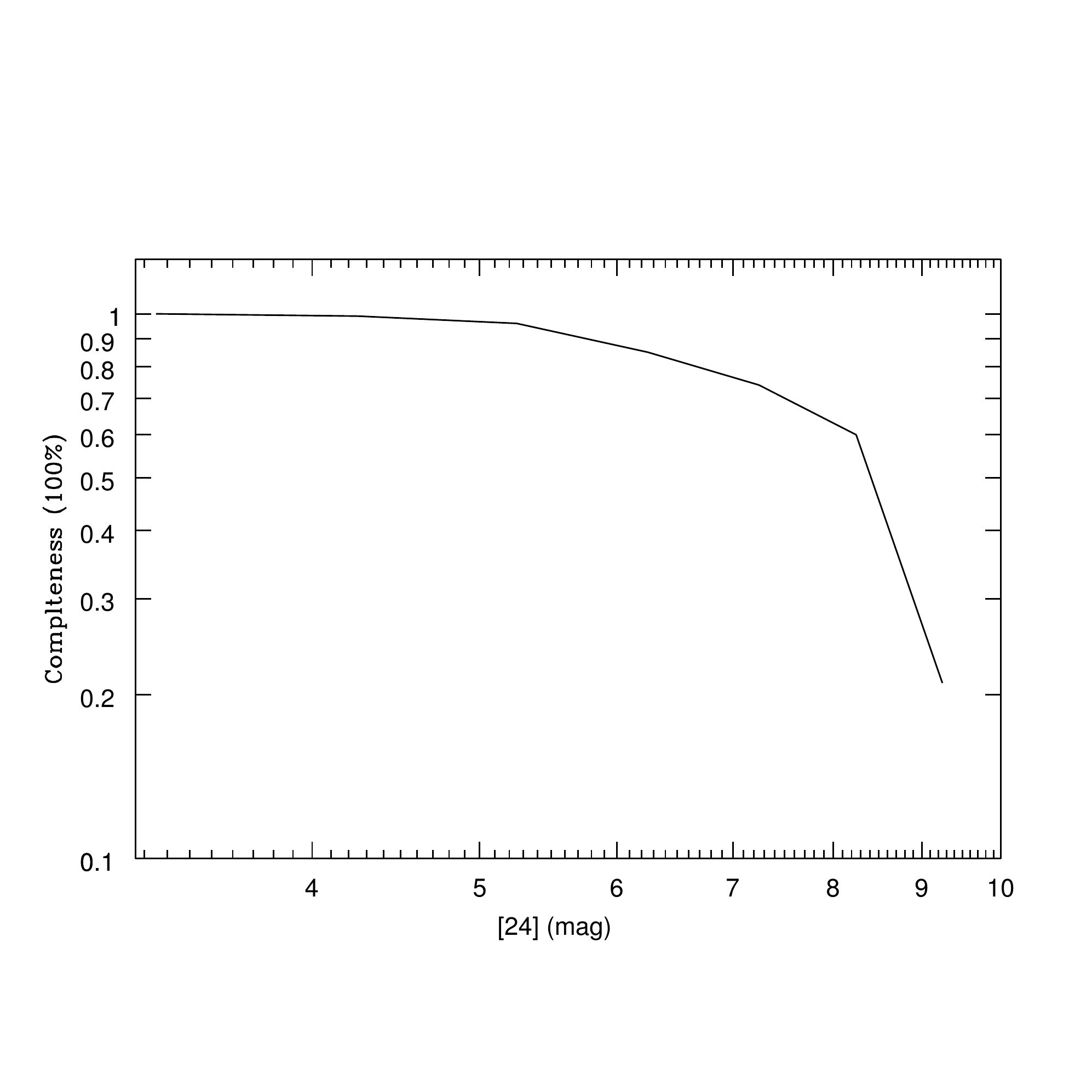}}
\caption{Completeness of the 24 $\mu$m band data for the IRDC G192.76+00.10 region from artificial stars experiment. }
\label{NGC6823_field.ps}
\end{figure}

\begin{figure}[ht]
\resizebox{8.5cm}{6.0cm}{\includegraphics[ clip=true,angle=0]{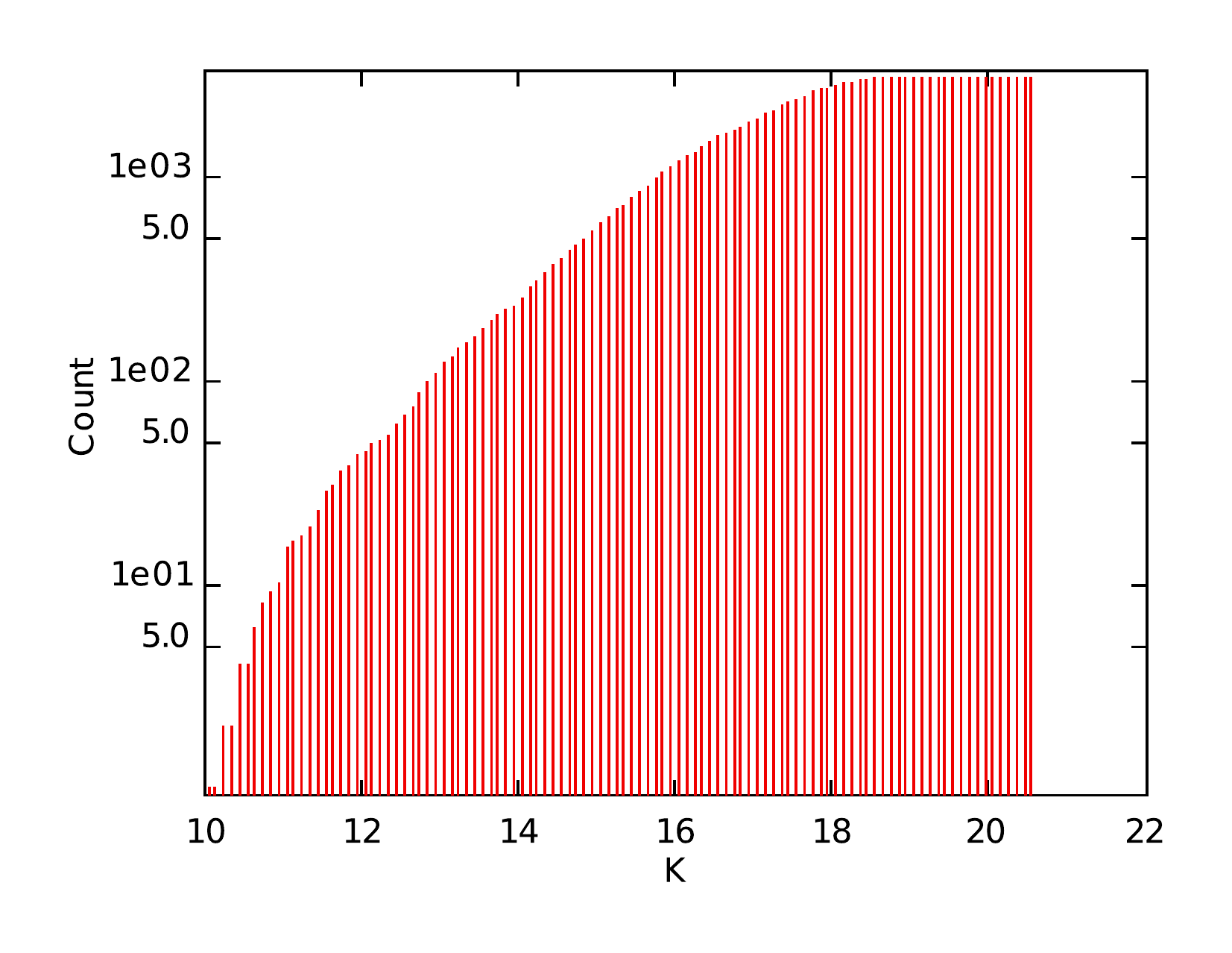}}
\caption{The cumulative distribution of all the K-band sources observed towards the IRDC G192.76+00.10 region. }
\label{NGC6823_field.ps}
\end{figure}

\subsection{Completeness limit}
The identification of YSOs in a star-forming region  using multi-band data sets is  a strong function of the  
band-pass sensitivities and extinction of the star-forming region. Our primary criterion to select YSOs 
in the present work is the detection in MIPS 24 $\mu$m band. Although the 5$\sigma$ detection limit of our 24 $\mu$m catalog is $\sim$
9.6 mag,  the completeness limit  of the catalog is  $\sim$ 6.7 mag.
We calculated completeness limit of our 24 $\mu$m catalog  using  artificial
stars experiment. In this approach, we insert  artificial stars of different magnitudes to the 24 $\mu$m image 
and then perform their detection and phototmetry to recover them. Figure 5 shows  the completeness fraction of artificial stars 
versus corresponding 24 $\mu$m magnitudes. 
This suggests that our 24 $\mu$m detection is 
80\% complete up to $\sim$ 6.7 mag.
Using SED models of \citet{robi07}, we roughly estimated that our 24 $\mu$m data is  actually more complete towards the high-mass ends. 
The additional YSOs obtained using the $K$ -[3.4] vs. [3.4]-[4.6] colour-colour diagram helped us to recover fainter 
YSOs. The WISE catalog is  95\% complete at [3.4] = 16.9 mag and [4.6] = 15.5 mag \citep{cutri12}.
We found that  90\% of the  3.4 $\mu$m and 4.6 $\mu$m sources have a K-band counterpart, thus the majority of the 
WISE sources have been detected in our K-band. We estimated the completeness limit of our K-band catalog by 
measuring the magnitude at which the cumulative logarithmic 
distribution of sources as a function of magnitude  
departs from a linear slope and begins to turn over.
Based on this distribution (see Fig. 6), we arrived that our K-band  catalog is largely complete 
down to K $\sim$ 16.5 mag. We tested this approach to our 24 $\mu$m detections, and the  resulting
completeness limit turns out to be $\sim$ 7 mag, 
comparable to the completeness limit found using artificial stars experiment.

One of the important properties 
of YSOs is the intrinsic luminosity. 
Because of extinction and/or youthfulness of the sources, we detected only $\sim$ 50\% of the total YSOs in the J-band in comparison to the K-band.
We thus used  
K-band magnitudes to estimate the luminosities of the YSOs.
K-band can be  affected by the excess emission from the circumstellar material. 
To  account for such excess emission, we used  average excess emission at K-band found
in  T-Tauri stars. \citet[][]{mey97} found that T-Tauri  
stars typically have a K-band excess between 0.1 and 1.1 mag, with a median value $\sim$ 0.6 mag. 
Considering  0.6 mag as the excess emission at the K-band for all our YSOs, and using the 
evolutionary model of \citet[][]{bar03} for an age of 1 Myr (see Sect. 4.1) at 2.5 kpc and at \av $\sim$ 10 mag, we estimated that our K-band completeness level (i.e., K $\sim$16.5 mag) 
corresponds  to a 0.14 \lsun~or 0.15 \msun~star. Thus, our YSO sample is expected to 
be largely complete above 0.14 \lsun~or~ 0.15 \msun. However, for very low-mass stars the adopted 
excess emission value of 0.6 mag may not be valid, as  in the substellar regime inner part of such disks
might not emit significant excess emission at K-band. Thus, if we consider the 
emission at K-band only from stellar
photospheres for low mass YSOs, our completeness limit would be $\sim$ 0.2 \lsun~or~ 0.2 \msun.

In  case of YSOs, it is  difficult to derive their true intrinsic luminosities  
without spectroscopic observations.  Therefore in the absence of
short wavelength fluxes or spectroscopic observations, the above 
approach appears to be reasonable to have approximate  luminosity (or mass) of such objects. 
Since the effect of excess emission in J-band is minimum, we also estimated
masses of YSOs using J-band magnitudes. For sources detected in both  J and 
K bands, we found that, for majority of our YSO  the difference in mass estimations is within $\sim$ 40\%.  
Considering tentatively that this is the uncertainty associated in 
the mass estimations, the absolute K-band 
magnitudes of majority of the YSO candidates  
suggest that the region is mainly composed of low-mass (< 2\msun) YSOs with a mean mass $\sim$ 0.3 \msun.

For the IRDC G192.76+00.10 region IRAC four-band data sets are unavailable. Thus to  
quantify the extra percentage of Class II and Class I YSOs, we might have detected if IRAC observation had been performed, 
we did analysis of a nearby cloud ``G192.75-0.08".
The G192.75-0.08 cloud  is located within the S254-S258 complex and has been 
studied by \citet{chava08} at {\it IRAC} bands. 
We have chosen the G192.75-0.08 cloud as it is devoid of bright PAH emission 
like the IRDC G192.76+00.10 region, and also at comparable extinction. To compare, 
we first selected  YSOs of the G192.75-0.08 region  using the [3.4] vs. [3.4]-[24] 
diagram as described in Sect. 3.3, and then compared the statistics  with the YSOs identified by \citet[][]{chava08}. 
We considered  only those YSOs  from the \citet[][]{chava08} catalog whose
spectral index is greater than -1.6, as done in the present work. 
This statistical comparison yields, if IRAC observation had been performed for 
the IRDC G192.76+00.10 region, we possibly would have 
detected 35\% (or 15) more YSOs of Class II, Flat-spectrum, and Class I nature. 
However, it is worth noting that we have already added 
17 extra YSOs to our MIPS identified YSO sample using 
$K$ - [3.4] vs. [3.4] - [4.6] diagram. 
This analysis  suggests that our YSOs  statistics
in terms of number is comparable, if Class II and Class I YSOs had been selected  based 
on only IRAC observations.

\subsection{Spatial distribution and separation of YSOs}
\begin{figure}
\centering
\resizebox{8.5cm}{6.5cm}{\includegraphics[trim=0cm 3cm 0cm 3cm, clip=true]{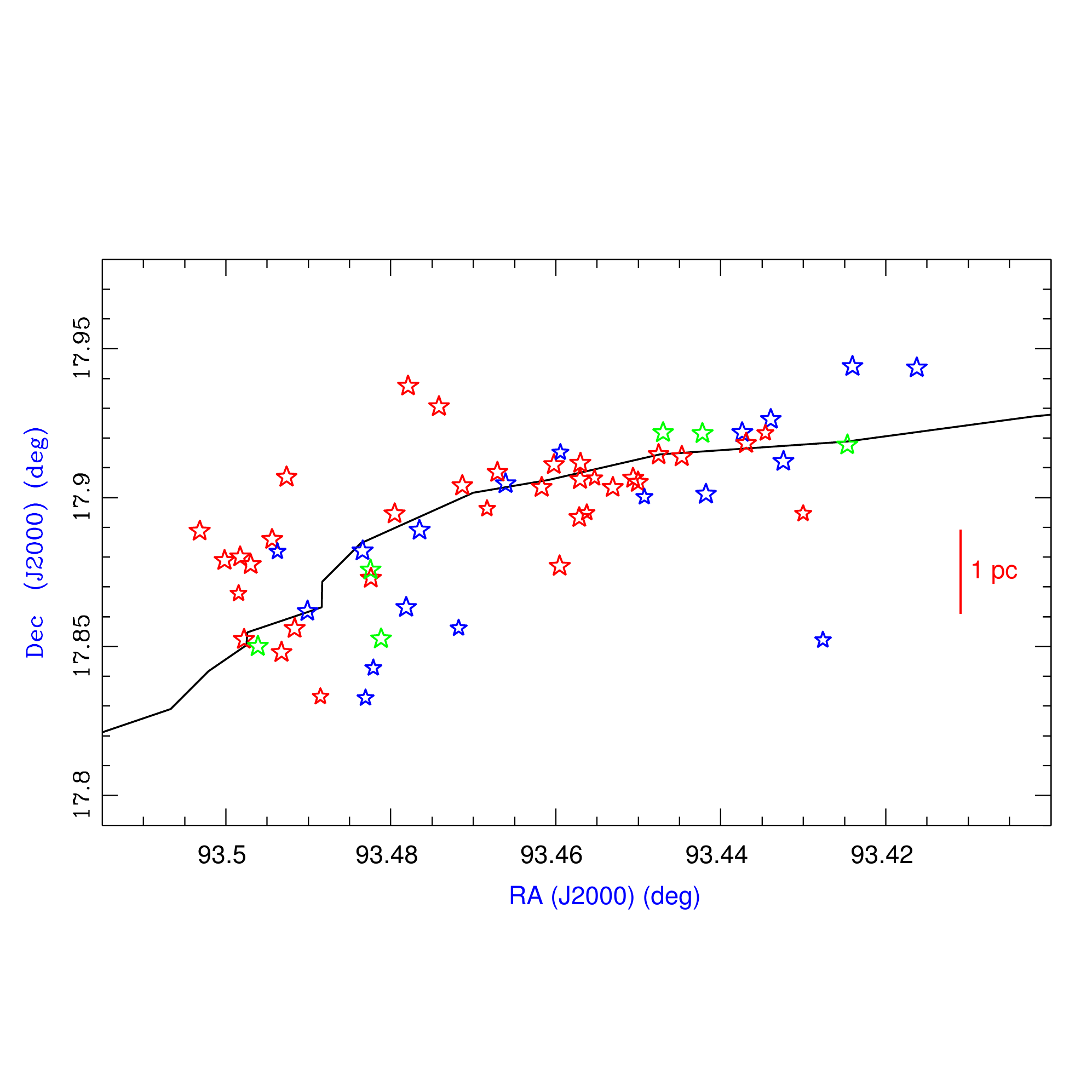}}
\caption{The spatial distribution of Class I (red stars), Flat-spectrum-spectrum (green stars), and Class II (blue stars) YSOs. The continuous 
line running from south-east to north-west  marks the 
highest H$_2$  column density line of the filament along its long axis.  
}
\label{NGC6823_field.ps}
\end{figure}
The spatial distribution of young stars is an useful tool to construct star formation scenario of molecular cloud complexes
\citep[e.g.,][]{kumar07,yun08,povich09,marchi11,jose12,pandey13,mallick13,gou14,neelam14,massi15}. In order to understand the star formation scenario of the 
IRDC G192.76+00.10 region, we have plotted the spatial distribution of its  YSO content in Fig.  7. 
In Fig.7, the continuous solid line is the highest H$_{2}$ column density line of the filament along its major axis, and the red, green and blue stars
are the Class I, Flat-spectrum-spectrum and Class II YSOs, respectively.
In order to define the column density line, we first created H$_{2}$  column density map from the $^{13}$CO column density map, and  
then traced its crests by applying a spatial filtering to its intensity distribution.
The H$_2$ column density map was constructed from the  $^{13}\mathrm{CO}$ column density map using the 
relations $n(^{12} \mathrm{CO} ) / n(^{13}\mathrm{CO}) = 45$ and $n(^{12}\mathrm{CO}) / n(\mathrm{H}_2) = 
8 \times 10^{-5}$ \citep[and references therein]{chava08}. In
 Fig.  7, one can notice that the majority of the YSO candidates are aligned closely with  the highest column density line of the filament. 
This close alignment  strongly suggests that the formation of the YSOs seems to be continuing in the dense regions of this filamentary cloud along its 
long axis. We find the mean H$_{2}$ column density of the highest density line is  $\sim$ 1.1 $\times$ 10 $^{22}$  cm$^{-2}$, 
which corresponds to a mean \av $\sim$ 10 mag \citep[using \nht = 0.94 $\times$ 10 $^{21}$  \av cm$^{-2}$ mag$^{-1}$;][]{bohlin78}. 
This is in agreement with the value 
(i.e,  \av  $\sim$ 8 mag) expected for filaments above which they are super-critical and capable of forming stars 
\citep{andre10,andre11}.  We find that the projected distances of the identified YSOs from the highest column density line  range from 0.05 to 1.5 pc,  with $\sim$ 70\% falling within 0.4 pc. 
This narrow separation is a strong indication of the fact that the core formation in this filamentary dark 
cloud is not random; their formation has occurred mainly along
the long axis of the filament. Also these YSOs are probably formed recently, because they 
possibly did not have  enough time  to move away from their birth locations. 
The typical velocity dispersion seen in relatively evolved clusters and association is 
roughly a few \kms~ \citep[][]{madsen02}. In a recent work,  \citet{foster15} showed that the 1-2 Myr old Class II stars of the NGC 1333 star forming region 
have an intrinsic velocity dispersion of $\sim$ 1 \kms, and the  average velocity dispersion of the dense cores of the region is around 0.5 km/s. Thus, the possibility
that few evolved YSOs might have migrated from their birth locations exists, however,
in general, protostars are the coldest objects, so expected to be embedded within the cores. 
For example, in OMC-3 region, \citet{taka13} observed eight out of sixteen {\it Spitzer} sources  are associated with the 
SMA continuum cores  (observed with a spatial resolutions $\sim$ 4.5\arcs, comparable to MIPS), which are identified as protostars, and 
six out of sixteen are categorized as T-Tauri stars having no counterparts in the SMA continuum emission. In the  
IRDC G192+00.10 region, most of the identified YSOs are  protostellar in nature. If we assume that these  protostars have been moved 
from their birth locations with a velocity close to  0.5 km/s,  their current locations ($\leq$ 0.4 pc) from the highest 
density line indicate that the formation of these protostars might have occurred in  the  last few $\times$ 10$^5$ yr.

In young filaments, the distribution of  cores  is thought to represent a possible preferred length-scale of the filament  fragmentation 
\citep[e.g.,][]{munoz07,jackson10,hacer11,miet13,taka13}. 
Since cores are the precursors to protostars, the  distribution and separation of young protostars is a good proxy to test the filament fragmentation theory. 
To do so, we determined the distribution of projected  separation and projected nearest-neighbor (NN) separation among the YSOs as shown in Fig. 8. 
Although the projected separation among the YSOs varies in the range 0.1-6.0 pc,  the NN separation of the majority of the YSOs shows a very 
narrow range (i.e., 0.1-0.5 pc) with a median $\sim$ 0.19 pc. 
 Identification  and distribution of cores at longer wavelengths are often limited by spatial resolution and sensitivity of instruments. 
So far only a very limited number of millimeter and submillimeter observational studies on filaments have been made
which achieve an angular resolution comparable to, or better than, existing IR data (i.e.,
a few arcsec resolution). For example,  with the high angular
resolution  observations a separation of cores $\sim$ 0.19 pc \citep[][beam $\sim$ 1.2\arcsec]{zhang09},  
$\sim$ 0.18 pc \citep[][beam $\sim$2\arcsec]{ragan15}, and $\sim$ 0.25 pc \citep[][beam $\sim$ 4.5\arcsec]{taka13} have been
observed in the filamentary clouds  IRDC G28.34+0.06, G011.11-0.12 and OMC-3, respectively. These results seem to be consistent 
with the median separation of the YSOs observed in the IRDC G192.76+00.10 region.

Theoretically, it is believed  that self-gravitating filaments are unstable to fragmentation and lead to formation of cores \citep[e.g.,][]{naga87,tomi95}.  Thus, it is
tempting to think that the formation of YSOs observed in the IRDC G192.76+00.10 region could
be the result of  the filament fragmentation.
In Sect. 4, we therefore discuss whether or 
not the observed distribution and separation of YSOs carry an imprint of fragmentation and  core formation
of the IRDC G192.76+00.10 filamentary cloud.

\begin{figure}
\centering{
\resizebox{4.4cm}{4.4cm}{\includegraphics{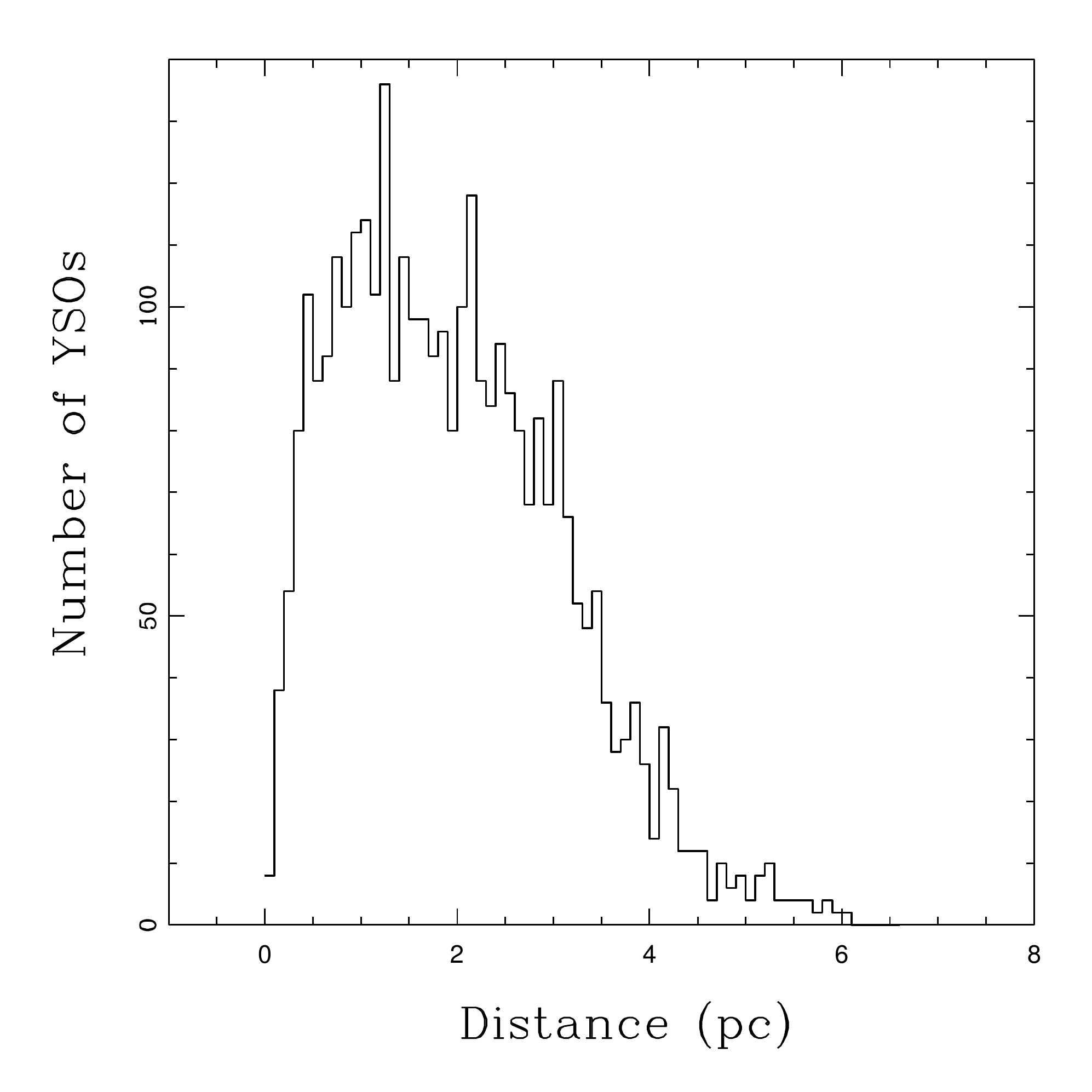}}
\resizebox{4.4cm}{4.4cm}{\includegraphics{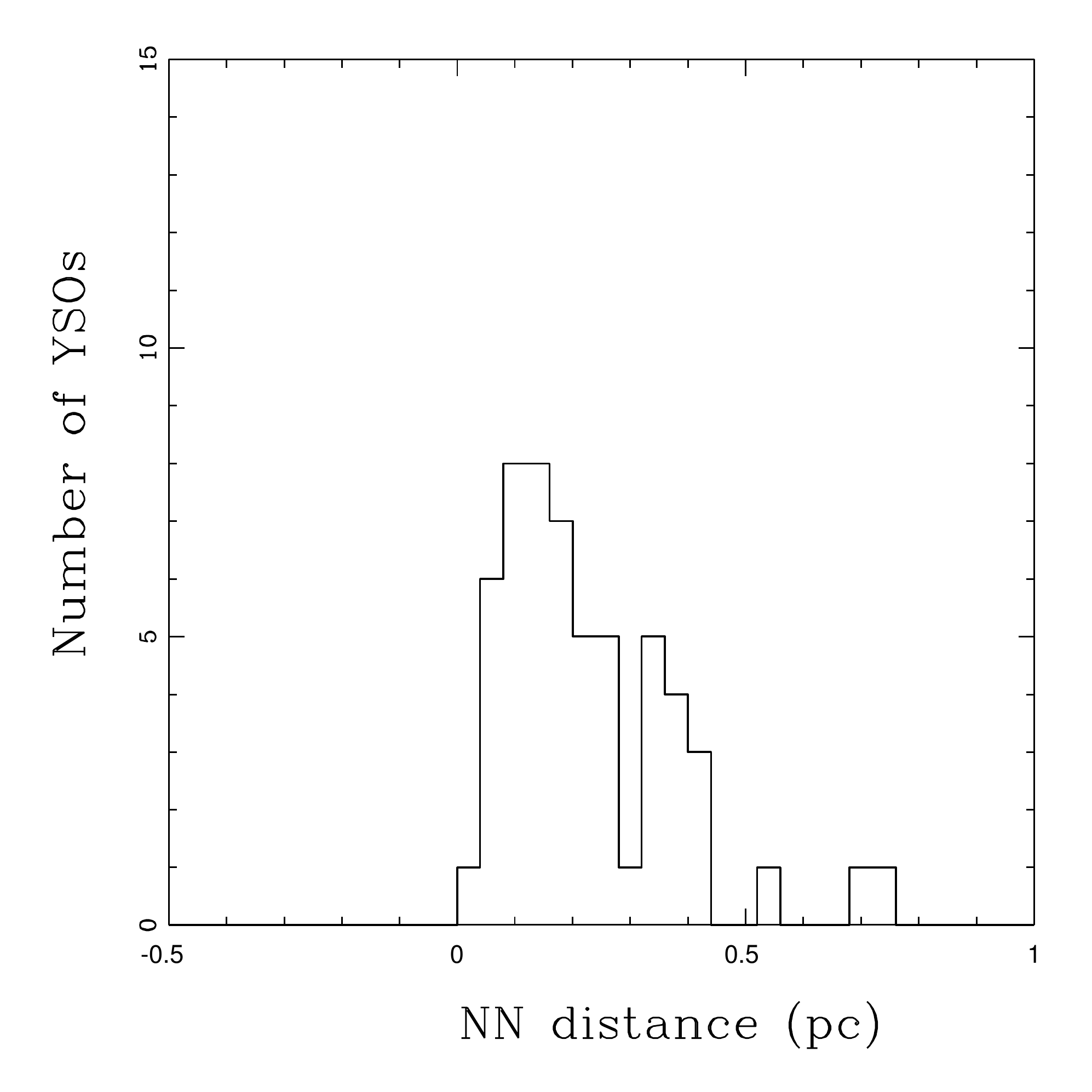}}}
\caption{Left:  Distribution of the projected distances among the sources detected at 24 $\mu$m. Right: 
Distribution of the nearest-neighbor separation among the YSOs.}
\label{NGC6823_field.ps}
\end{figure}

\subsection{Properties of the YSOs}
 In the case of young protostars, it is very difficult to infer their stellar and circumstellar properties  based on photometric observations alone.
Theoretical  models that reproduce the observed SEDs provide good  representations of the underlying source properties.

 To 
get deep insight into the nature  of the detected YSOs, we fitted radiative transfer models of \citet{robi06,robi07}  
to their observed SEDs.  Interpreting SEDs using the radiative transfer code is subject of degeneracy 
and spatially resolved multi-wavelength observations can reduce the degeneracy. 
We thus compiled counterparts of the YSOs  at NIR (J, H, and K; this work), 
IRAC \citep[3.6, 4.5, 5.8, and 8.0 $\mu$m;][]{chava08}, WISE \citep[3.4, 4.6, 12.0, and 22.0 $\mu$m;][]{cutri12}, MIPS (24 $\mu$m 
and 70 $\mu$m; this work) and AKARI \citep[65, 90, 140 and 160 $\mu$m;][]{Yamam09} bands, wherever available. 
We  fit the models to only those sources for which we have flux values at least in five bands in 
the wavelength range from 1 $\mu$m to 24 $\mu$m.  While fitting  models to the observed fluxes we adopt the following  approaches: 
i) when IRAC 3.6 $\mu$m and 4.5 $\mu$m fluxes were available, the WISE fluxes at 3.4 $\mu$m and 4.6 $\mu$m
were not used; ii) for sources not detected at 70 $\mu$m or 65 $\mu$m, we set the
upper limits at these bands by assigning the minimum 70 $\mu$m or 65 $\mu$m detection flux obtained for the point sources detected in the 
region. The 70 $\mu$m flux upper limit is assigned to only  those sources that are not detected in the 
70 $\mu$m observed area. Sources that are located outside the 70 $\mu$m observed area and 
not detected in 65 $\mu$m band, we  assigned the minimum 65 $\mu$m flux obtained for the point sources 
observed in the region from the AKARI survey; 
iii) we scaled the SED models to the distance of the filament (i.e., 2.5 $\pm$ 0.2 kpc; \citet{chava08}) and allowed  
a maximum \av value determined by 
tracing back the YSOs' current location on J vs. J-H or K vs. H-K or 
K vs. K-[3.4]   diagram to the intrinsic 
dwarf locus along the reddening vector \citep[see e.g.,][]{samal10}.

 Figure  9 shows the SEDs of 38 sources that satisfy our five data point 
flux criteria.  Due to lack of optical, far-infrared, and millimeter data points, it is quite 
apparent that the SED  models show high degree of degeneracy; nonetheless, the SEDs 
clearly indicate that the majority possess IR-excess emission, possibly 
emission from circumstellar disk and envelope. While looking  at Fig. 9, we found that the observed SEDs 
of four  sources (IDs. 4, 18, 19 and 29 in Fig. 9) are fitted  well by reddened stellar photosphere models. We find that they correspond
to those four sources that have been rejected as YSO candidates in Section 3.2 on the basis of their location on 
the [3.4] vs. [3.4]-[24] diagram (i.e., sources with [3.4]-[24] colour $<$ 1 mag). 
Thus, the SED models confirm our previous results that the bright 3.4 $\mu$m sources found in the 
field-star/MS zone of [3.4] vs. [3.4]-[24] diagram are indeed  disk-less stars.

 It is not possible
to characterize all the  SED parameters from the models due to limited observational data points. 
However, as discussed in \citet{robi07}, some 
of the parameters  can still be constrained
depending on the available fluxes. For example, in the present study, for 
majority of the sources, the SED models between 1 $\mu$m to 70 $\mu$m 
represent fairly well the data points, hence,  the disk parameters 
are expected to be better constrained. However, it is worth noting that precise 
determination of disk parameters using SED models, as demonstrated 
by \citet{spe13}, requires data from optical to millimeter 
bands, as well as good knowledge of the physical parameters 
of the central star. 
 Similarly, as discussed in \citet{robi06}, in the case of young protostars, the disk is  expected to be deeply embedded inside the envelope and the 
relative contributions of the disk and envelope to the SED are difficult to disentangle.
Because of the above reasons, in the present case, even though the uncertainty in the disk  parameters 
is expected to be high, it should nonetheless give a good proxy of the nature of the sources. In Table 2,  the disk parameters (disk mass ($M_{\rm disk}$) and 
disk accretion rate ($\dot{M}_{\rm disk}$)) are  tabulated.
Since our SED models are highly degenerate, the
best-fit model is unlikely to give an unique solution. So, 
the tabulated values in Table 2 are the  weighted mean and standard deviation of the physical parameters
obtained from the  models that satisfy $\chi^2 - \chi^2_{\rm min} \leq 2N_{\rm data}$ weighted by e$^{({{-\chi}^2}/2)}$ of each model, where $\chi^2_{\rm min}$ 
is the goodness-of-fit parameter for the best-fit model and $N_{\rm data}$ is the number of input observational
data points \citep[see][]{samal12}. 
From Table 2, we found that the disk masses and  
disk accretion rates of $\sim$ 80\% YSOs are in the range 0.001 - 0.021 \msun~ and 0.01 - 0.95 $\times$ 10$^{-6}$ \msun/yr, 
respectively.  We note that in addition to the limitations outlined above, the absolute 
uncertainties associated with the disk parameters  are possibly in a range of 1-3 orders of magnitude \citep[see][]{robi07}. 
Therefore, we stress that the derived disk parameters must be considered 
as representative values, and should be treated with caution.

Due to several limitations, it is not possible to comment on individual
objects, but if we take  the SED results for statistical purpose,  the derived disk accretion rates  suggest that the 
YSOs of the IRDC G192.76+00.10 region are mainly  low-mass in nature \citep[e.g.,][]{dahm08}, consistent with
the nature of the sources derived from the photometric data.

\begin{figure*}
\centering{
\includegraphics[width=3.6cm] {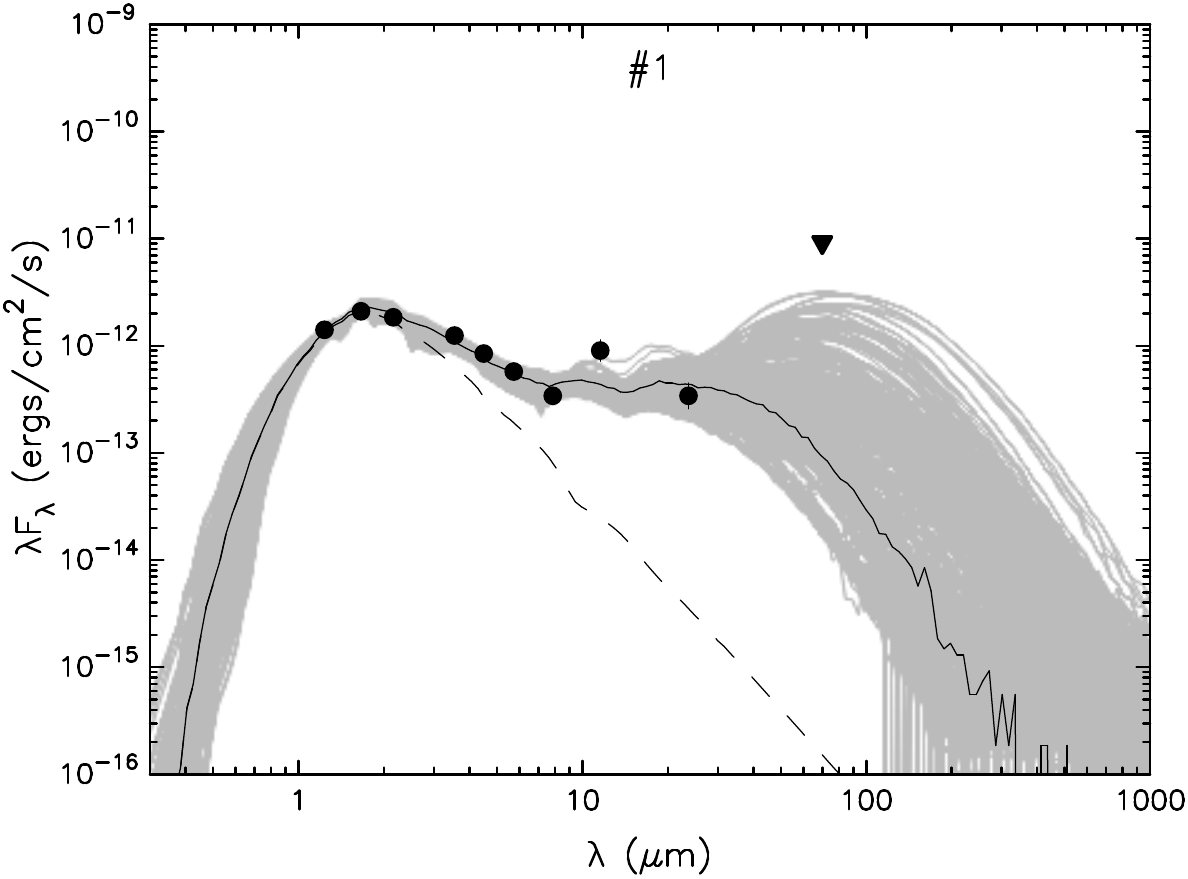}
\includegraphics[width=3.6cm] {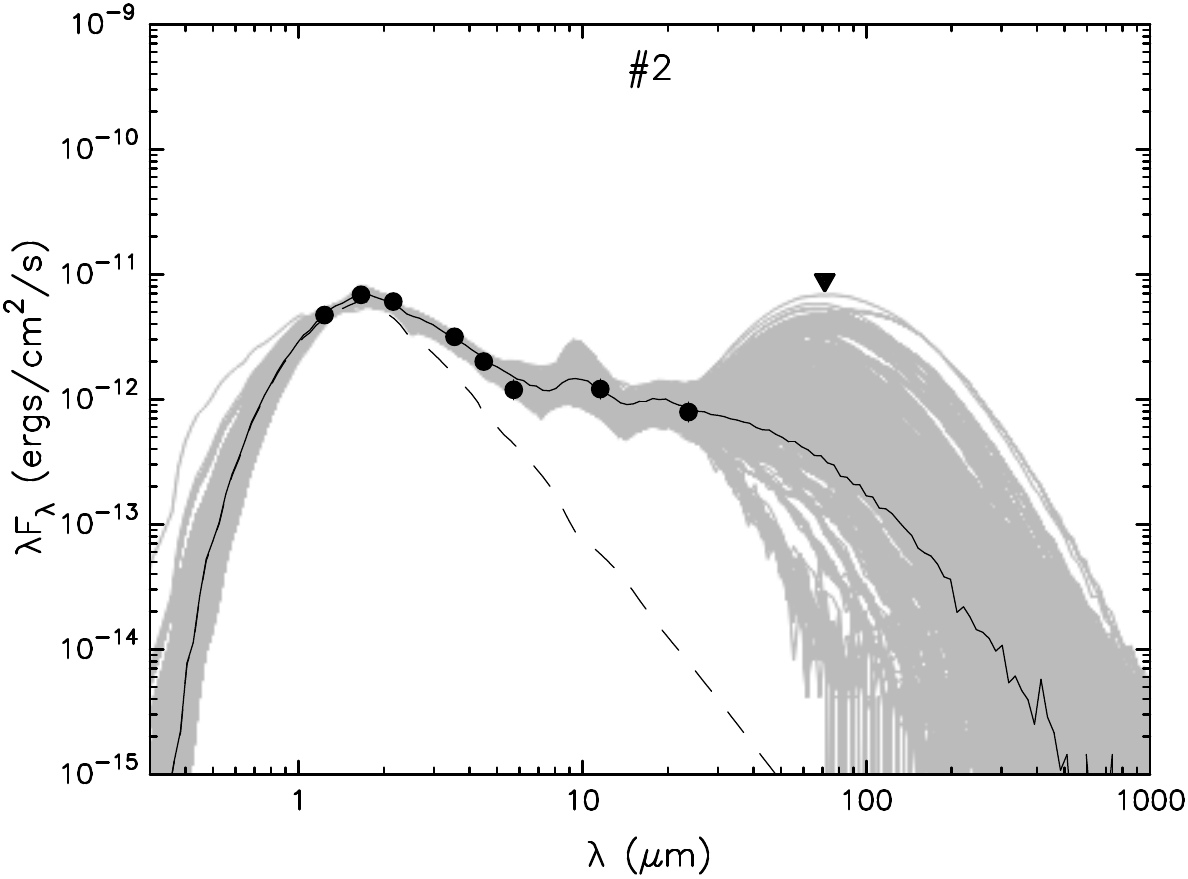}
\includegraphics[width=3.6cm] {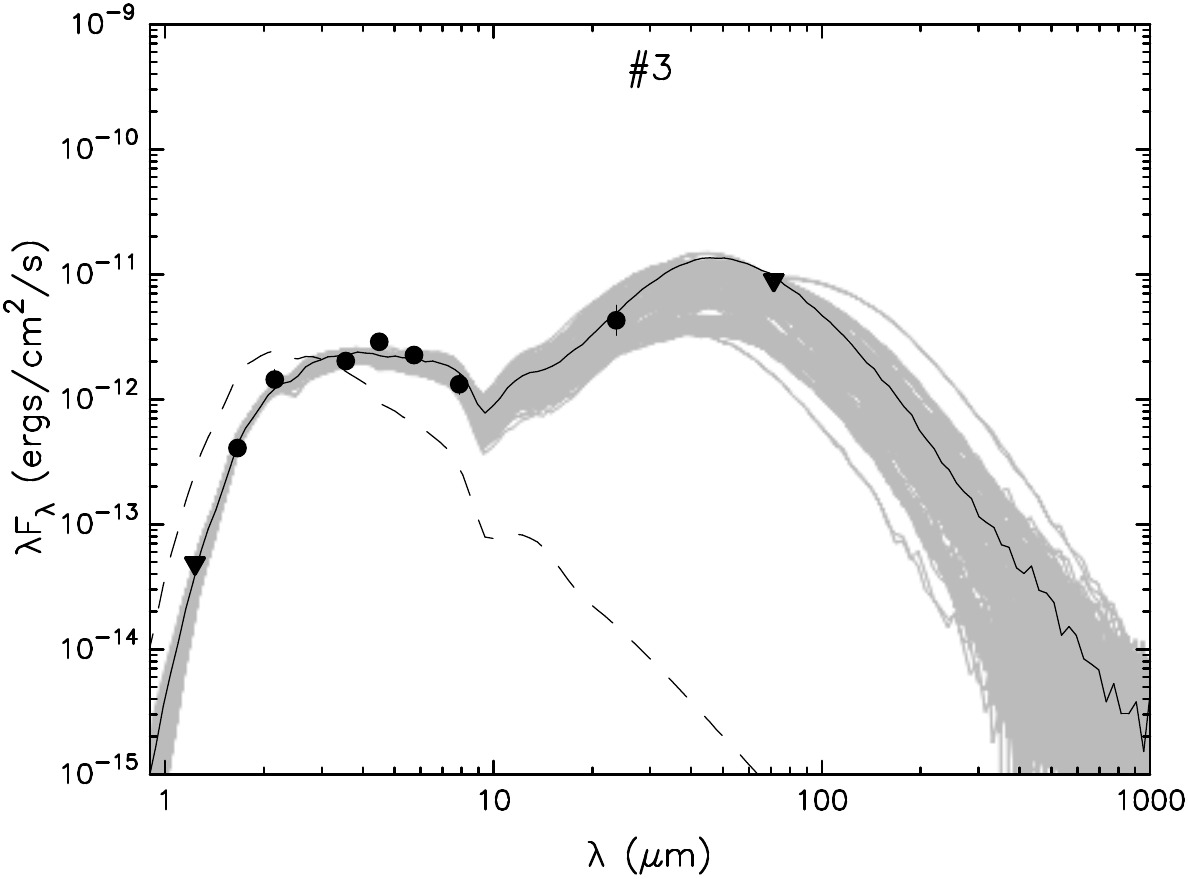}
\includegraphics[width=3.6cm] {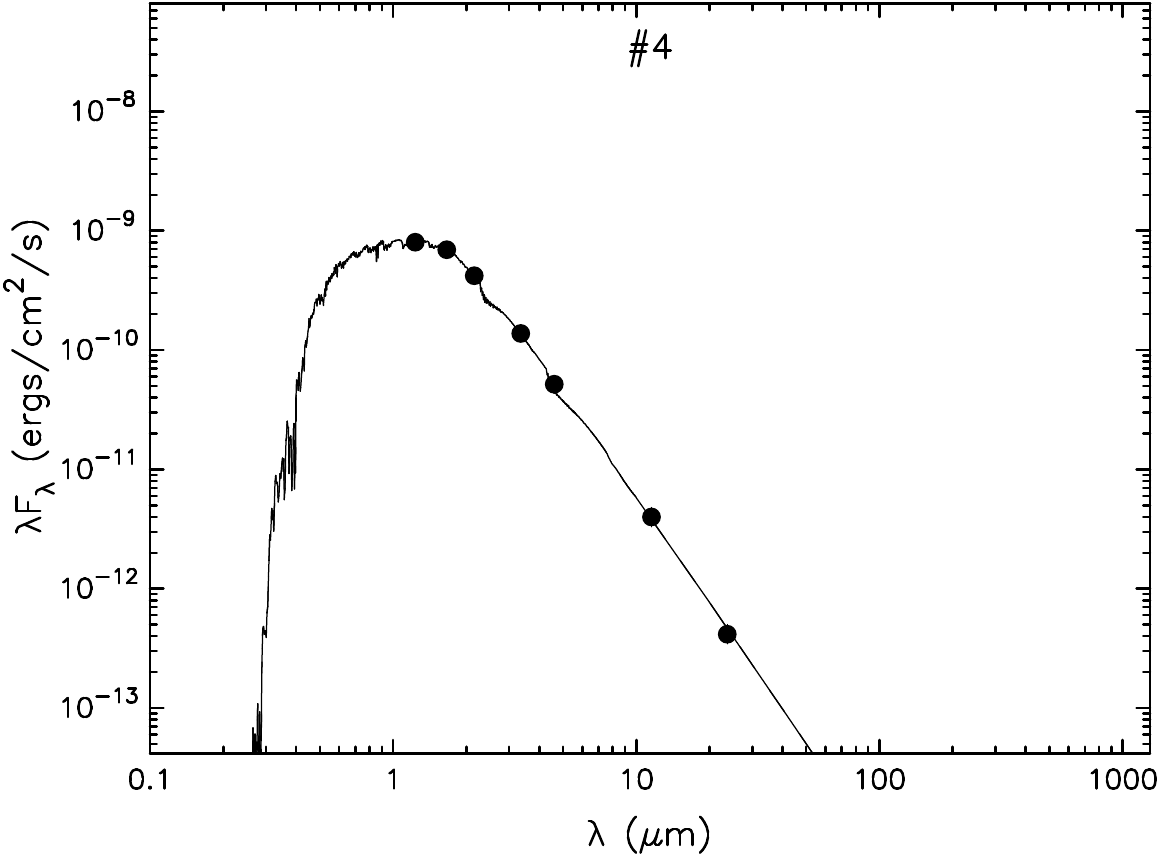}
\includegraphics[width=3.6cm] {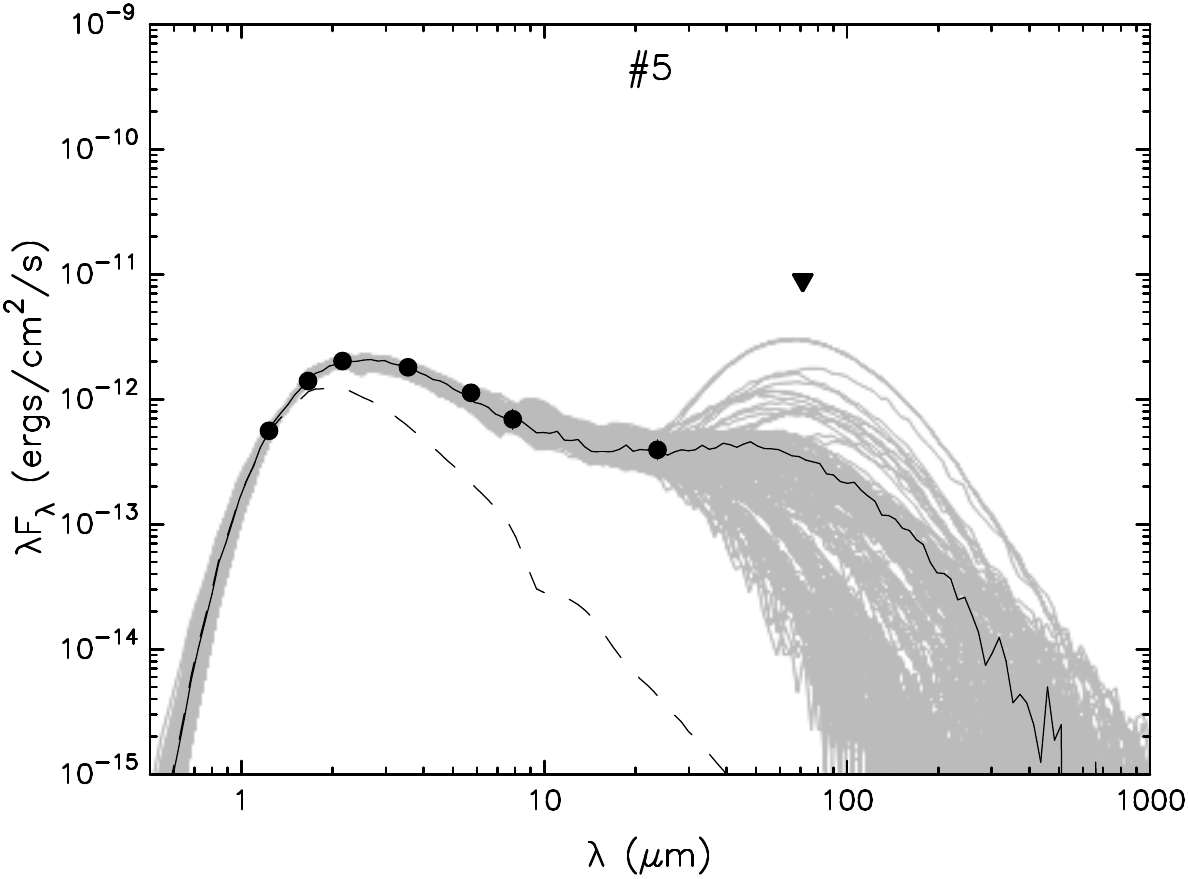}
\includegraphics[width=3.6cm] {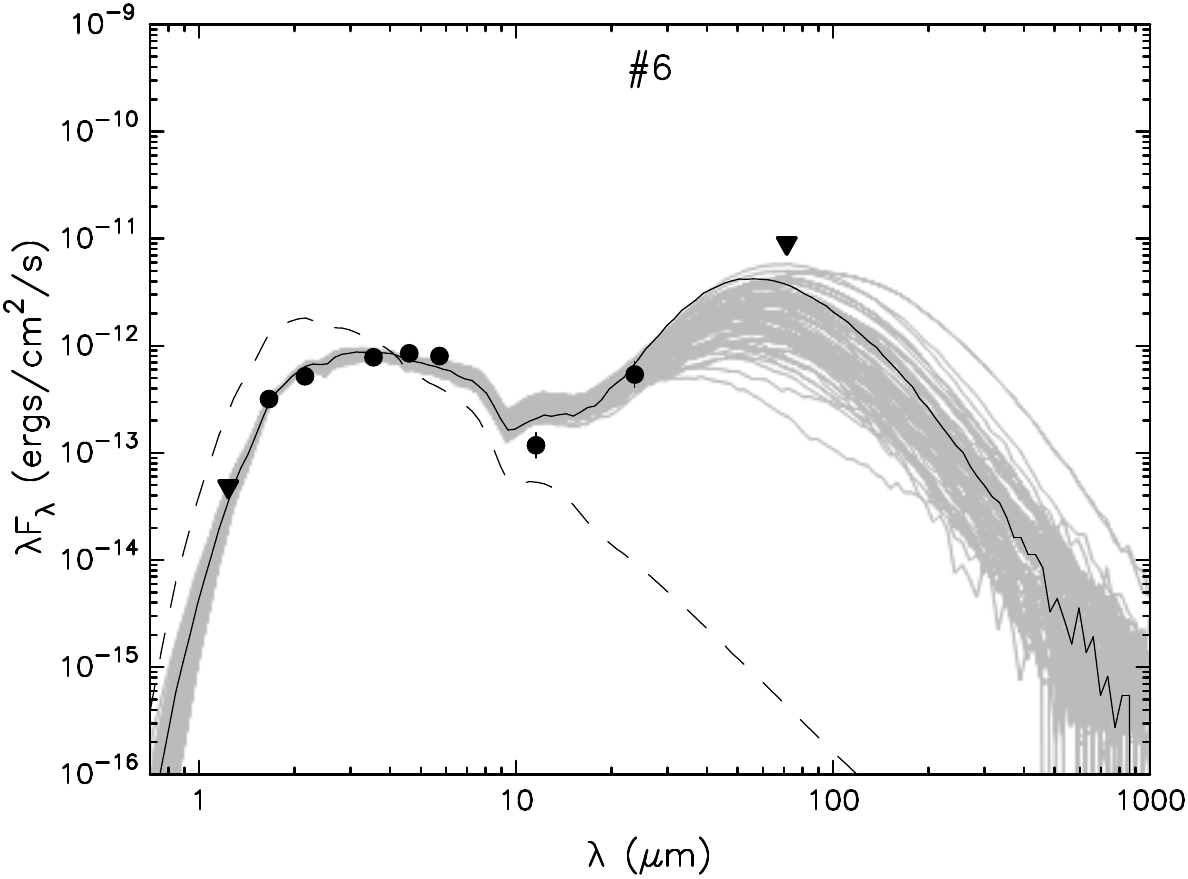}
\includegraphics[width=3.6cm] {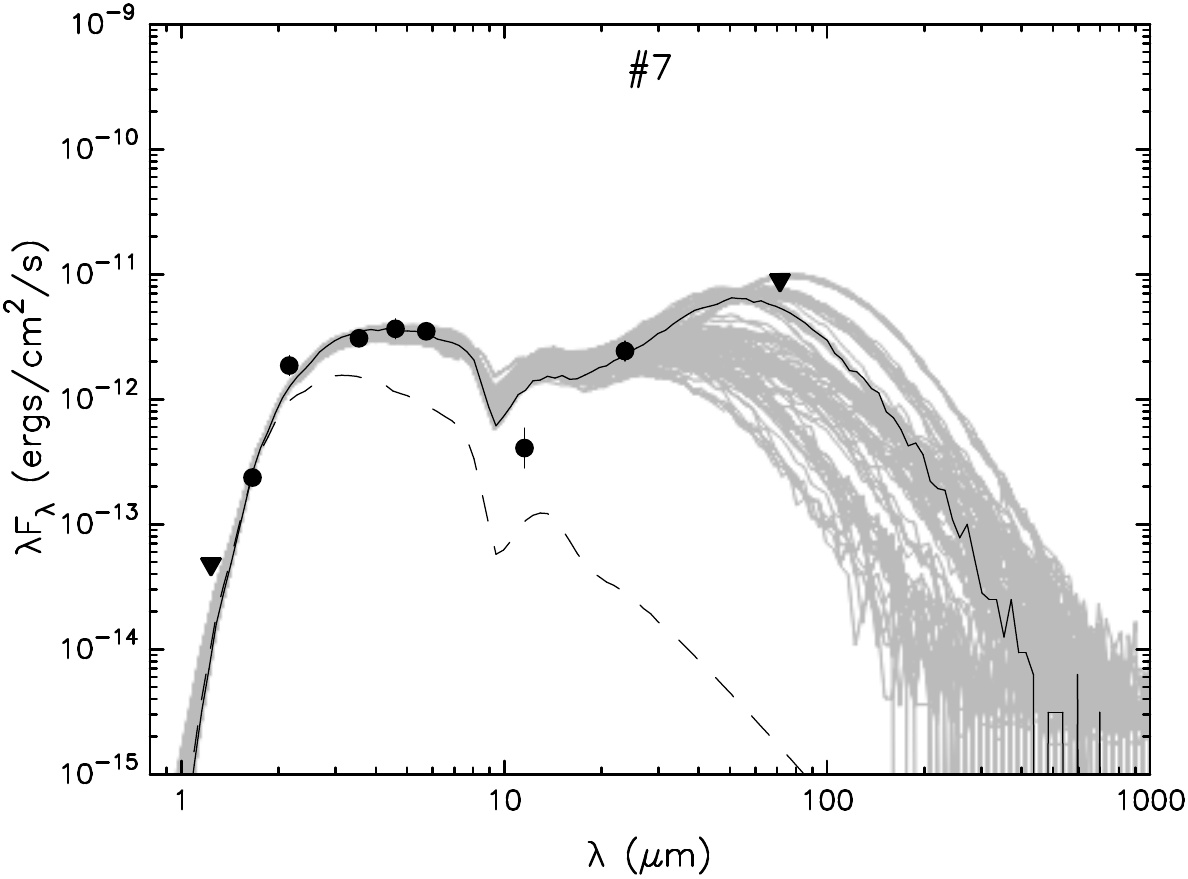}
\includegraphics[width=3.6cm] {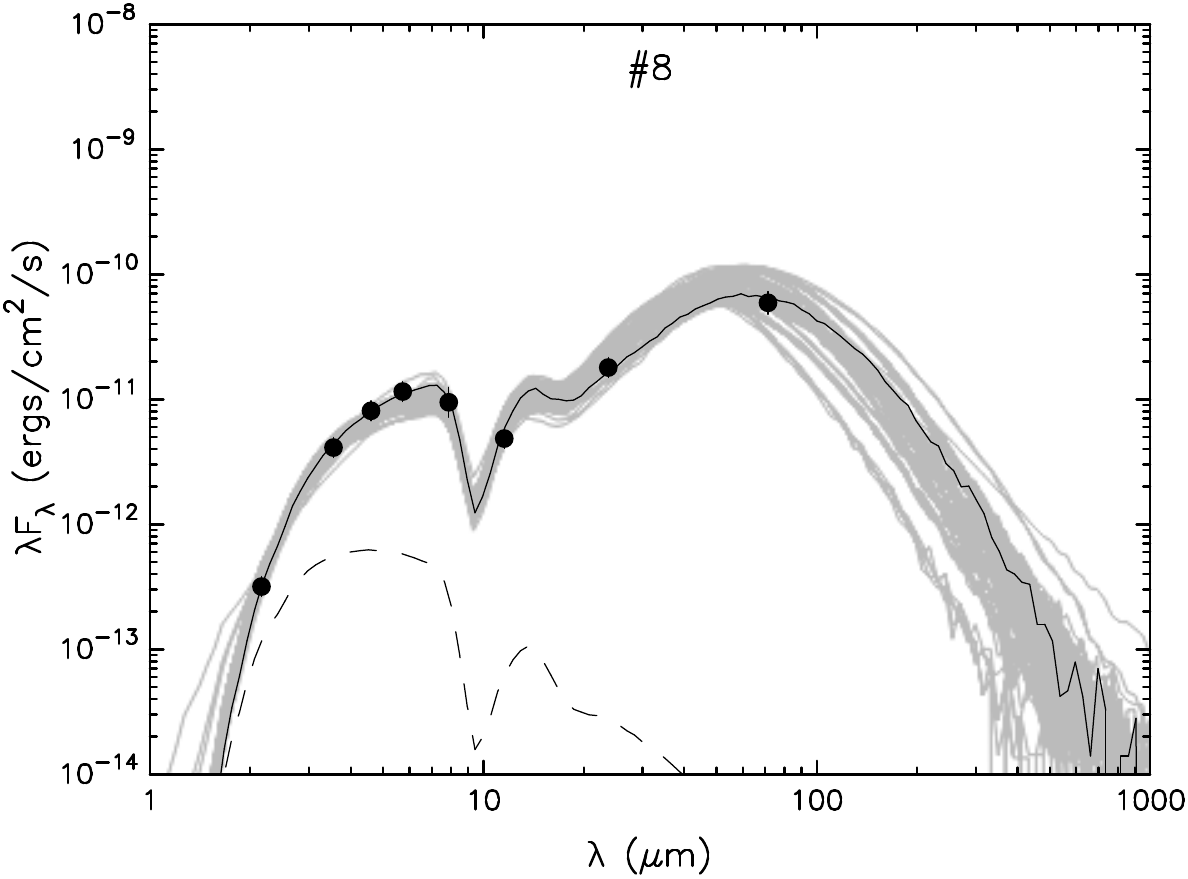}
\includegraphics[width=3.6cm] {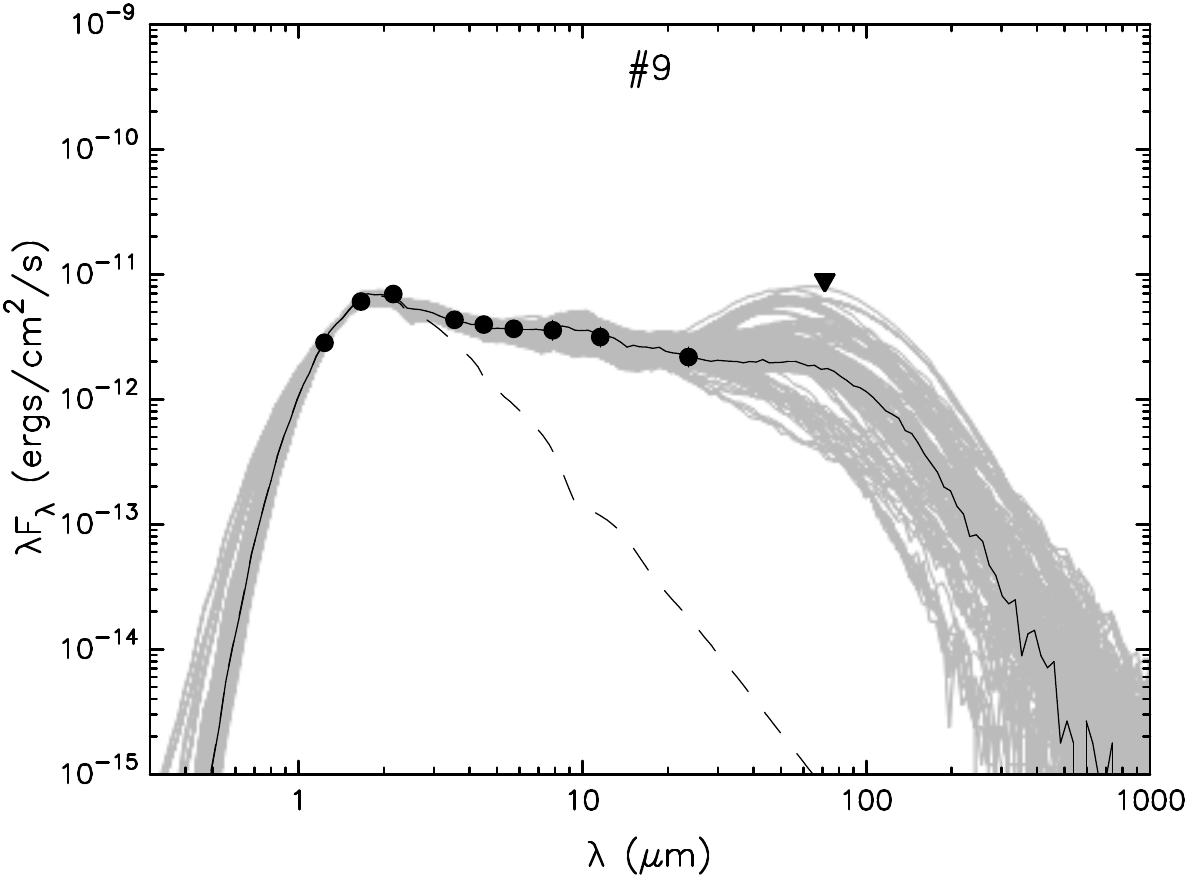}
\includegraphics[width=3.6cm] {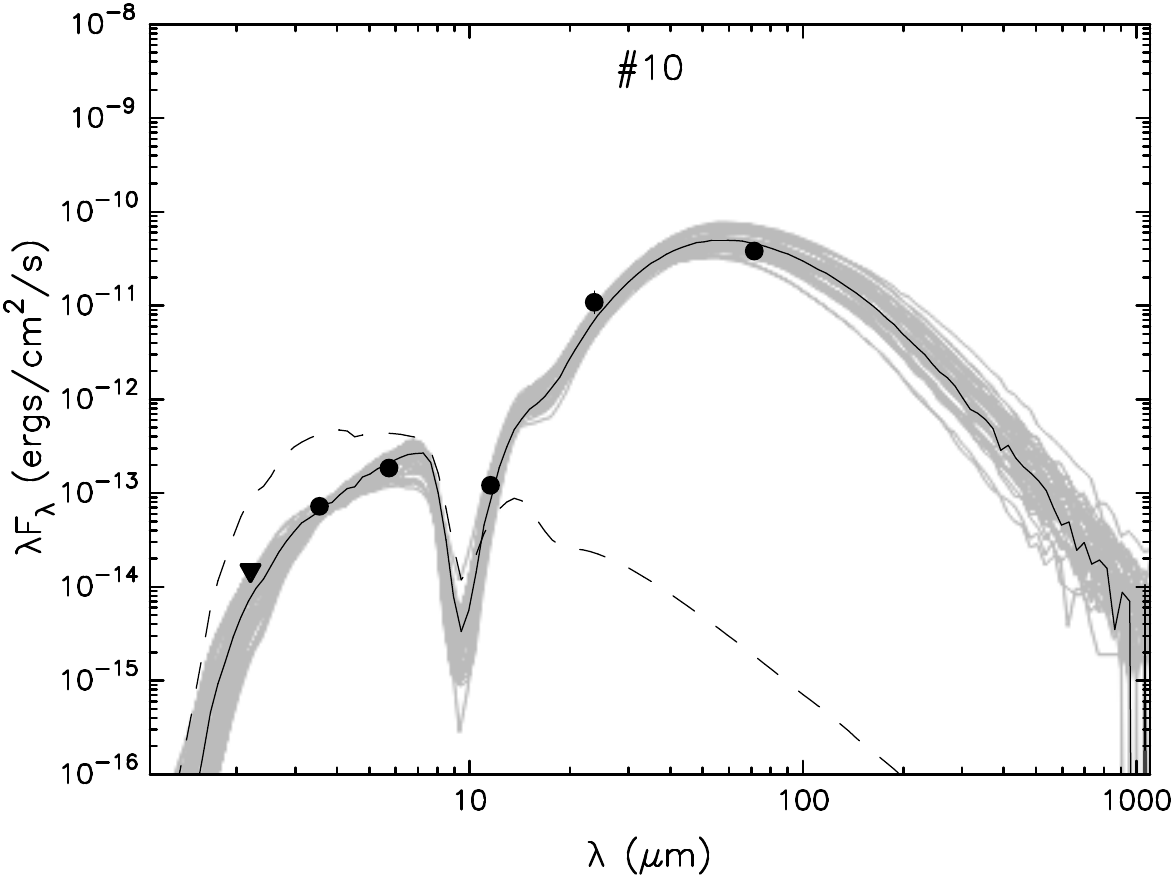}
\includegraphics[width=3.6cm] {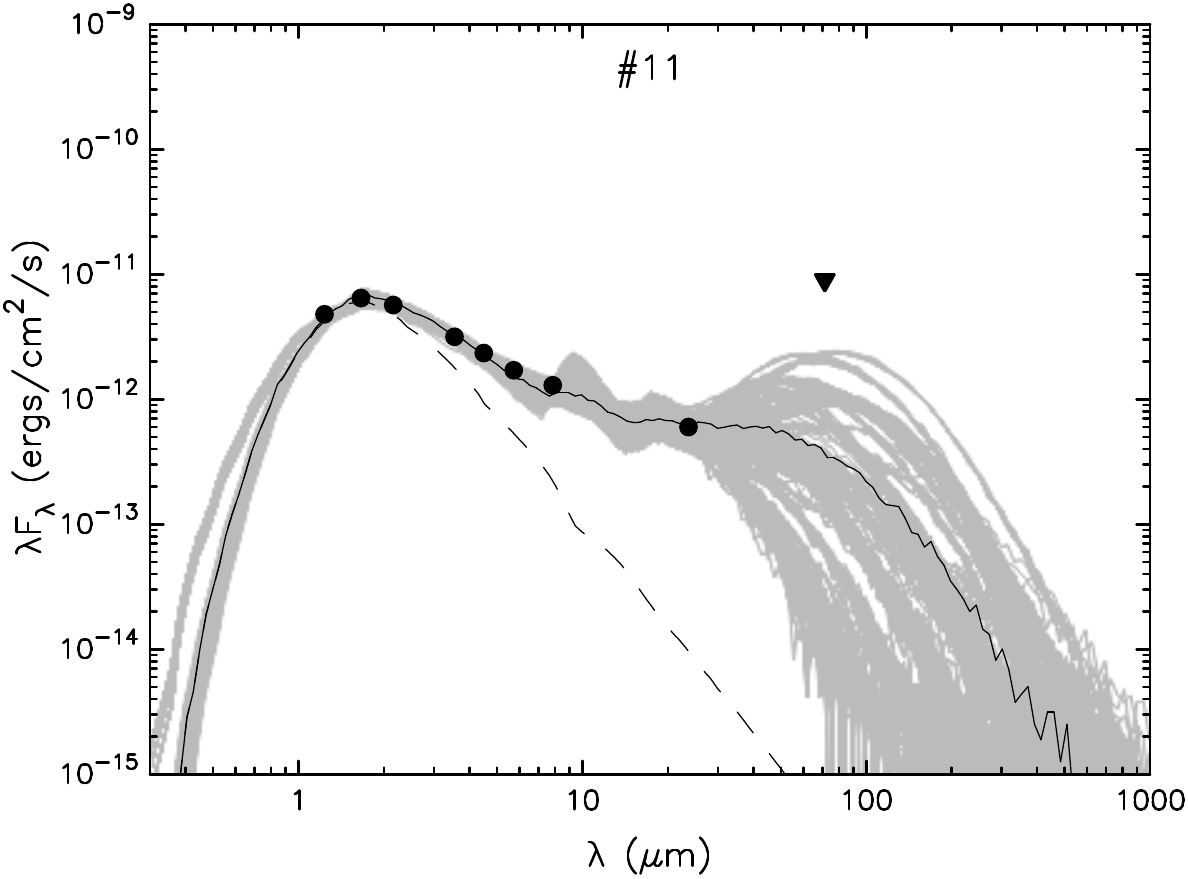}
\includegraphics[width=3.6cm] {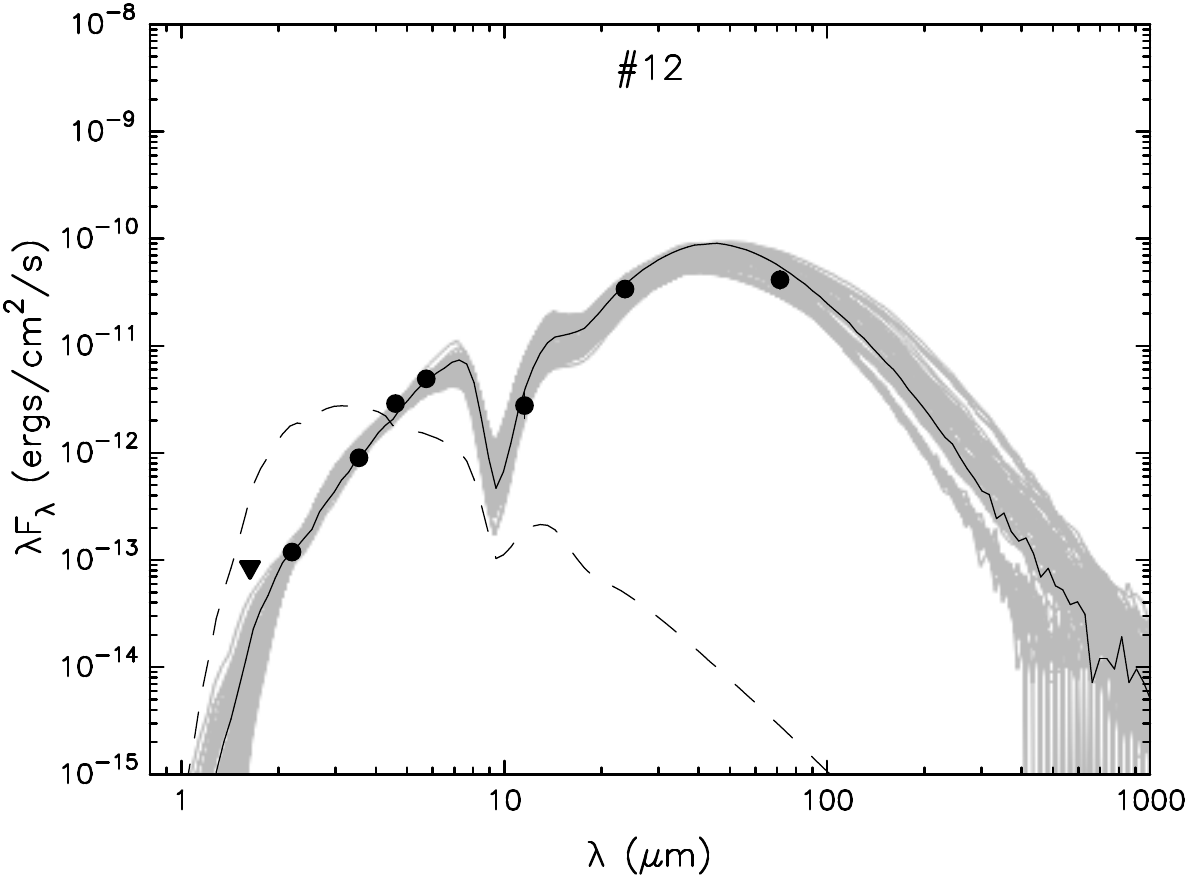}
\includegraphics[width=3.6cm] {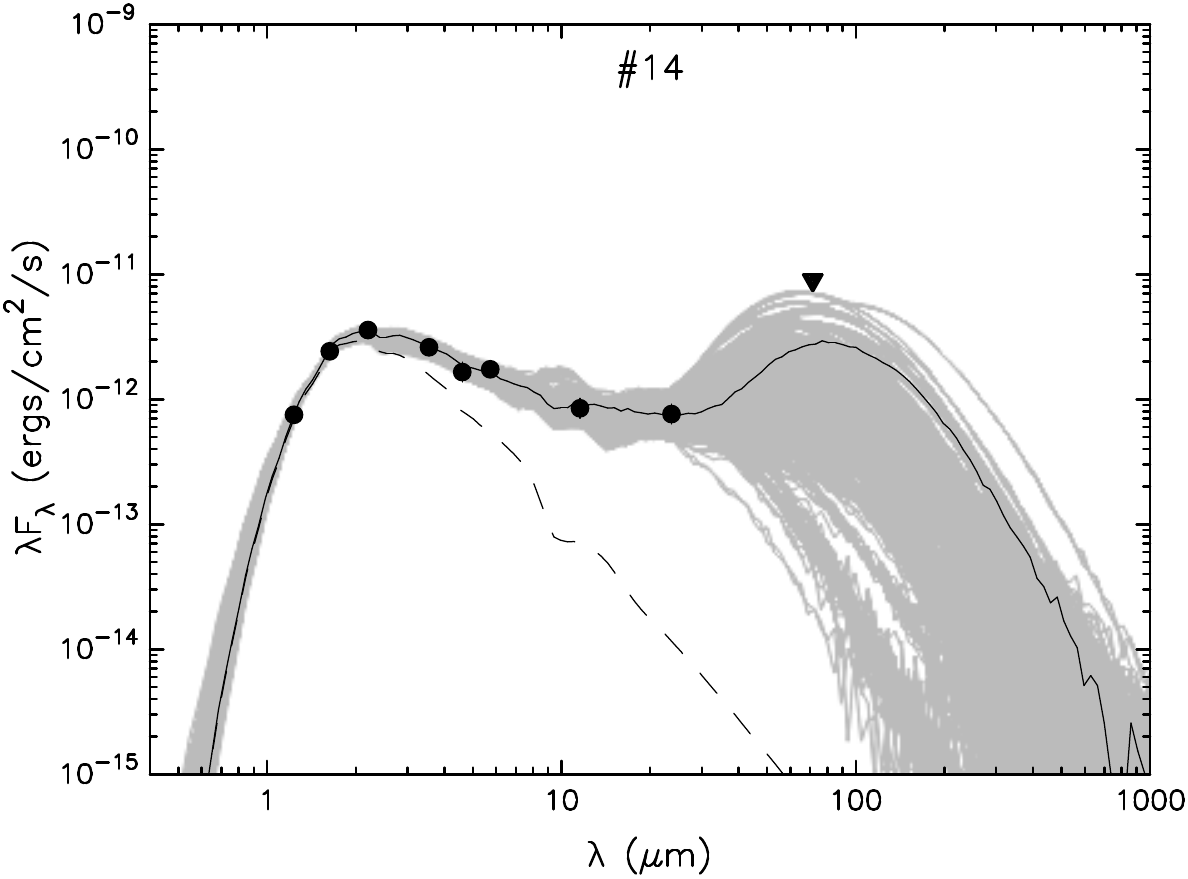}
\includegraphics[width=3.6cm] {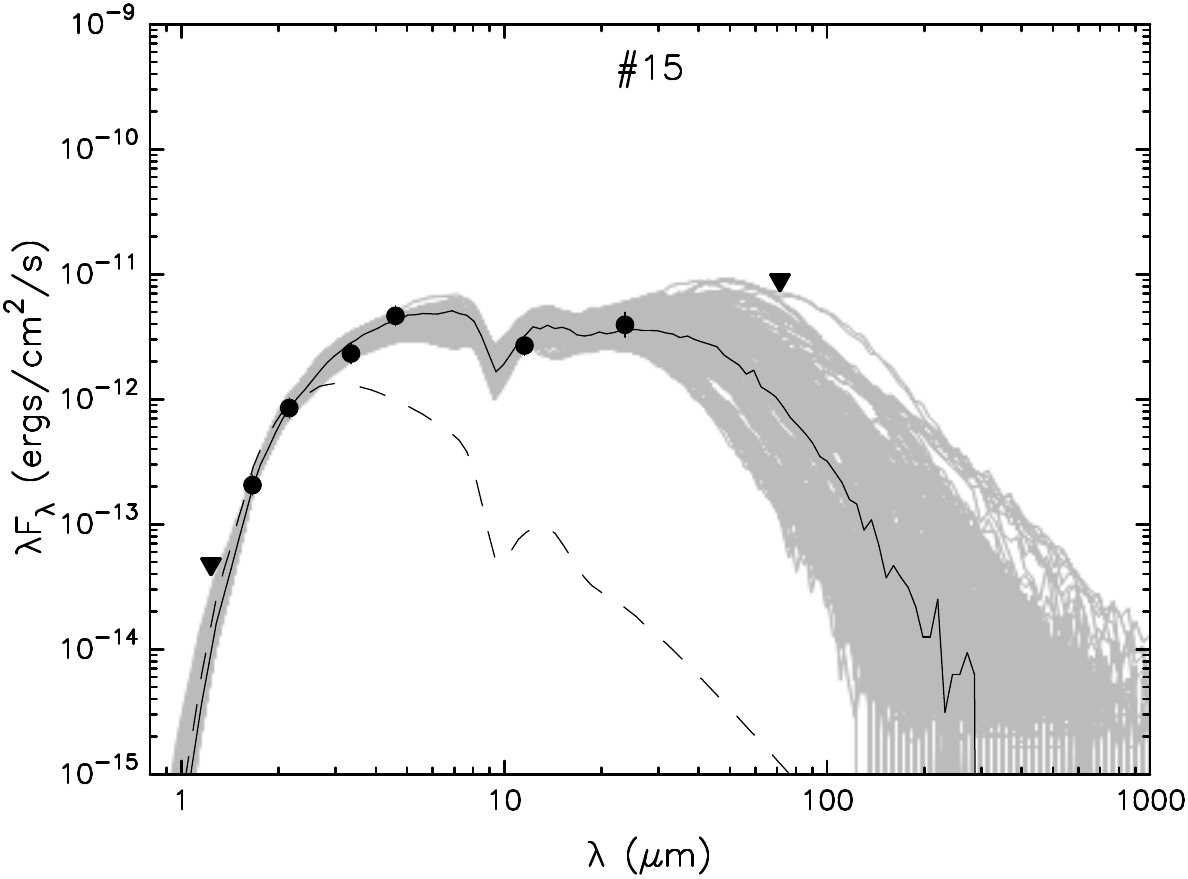}
\includegraphics[width=3.6cm] {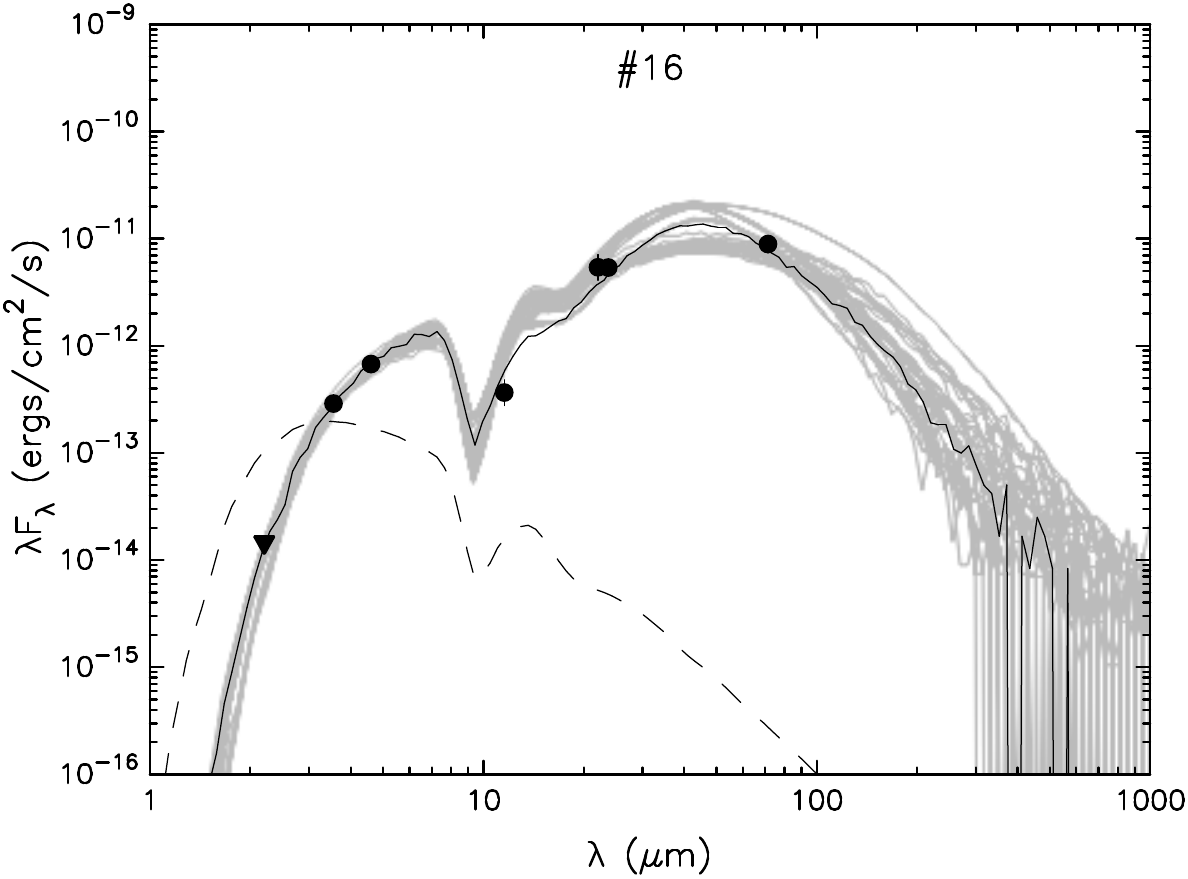}
\includegraphics[width=3.6cm] {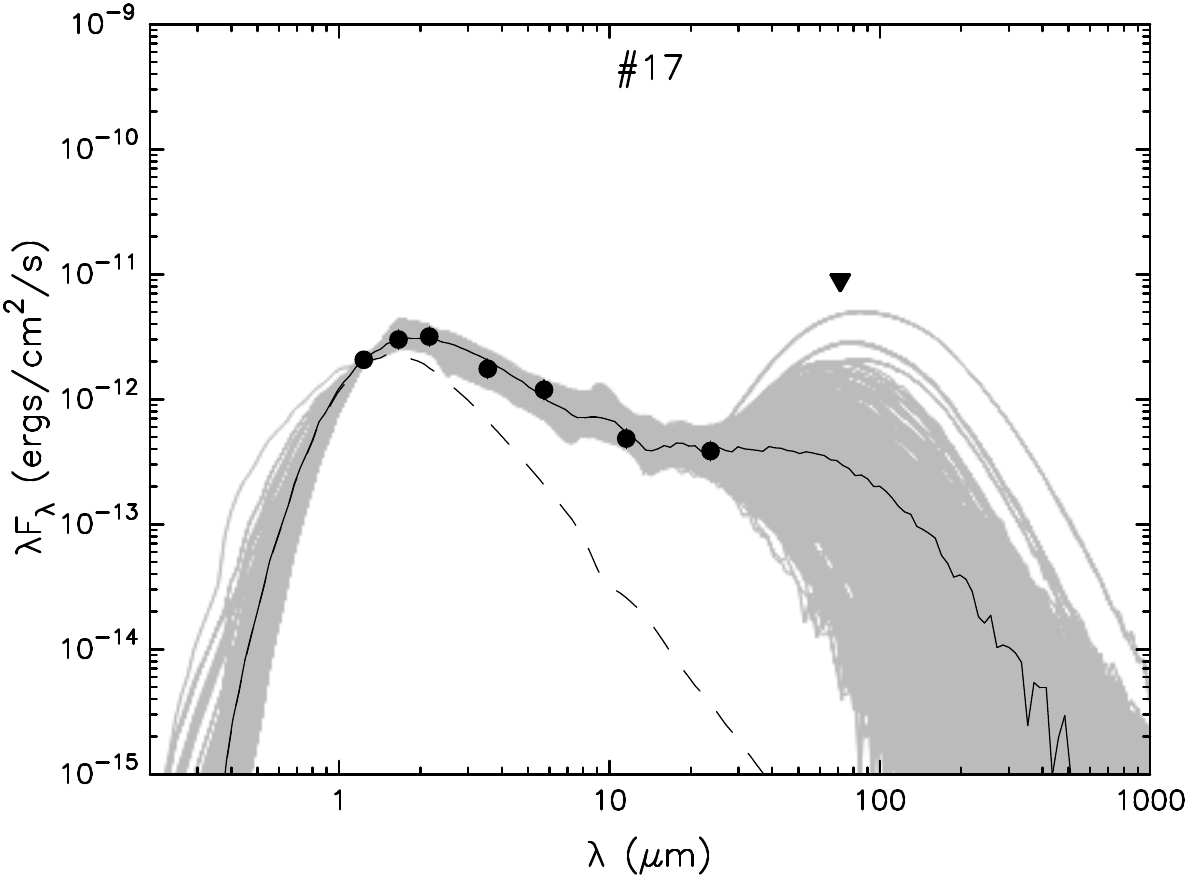}
\includegraphics[width=3.6cm] {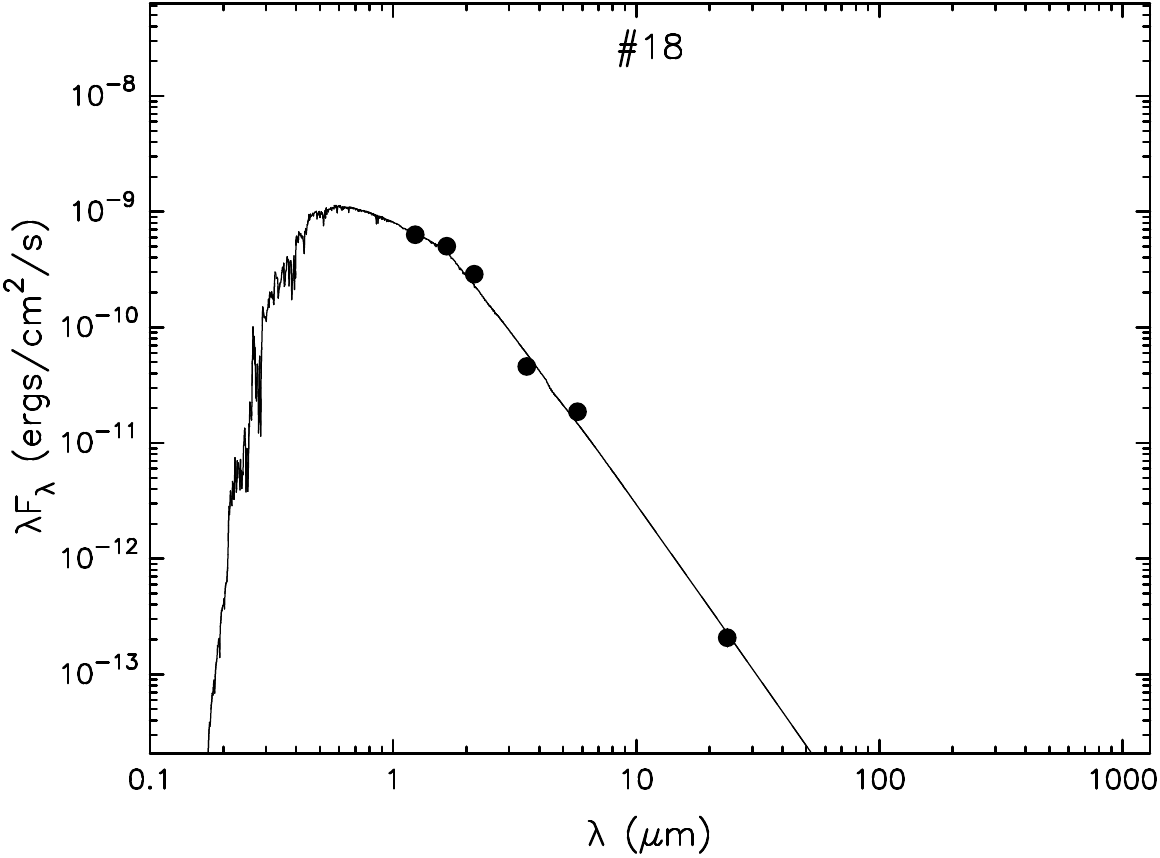}
\includegraphics[width=3.6cm] {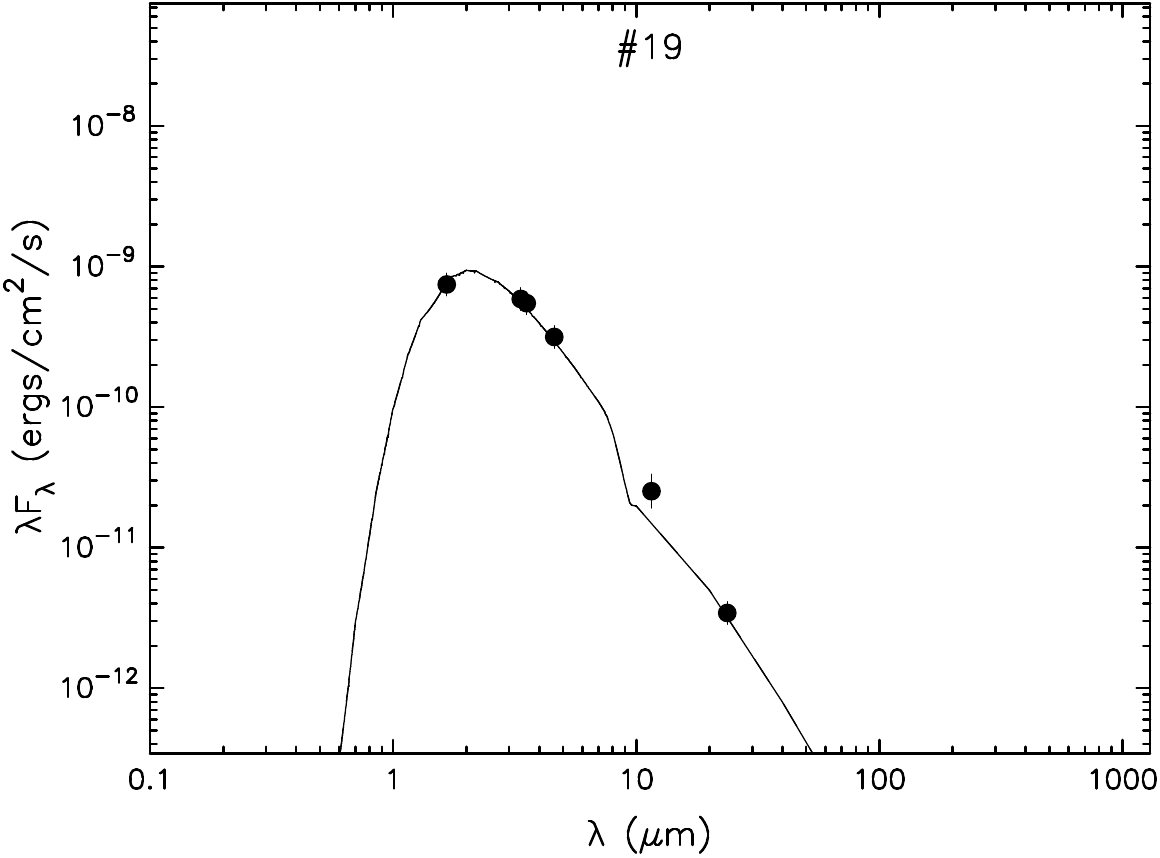}
\includegraphics[width=3.6cm] {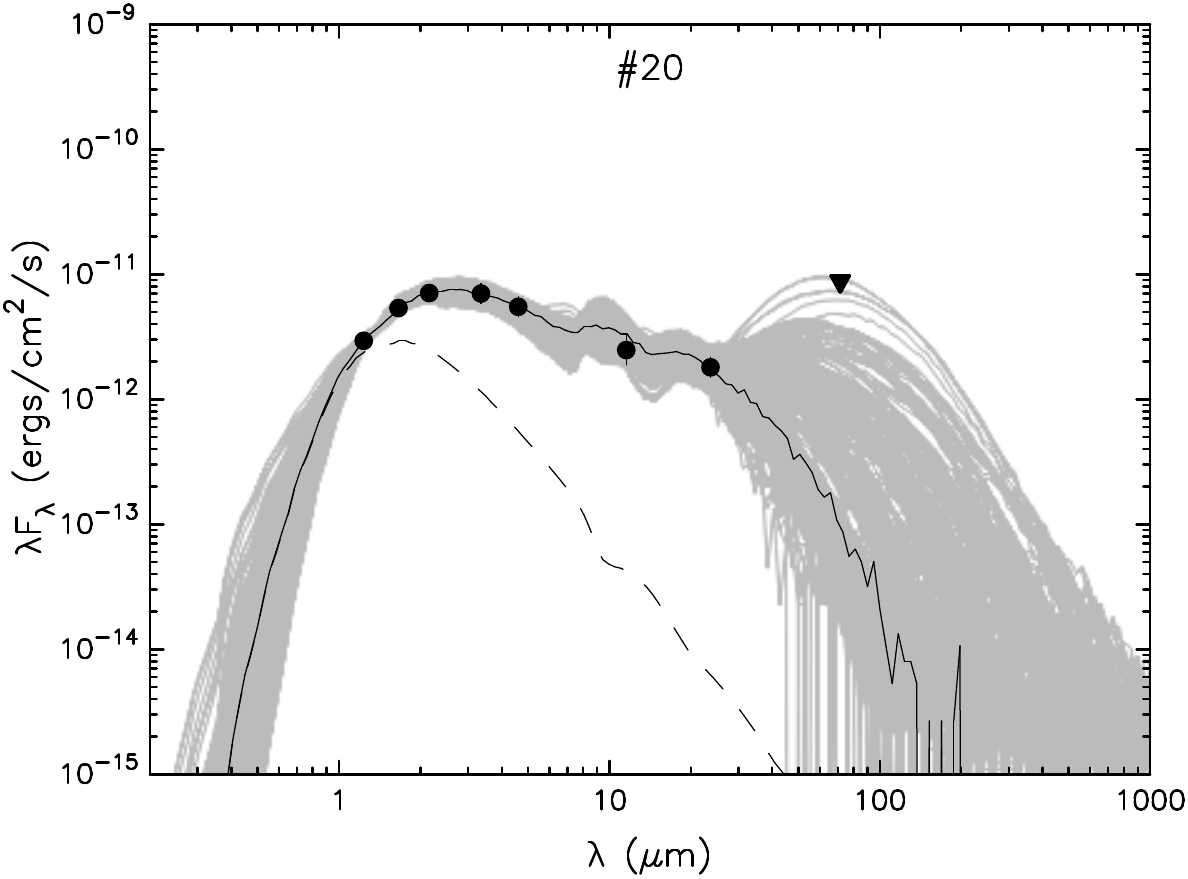}
\includegraphics[width=3.6cm] {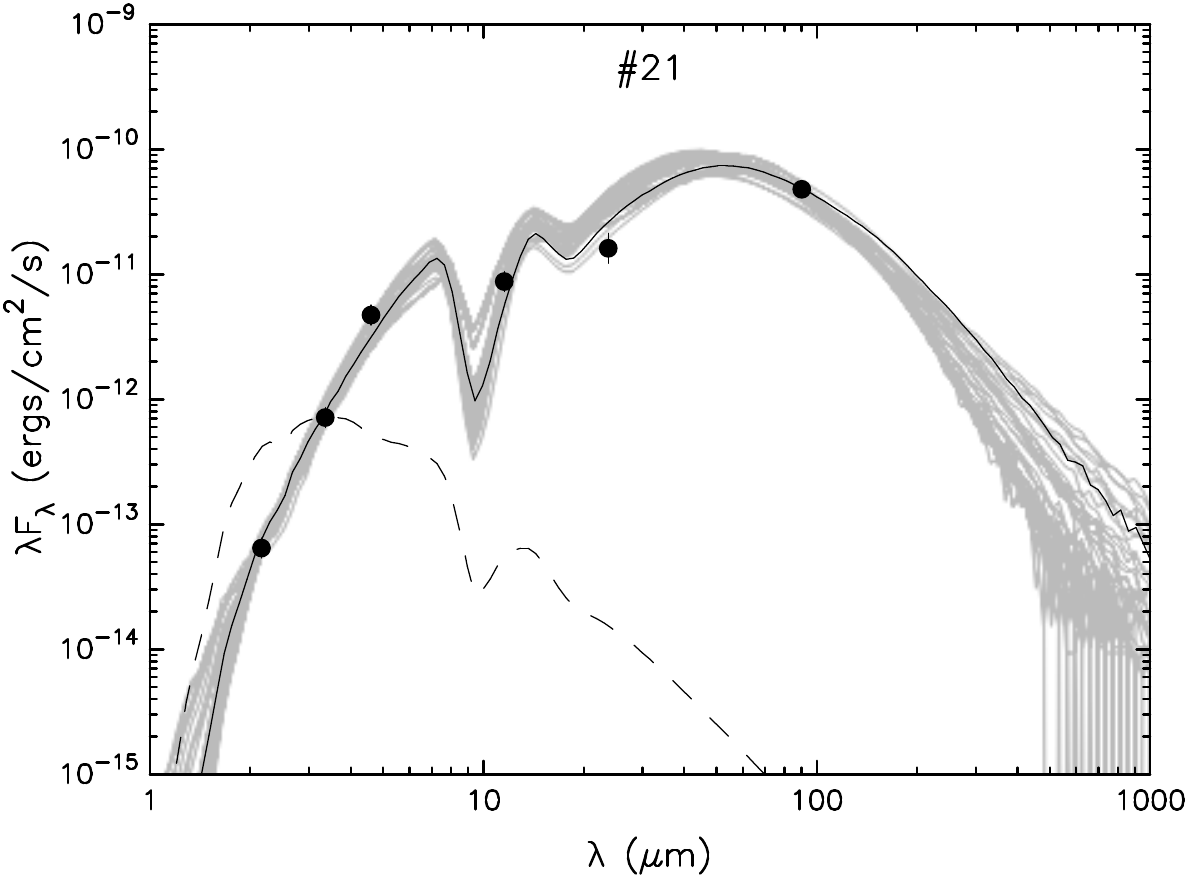}
\includegraphics[width=3.6cm] {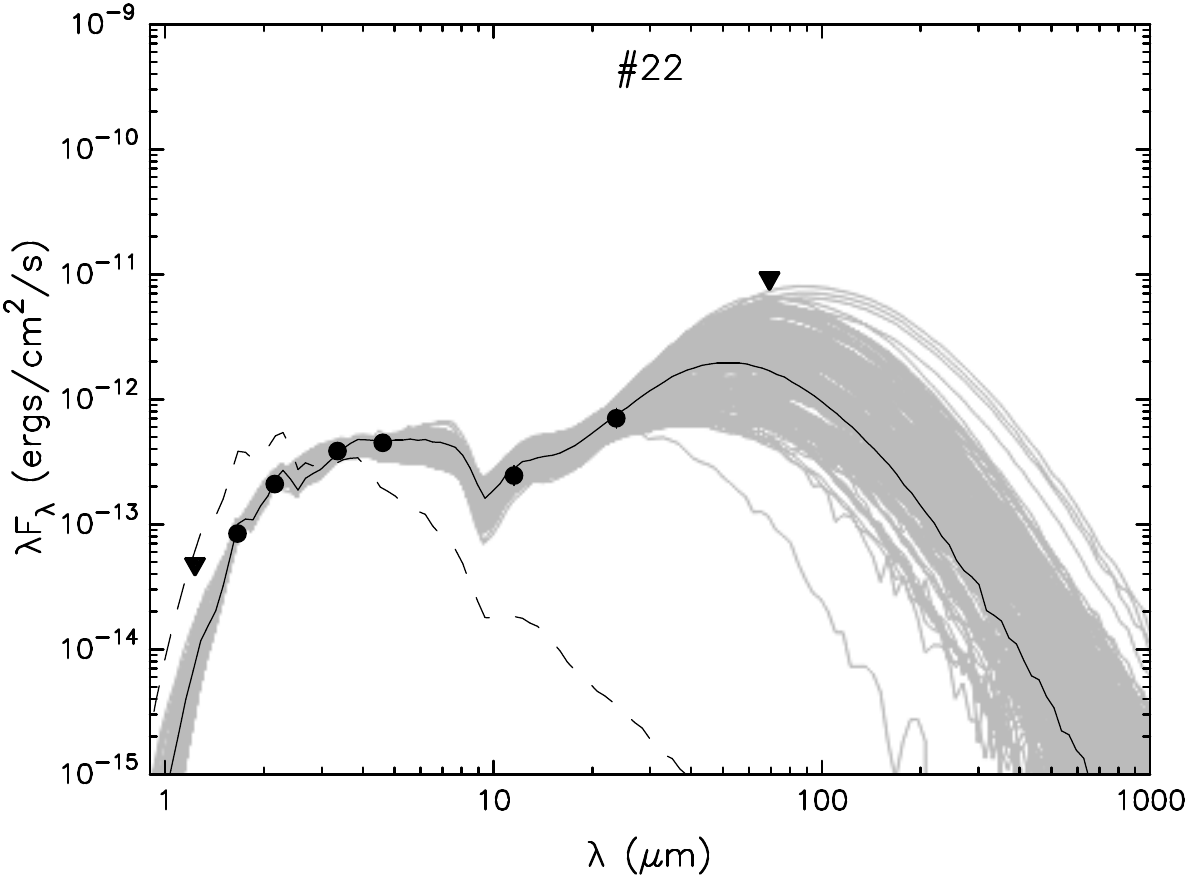}
\includegraphics[width=3.6cm] {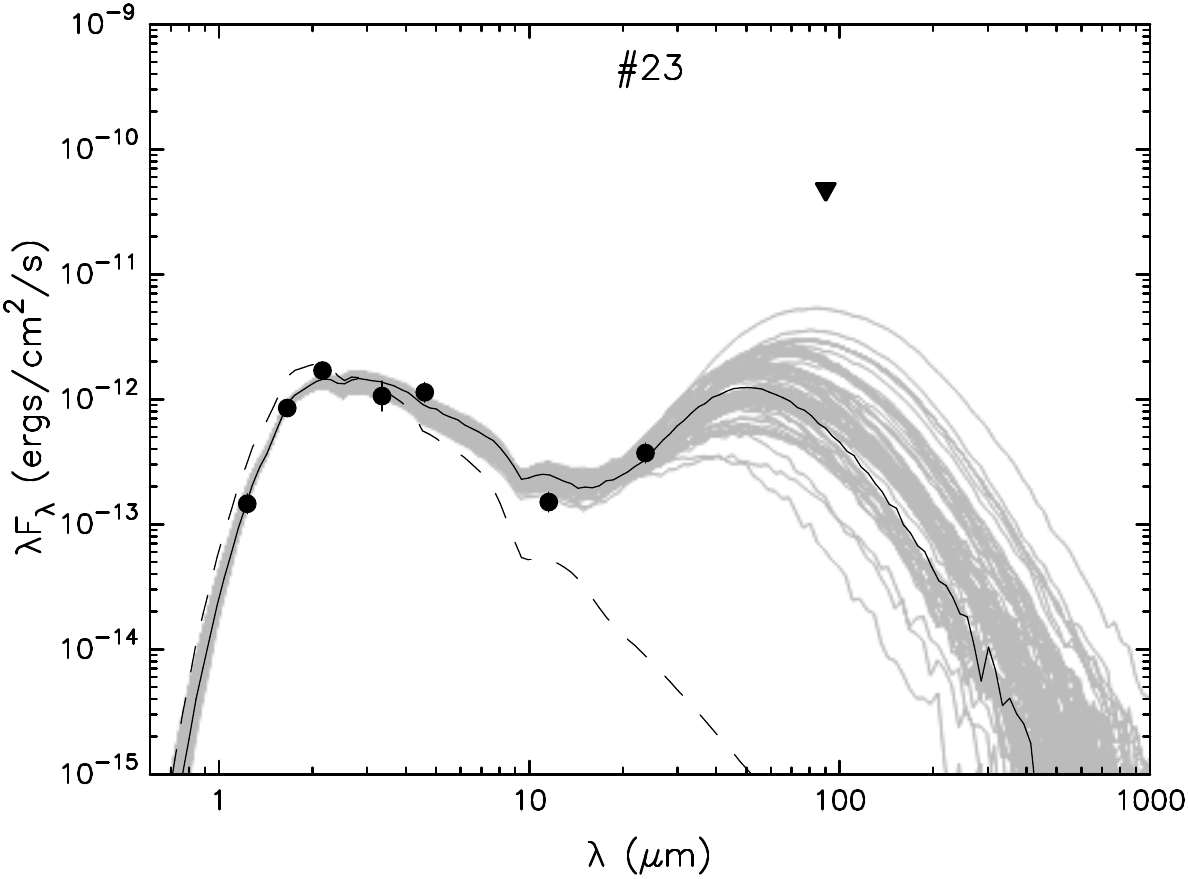}
\includegraphics[width=3.6cm] {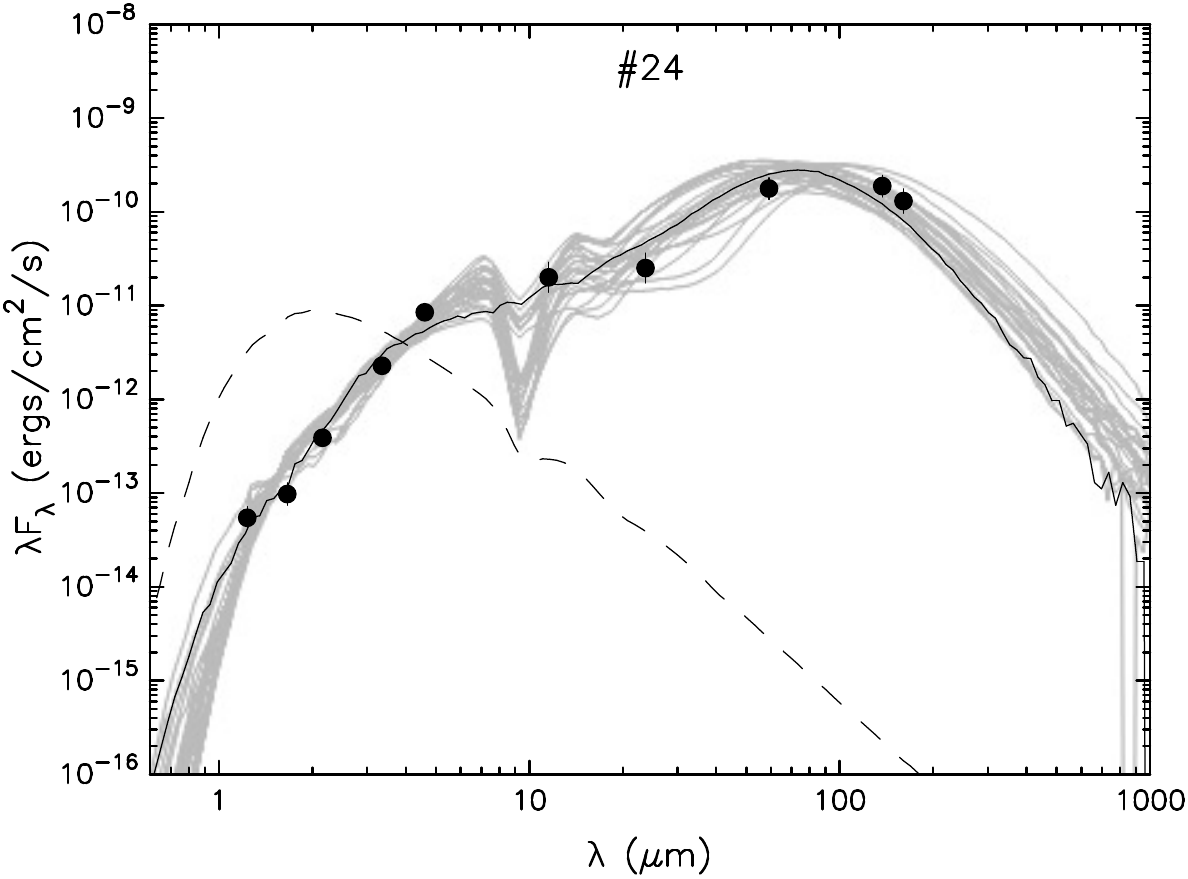}
\includegraphics[width=3.6cm] {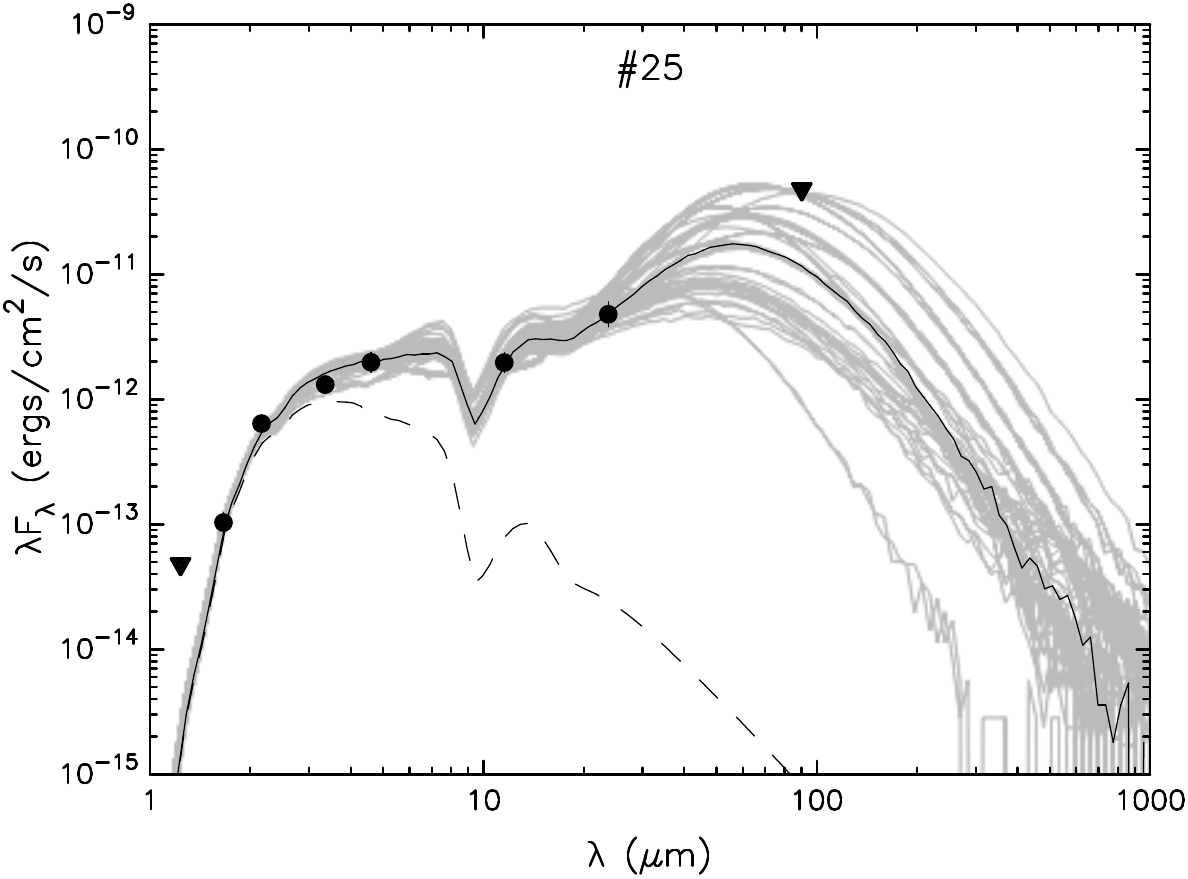}
\includegraphics[width=3.6cm] {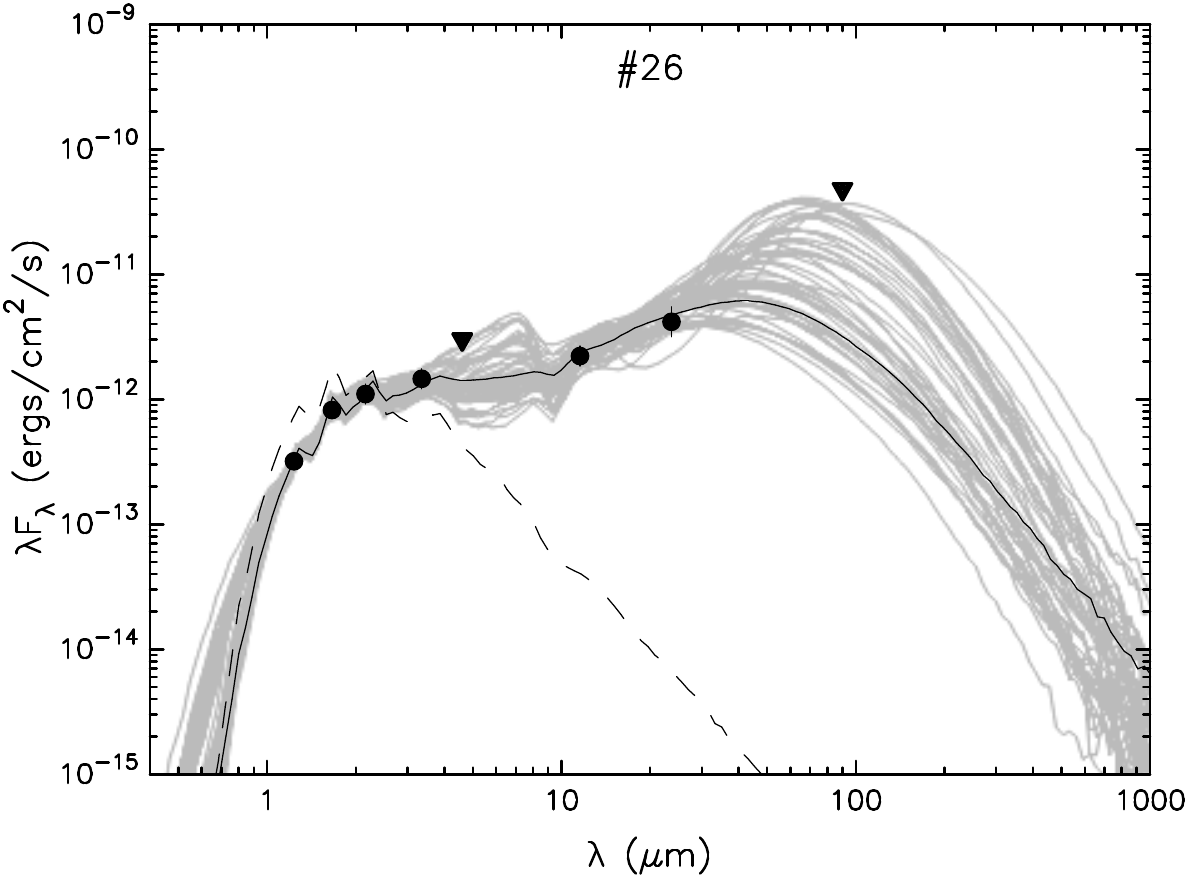}
\includegraphics[width=3.6cm] {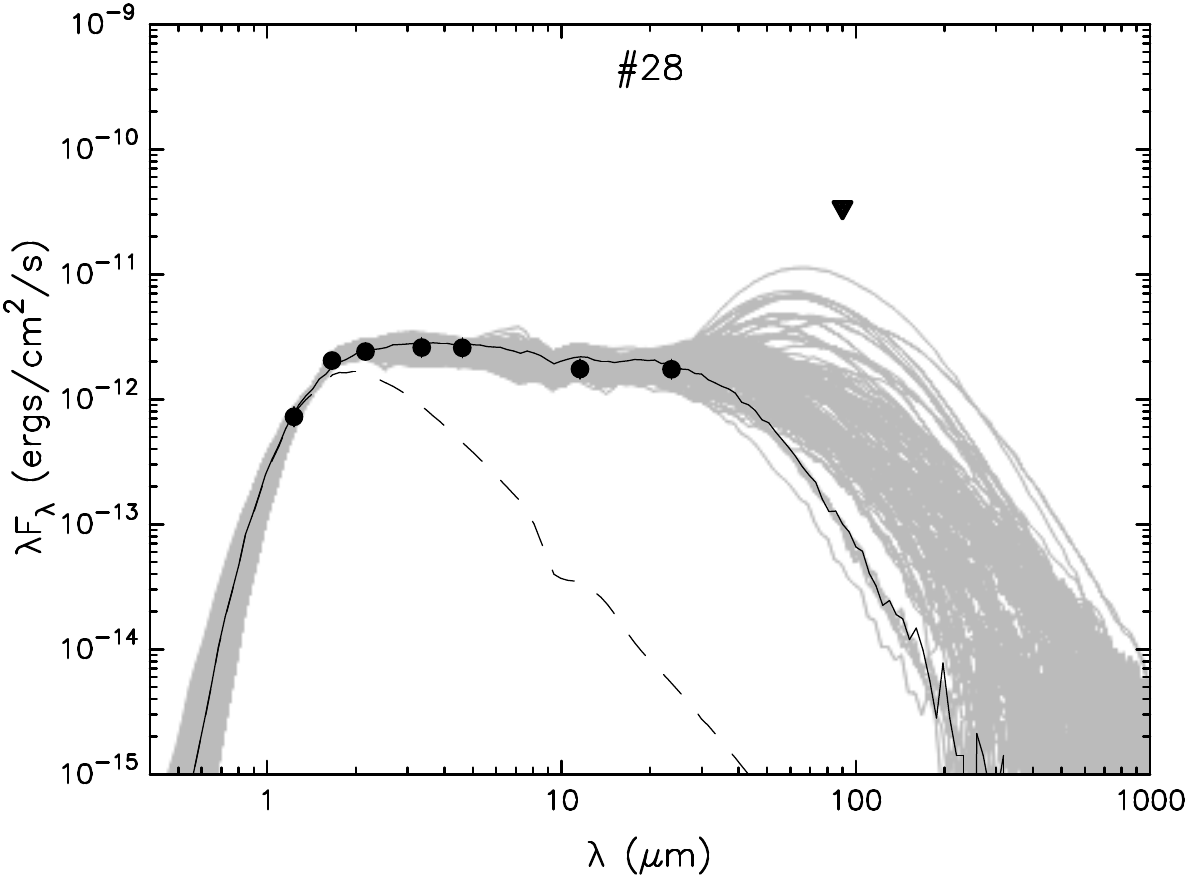}
\includegraphics[width=3.6cm] {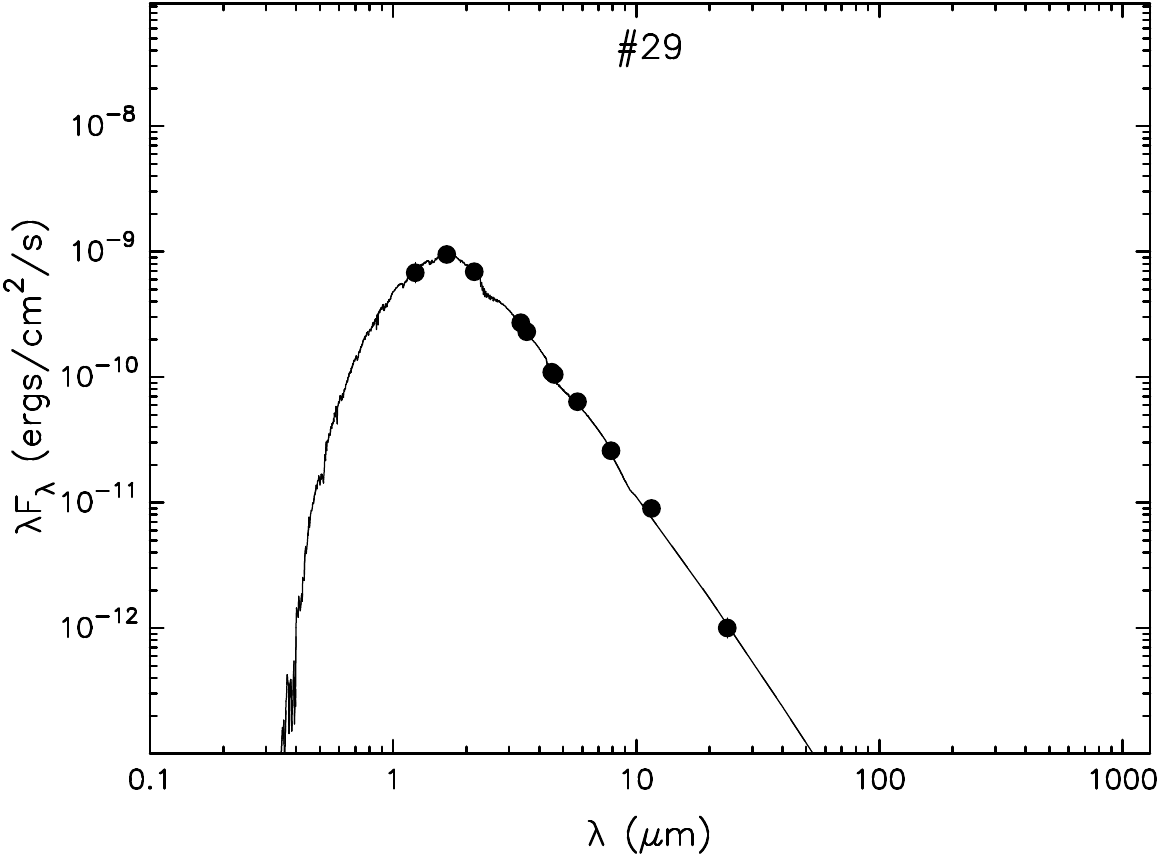}
\includegraphics[width=3.6cm] {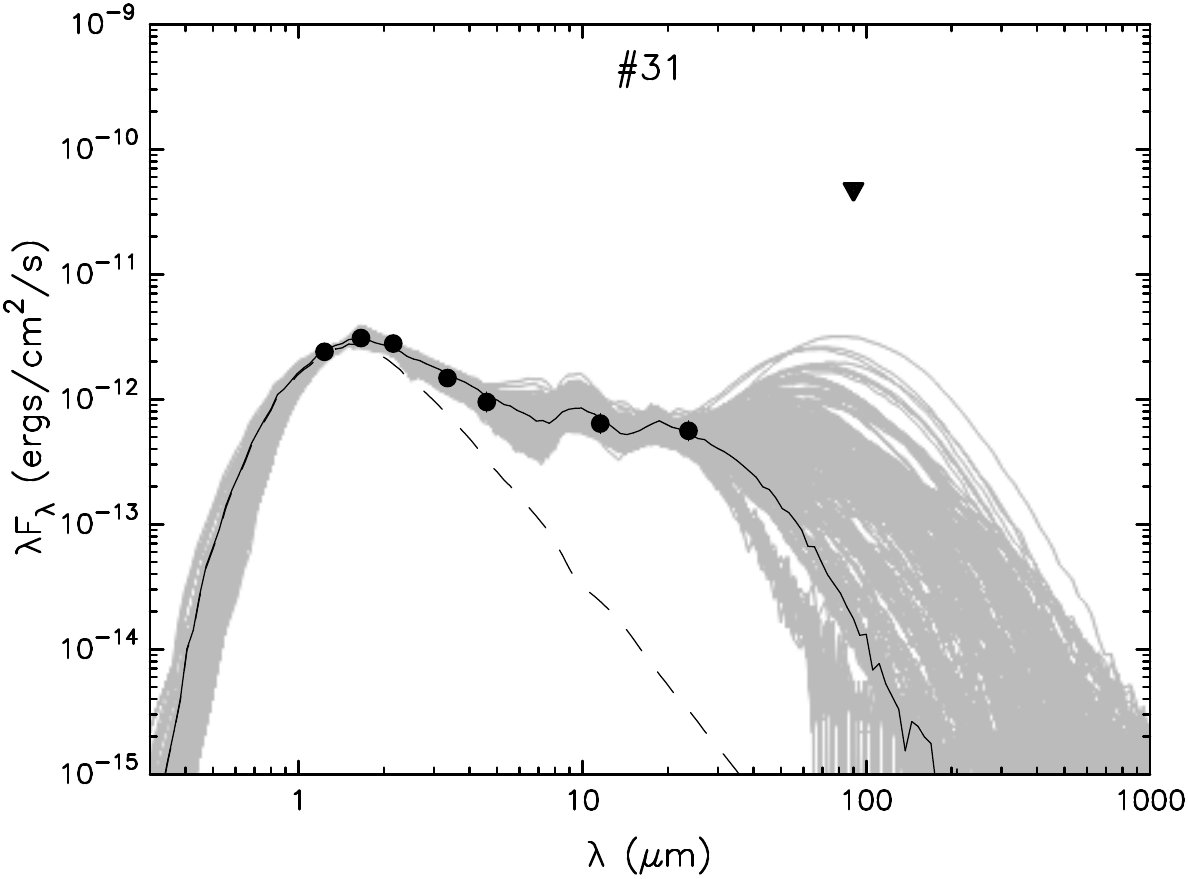}
\includegraphics[width=3.6cm] {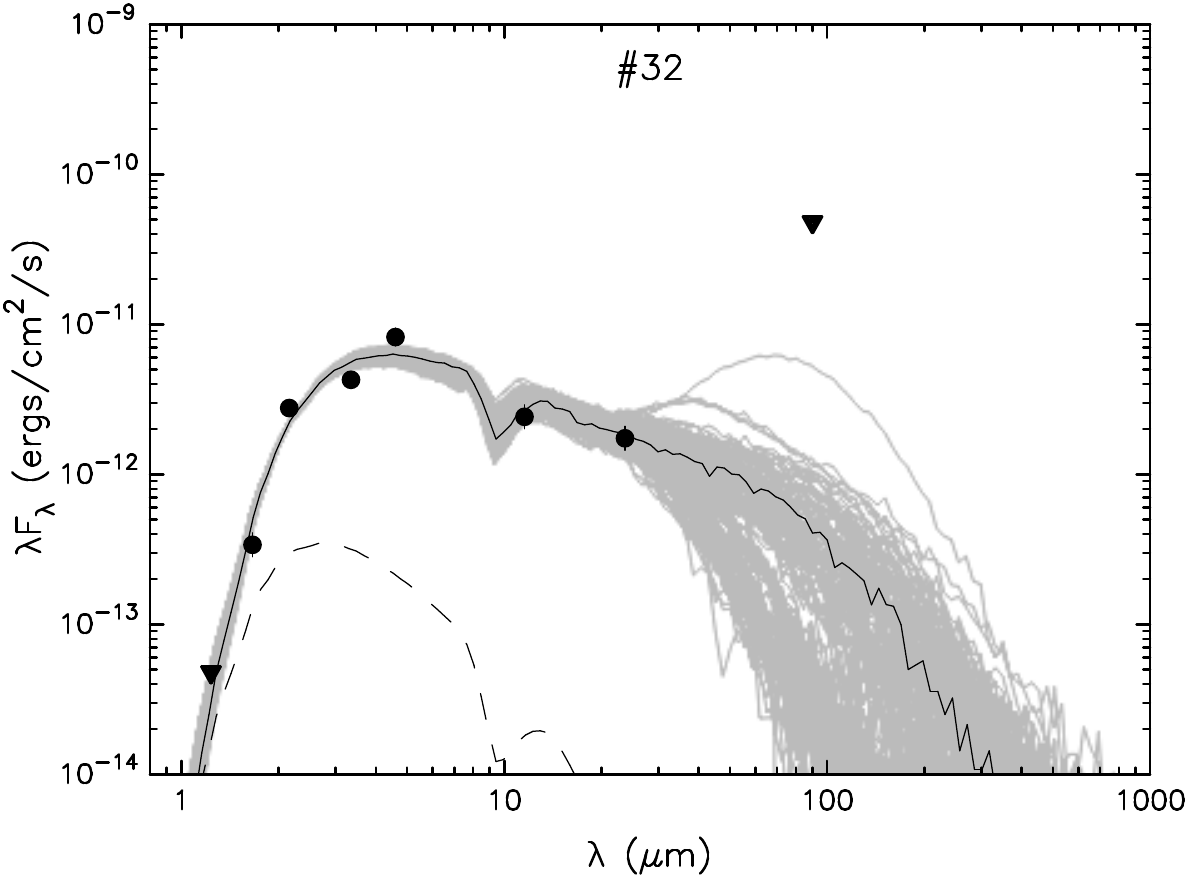}
\includegraphics[width=3.6cm] {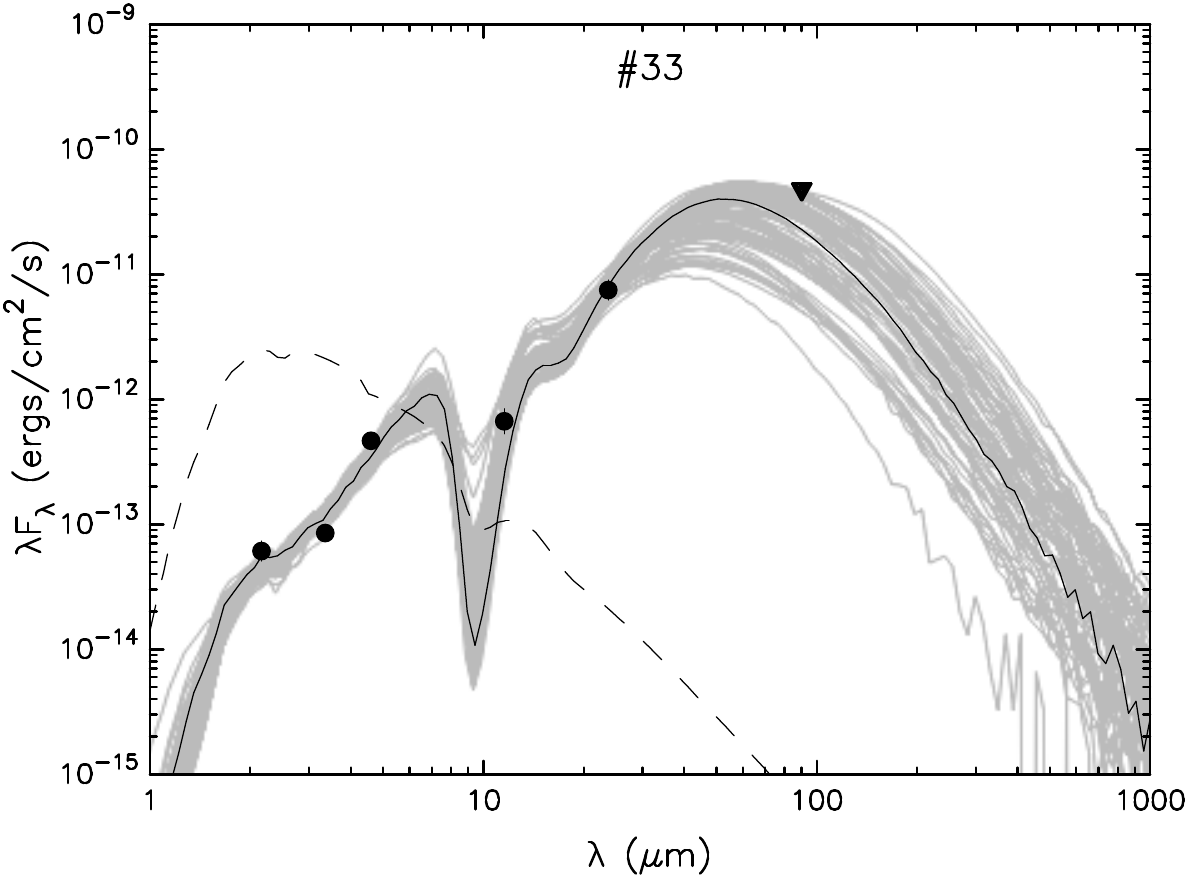}
\includegraphics[width=3.6cm] {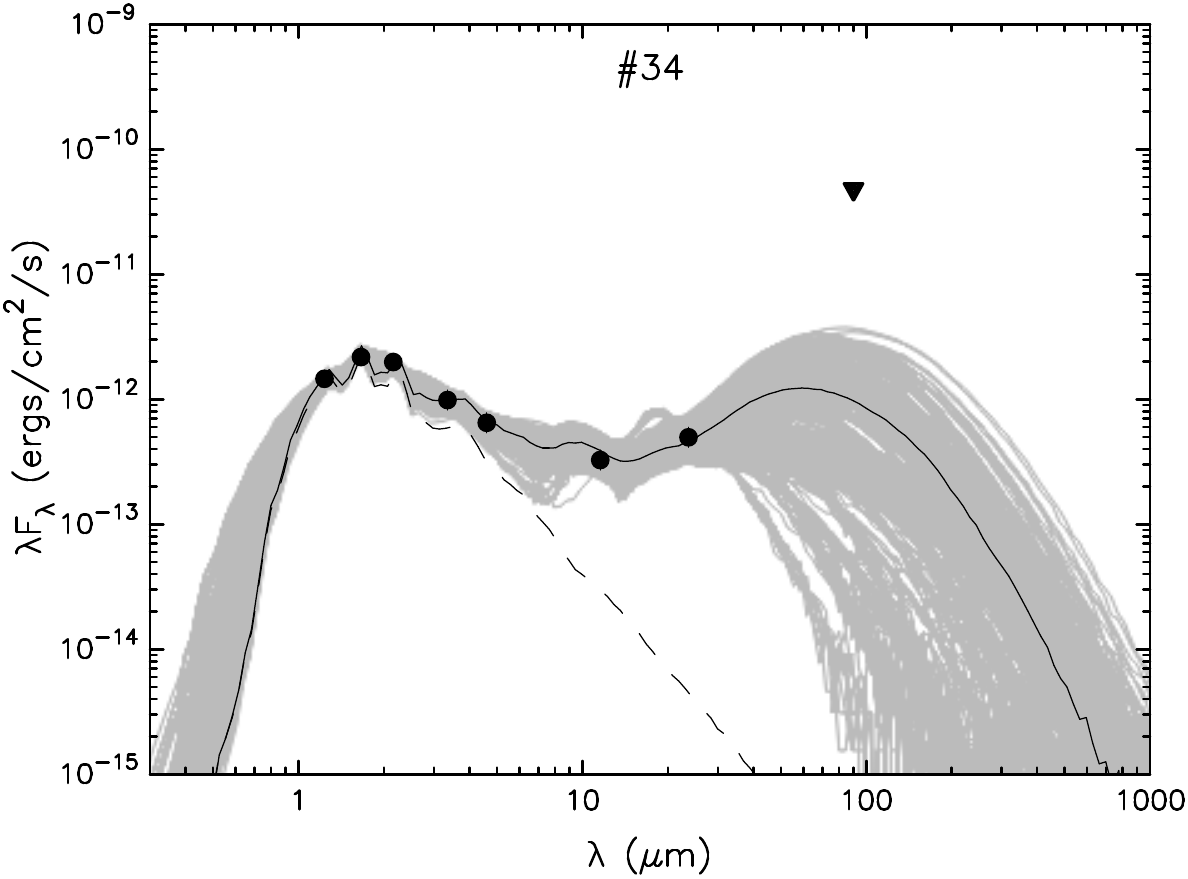}
\includegraphics[width=3.6cm] {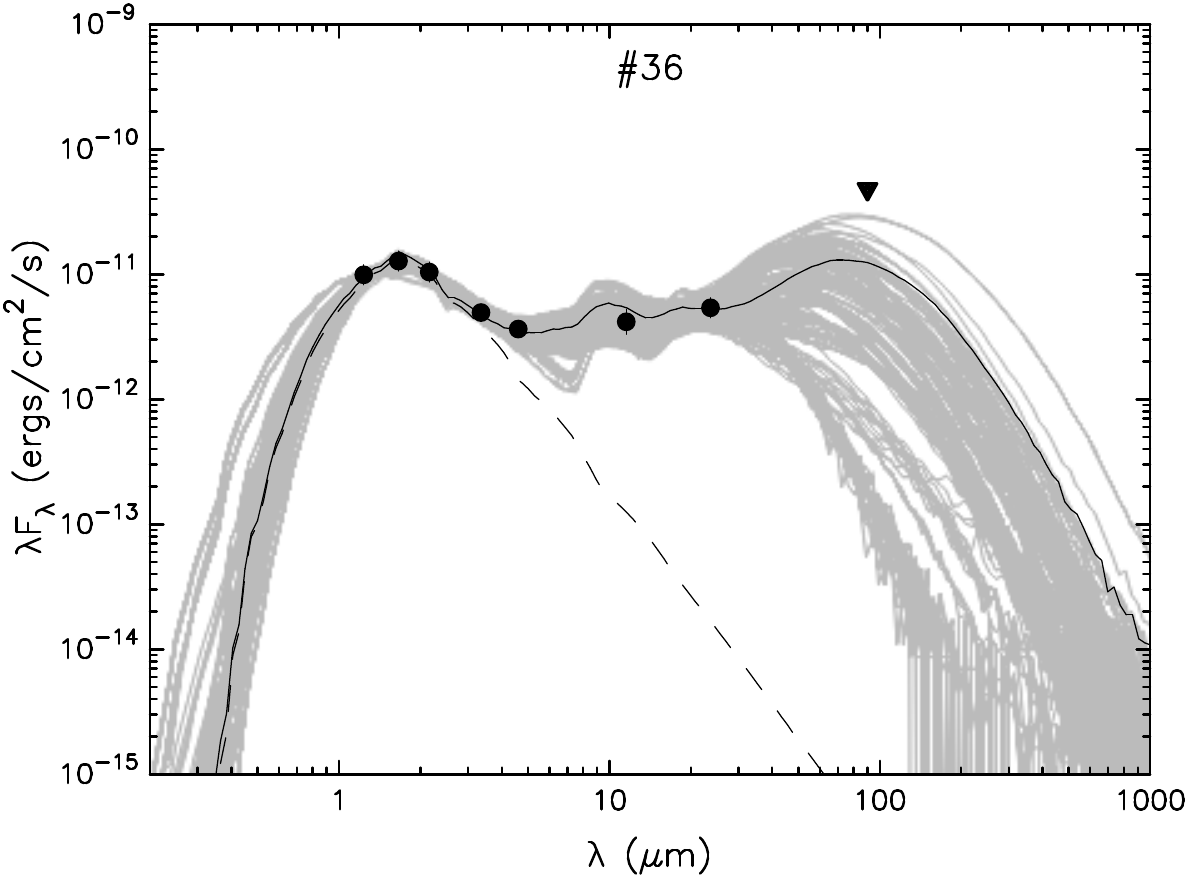}
\includegraphics[width=3.6cm] {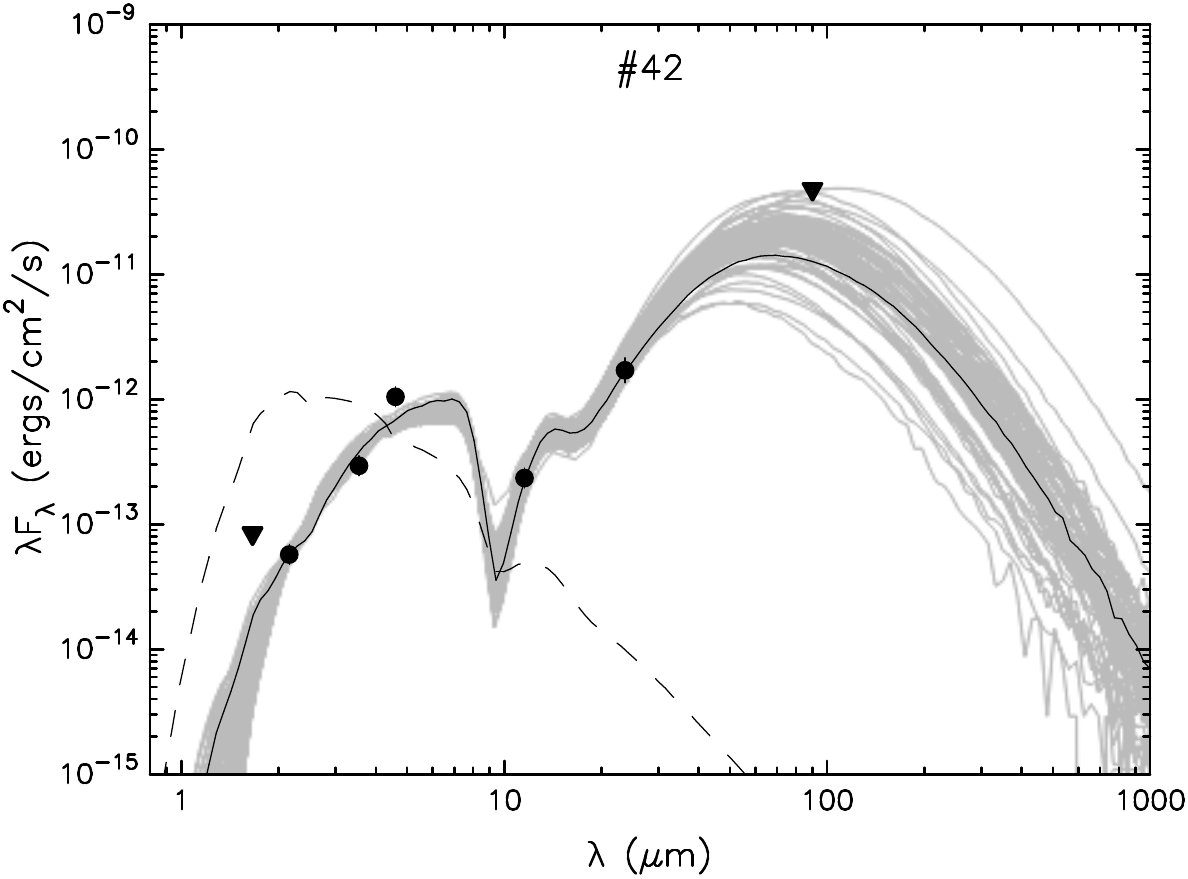}
\includegraphics[width=3.6cm] {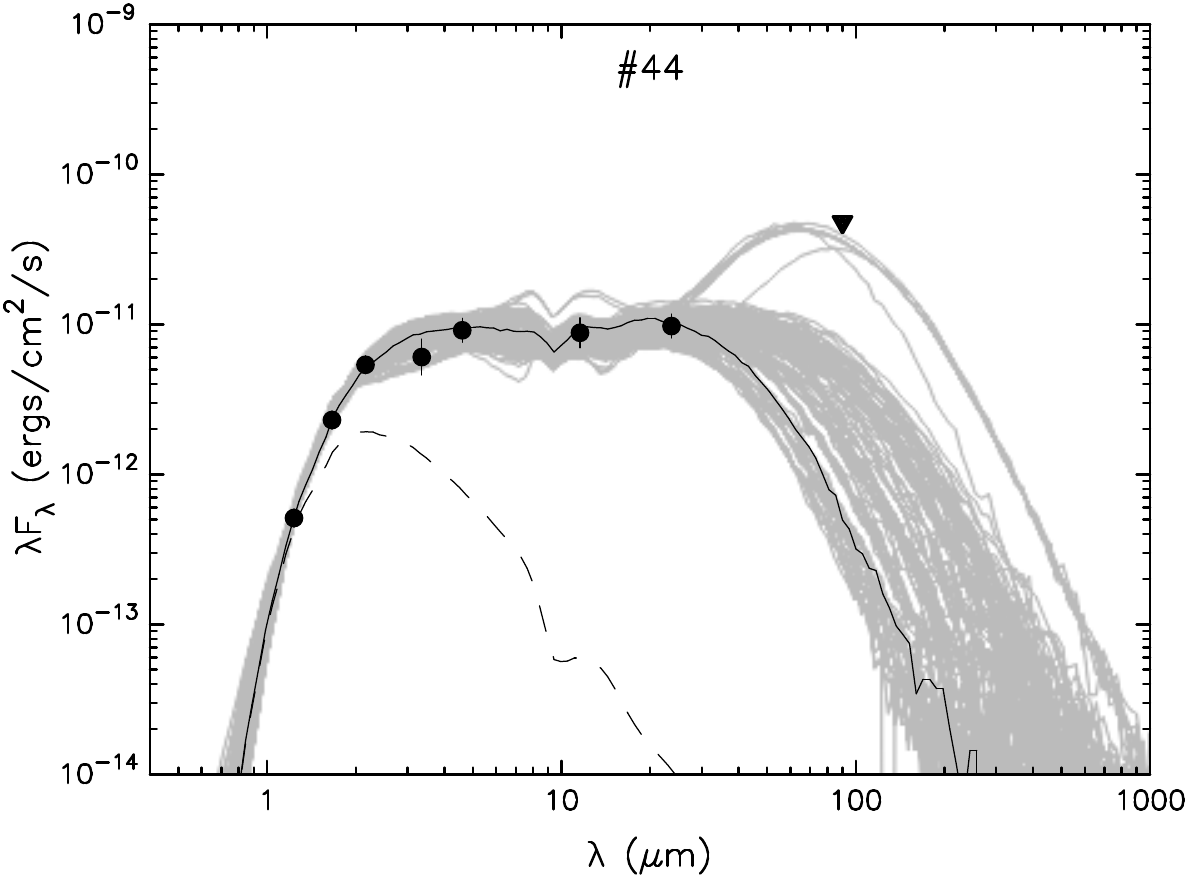}
\includegraphics[width=3.6cm] {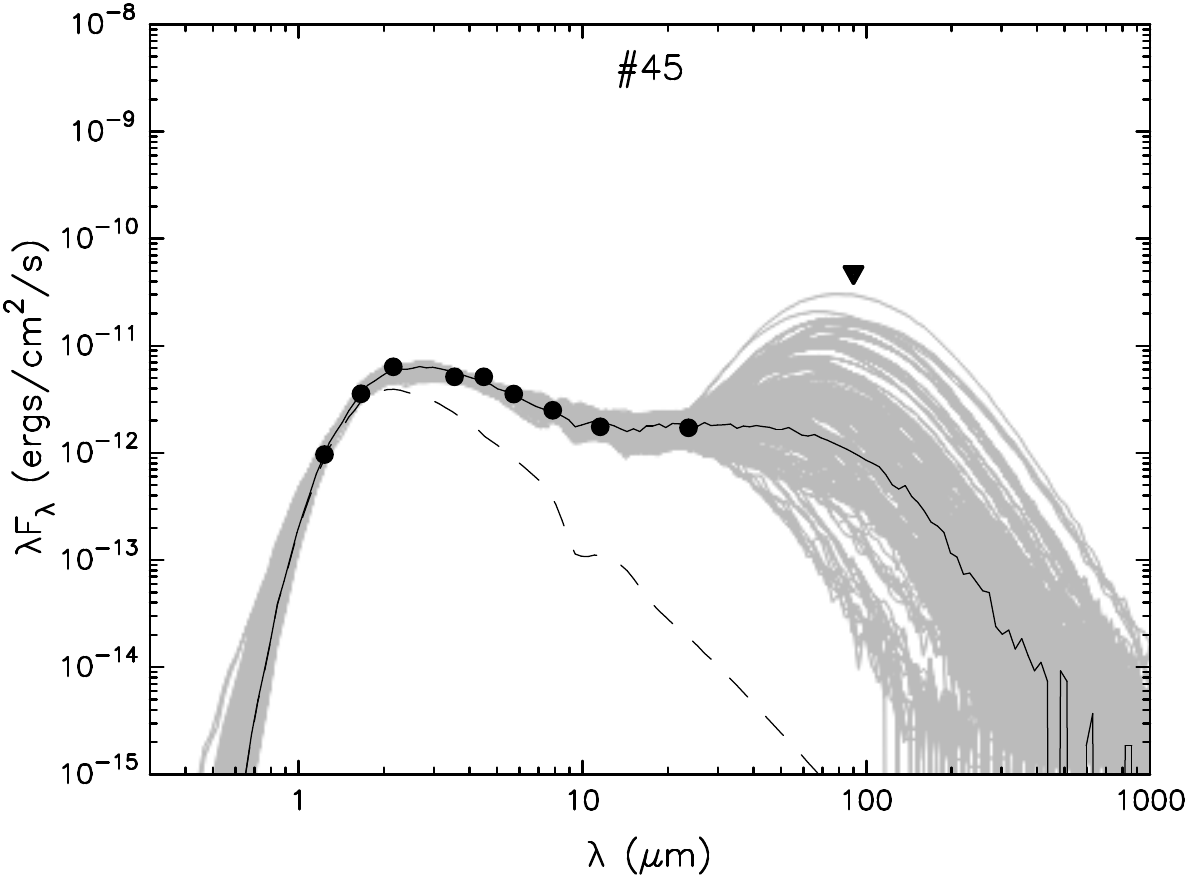}
\includegraphics[width=3.6cm] {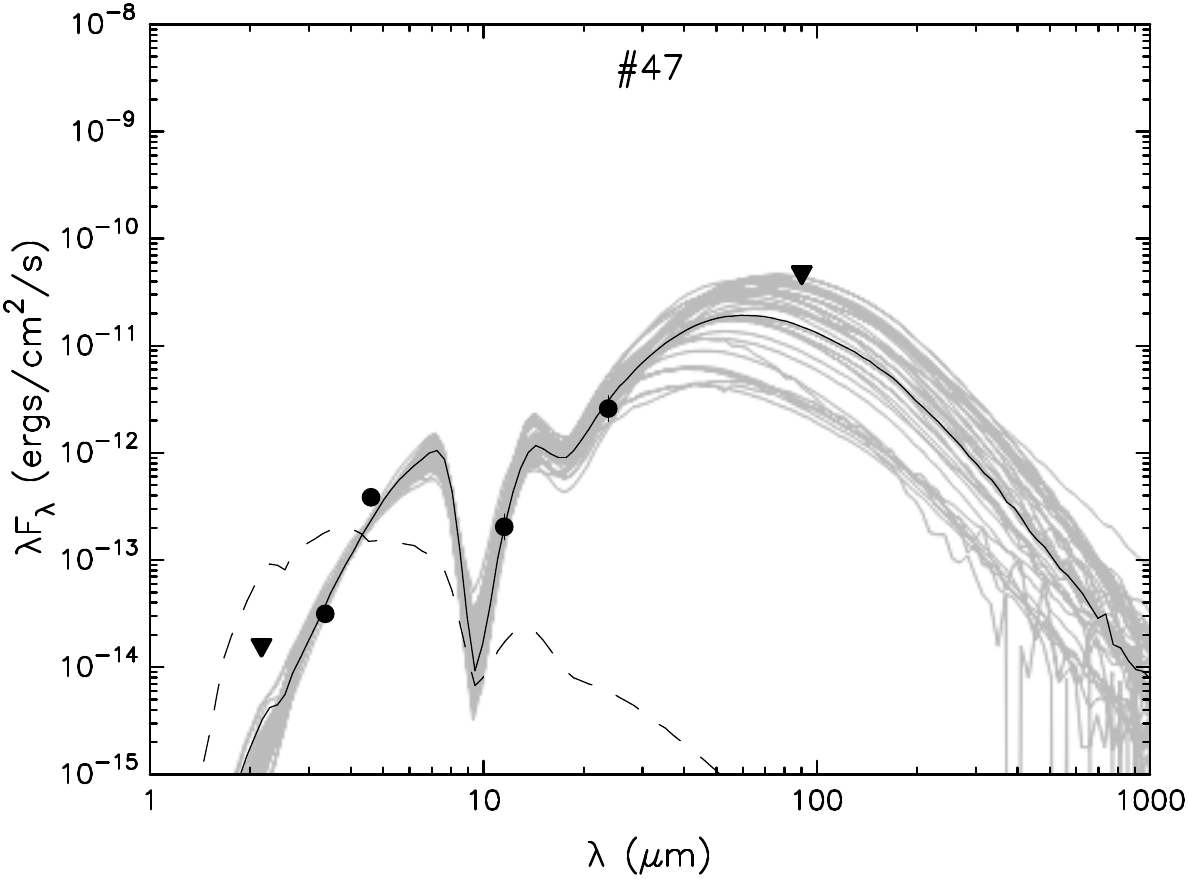}
\includegraphics[width=3.6cm] {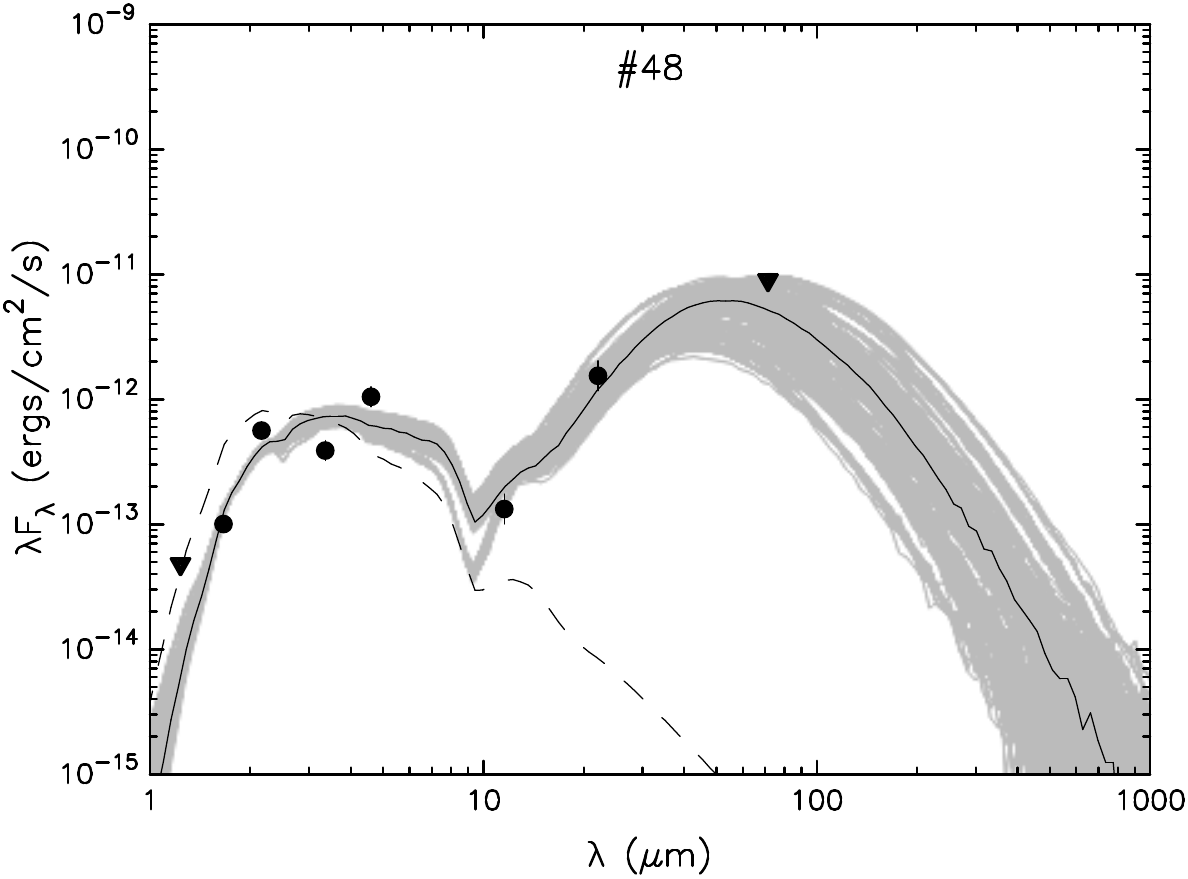}
\includegraphics[width=3.6cm] {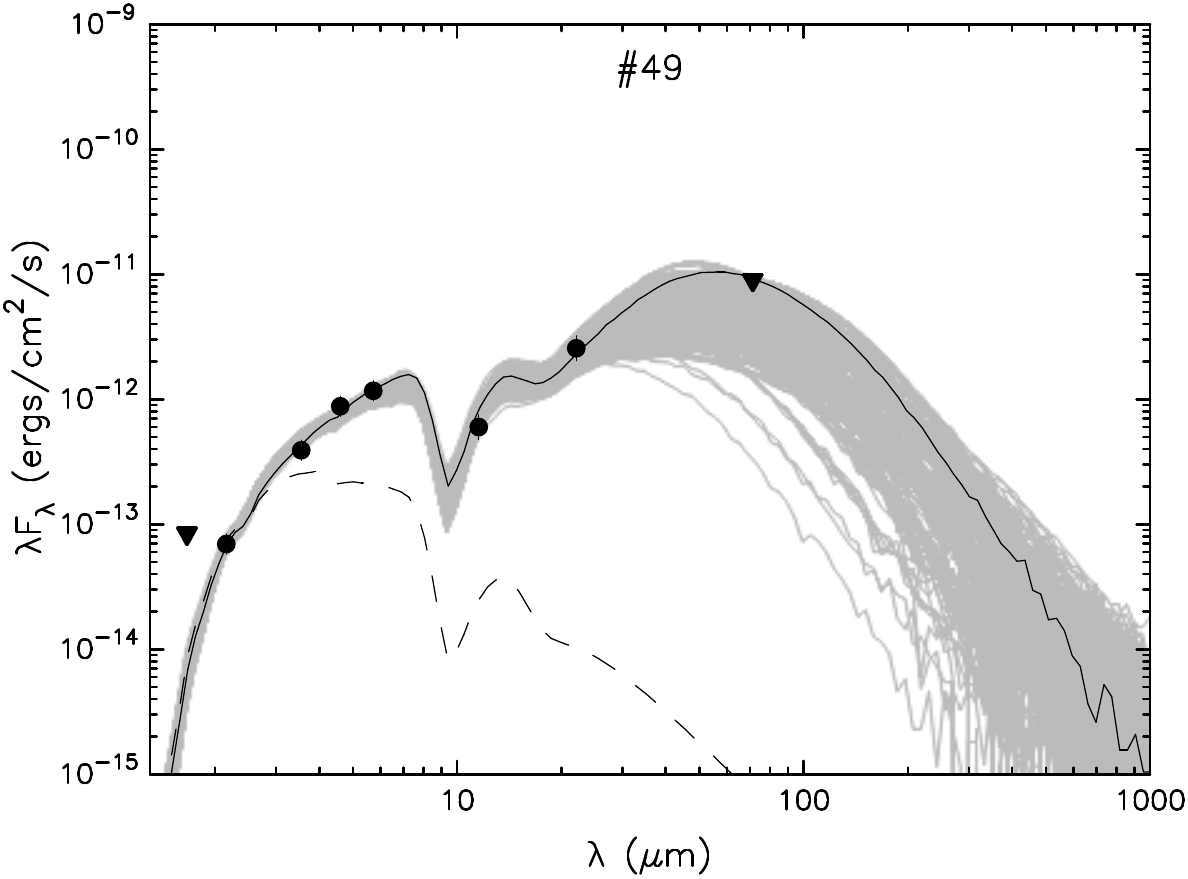}}

\caption{SEDs of the YSOs discussed in the text.
The black line shows the best fit model, and the grey lines show subsequent models 
that satisfy $\chi^2 - \chi^2_{\rm min} \leq 2N_{\rm data}$ criteria. The
dashed line shows the stellar photosphere corresponding to the
central source of the best fitting model. The circles denote the
 observed flux values. The IDs correspond to the source given in Table. 2.}
\label{FigVibStab}
\end{figure*}

\section {Discussions and Conclusions}
\subsection{Evolutionary status}

 Protostar fraction (i.e., number of protostars (Class I + Flat-spectrum) of the total number YSO population) in young clusters  is a good tracer of age. For example, 
the  protostar fraction derived from the data involving 2.2-24 $\mu$m  is 14\% in the IC 348 cluster of age $\sim$ 2-3 Myr, $\sim$ 16$\%$ in Chamaeleon II \citep{alc08}  of age $\sim$ 2-3 Myr \citep{sciortino07,spe08},
and $\sim$ 36$\%$ in the NGC 1333 cluster of age $\sim$ 1-2 Myr \citep{jorgen06}. This
suggests an evolutionary difference, with NGC 1333 being young consists of more younger population of
YSOs compared to IC 348  and  Chamaeleon II.  In the IRDC G192.76+00.10 region, we have identified a total of 62 YSOs, of which 19 are Class II YSOs and 43 are Class I plus Flat-spectrum YSOs.
This indicates that this dark cloud is unusually rich  in protostars with  fraction $\sim$ 70 \%. 
This high fraction strongly suggests that the IRDC G192.76+00.10 region is too young for most YSOs to have reached the Class II stage.
In young clusters, the error in the Class ratio in general is dominated by the detection of the actual number
of Class I, Flat-spectrum and Class II objects. If we consider all the YSOs above our completeness limit,  we find that 
the protostar fraction is still high (i.e., $\sim$ 55 \%), indicating that the region indeed contains a high percentage of  protostars.
A comparison of the protostar fraction of the IRDC G192.76+00.10 region with the IC 348, NGC 1333, and  Chamaeleon II suggests that 
the IRDC G192.76+00.10 is indeed young, possibly younger than 1 Myr.

Another possible way to assign age to a star-forming region is to use the lifetime of different phase of YSOs associated in the region.  
 \citet{evans09} derived  lifetime of different YSO phases from a large sample of YSOs collected from five  nearby molecular clouds, 
 assuming that in these clouds star formation has proceeded at a constant rate over 2 Myr period of time and  all the prestellar cores have evolved into 
 Class 0/I YSOs, then into  Flat-spectrum YSOs and then to Class II YSOs. 
Although the  derived lifetimes are still subject to large uncertainties \citep[see][and discussion therein]{evans09},  the above work  
involving a large set of YSOs suggests that in general a lifetime of  Class II, Class I, and Flat-spectrum  stages of a YSO is  roughly $\sim$ 2, $\sim$ 0.44, and 
$\sim$ 0.35 Myr, respectively.  In  the IRDC G192.76+00.10 region, we find that most of the identified YSOs  are  Class I, Flat-spectrum, and Class II in nature. 
If we use the above lifetimes, then the YSOs class statistics suggest a mean age  of $\sim$ 1 Myr for IRDC G192.76+0.10.

\subsection {Possible fragmentation process of the IRDC region}
The S254-S258 complex harbors a 20 pc long filament.  
Numerical simulations suggest that there are various processes by which filaments can be formed.
Simulation shows that the fragmentation of clouds into  sheets and filaments are the natural consequence of  supersonic turbulence 
present in  the inter-stellar medium   \citep{klessen00,padoan01,ostriker01,bate09}. 
The driving sources for large-scale turbulence could be  flows of atomic gas \citep[e.g.,][]{hennebelle08, banerjee09} or waves from supernova 
explosions and superbubbles \citep[e.g.,][]{matzner02,dale11} or  collisions between
molecular clouds \citep[e.g.,][]{tasker09}. Once filamentary 
structures formed, theoretical models suggest that they are subject to fragmentation \citep [e.g.,][]{larson85}.

For 
an  infinite and isothermal filament, the filament  is unstable to axisymmetric perturbation, 
if its line mass M$_{line}$ (i.e., mass per unit length) value exceeds its  critical equilibrium mass value 
M$_{crit}$ ~=~ $2\, c_s^2 /G$ \citep[e.g.,][]{ostriker64}, where c$_s$ is the sound speed of the medium 
and $G$ is the gravitational constant. The  critical line mass only depends on 
the gas temperature. 
The temperature of the IRDC G192.76+00.10  cloud 
is $\sim$ 14 K (discussed below), which corresponds 
to M$_{crit}$ $\sim$ 25 \msun~ pc$^{-1}$.

The projected size of the  IRDC G192.76+00.10 cloud along its long axis is about 5.7 pc.  To estimate M$_{line}$, we first 
 estimated the radial profile  of the filament from the H$_{2}$ column density  map. To do so, at several positions along the filamentary structures, 
we extracted perpendicular column density profiles and established the radial extent of each
profile using  Gaussian fitting  similar to other works 
\citep[e.g.,][]{arzoum11, smith14}. 
We then estimated M$_{line}$  by dividing the length of the filament to the 
mass estimated over the mean radial extent of the entire filament, which turns out to be  $\sim$ 120 \msun~ pc$^{-1}$. 
The observed line mass is $\sim$ 5 times larger than the critical line
mass, indicating that the filament
is supercritical, hence  susceptible to fragmentation.
Supercritical filaments are believed to be  globally unstable to radial gravitational collapse and fragment into prestellar 
clumps or cores along their major axis. 
This is supported by the roughly linear sequence of protostars observed along the long axis 
of the IRDC G192.76+00.10 region.  Recent  {\it Herschel}  observations have also shown that  supercritical filaments usually harbor several 
prestellar cores and Class 0/Class I protostars along their length \cite[e.g.,][]{andre10}.

Numerical simulation also suggests  that gravitational fragmentation is possibly  the dominant mechanism of formation of cores 
and stars in the dense regions of the filaments \citep[e.g.,][]{klessen04}. 
To further validate the above hypothesis, 
we compare our results with the simple model of gravitational fragmentation, given by \citet[][]{hartmann02}.  
In the model of \citet[][]{hartmann02}, the fragmentation length ($\lambda_c$) of a cylinder due to gravitational  instability is given by:
\be
\lambda_c ~=~ 3.94 \, c_s^2 / (G \Sigma) 
 ~ =~ 1.5 \,  T_{10}\,  A_V^{-1}\,{\rm pc} \, \label{eq:lambdac}
\en
where, $A_V$ is the visual extinction corresponding to the surface density ($\Sigma$) through
the center of the filament and  $T_{10}$ is the gas temperature 
in units of 10 K. The corresponding 
collapse timescale is given by:
\be
\tau ~\sim~ 3.7 \, T_{10}^{1/2} \, A_V^{-1}\, {\rm Myr}\,. \label{eq:tgrowth}
\en

{\it Herschel} estimates of dust temperatures  indicate 
that the temperature of the quiescent and star-forming filaments are in 
the range of 10 K to 15 K {\citep{arzoum11}.
Observations of the ammonia (1, 1) and (2, 2) inversion transitions suggest that the kinetic temperature in the dense part of the IRDC G192.76+00.10 filament is in the range of 13 to 15 K \citep{dunham10}.
Considering  14 K as the temperature of the filament and 10 mag  as the 
mean visual extinction, the above equation predicts 
$\lambda_c$ $\sim$ 0.21 pc and $\tau \sim 0.45$~Myr. 
For  randomly oriented filament, the line of sight median inclination 
angle could be $\sim$ 60\degree~ \citep{genzel89}. If this is the case, 
then this would decrease $\lambda_c$ to a value 0.19 pc.  
Most of the YSOs in the filament are protostellar in nature.  The lifetime of embedded protostars is around 0.5 Myr. 
The predicted  fragmentation length  and  collapse time  for the
region  are comparable to the  observed  median  NN separation ($\sim$ 0.2 pc)
and age of majority of the YSOs; thus the model predictions are broadly in agreement with the observations.

\subsubsection {Limitations}
The scenario presented above is roughly consistent  with the results  discussed in \citet{Ball04}  and with the scenario preferred for 
the YSOs formation in the Taurus  \citep{hartmann02}. 
However,  there are number of issues yet to be addressed.  For example,  
the exact  role of turbulence prior to fragmentation is not known to us; its presence would increase the effective 
sound speed, thus would increase the critical mass per unit length. 
However, it is suggested that pre-stellar cloud is most-likely stabilized against global collapse by
interstellar turbulence, but this support is quickly removed on small spatial scales  \citep[e.g.,][]{elmegreen00}.  
Recent observations  suggest that the core formation in filamentary cloud is possibly a two-step process, in which
first, the subsonic, velocity-coherent filaments  condense
out of the more turbulent ambient cloud. Then, the cores  fragment 
quasi-statically, where turbulence seems to play a little or no role 
in the formation of the individual cores  \citep[e.g.,][]{Ball04,andre11,hacer11}. 
Based on ammonia observations,  \citet{dunham10} observed non-thermal velocity dispersion of $\sim$0.43 \kms~ in the central region
of  IRDC G192.76+00.10. However, non-thermal dispersion in clouds could be of various origin. For example,  non-thermal
motions can be driven by the accretion flows to the filament potential \citep[e.g.,][]{kle10}, 
if the IRDC G192.76+00.10 filament has undergone dynamical evolution after fragmenaion.
 Similarly, protostellar feedback can also drive supersonic turbulence at 
sub-parsec size scales \citep[e.g.,][]{ana12}. If the non-thermal motion (corresponding to the velocity dispersion) is to be considered in addition to the thermal 
component  to support the filament, this would then increase the effective sound speed of the filament  \citep[e.g., see][]{mac88}. As a consequence of this, the  M$_{crit}$ of the 
filament would increase to  $\sim$ 75 \msun~ pc$^{-1}$, which is still lower than than the observed M$_{line}$,  implying that the total amount of support available
is still insufficient to keep the filament in equilibrium, hence the filament should be radially contracting.

Other physical processes that might affect the model parameters discussed above, are helical magnetic field \citep{fiege00} and star formation due to dynamical effects 
such as   flows from surrounding sub-filaments  \citep[e.g.,][]{smith14,schneider10}. 
Helical magnetic fields are believed to  
decrease the critical length of fragmentation  of a cylinder \citep{fiege00},
whose effect in the present case is  difficult to assess. However,
in case of star formation due to converging flows, higher  concentration of dense gas is expected at the junction point of the converging 
sub-filaments and the main filament, which tend to increase star formation activity. For example, in the DR21 region several low-density striations
or sub-filaments were observed perpendicular to the main filament
and apparently feeding matter to the main filament \citep[e.g.,][]{kumar07,schneider10,henn12}.
Considering that the IRDC G192.76+00.10 region 
is slightly structured at its eastern side, the possibility that few stars might have formed by other dynamical processes 
can not be ignored. However, the morphology of the IRDC G192.76+00.10 area 
is largely linear in CO and we do not see strong morphological signature of 
such structures (e.g.,  thin long sub-filaments) that 
are radially attached to the IRDC G192.76+00.10 region, though this cannot be warranted without high resolution and sensitive observations.
Nonetheless, to limit the impact of dynamical processes on our estimations, 
we calculated again the model parameters, without accounting for the group of stars seen  around $\alpha_{2000}$=93.50 \& $\delta_{2000}$= +17.87 (i.e., the group of stars seen distinctly away from the filament's long axis), as well as a few stars located far away from the 
highest density line (see Fig. 7). Considering tentatively that these are  stars that might have 
formed by some other processes, we found that the model parameters are changed by 20\% only.

We also stress that even though smooth cylindrical models  are useful to understand
star formation processes in filaments, they can only represent a first order approximation because 
real clouds are likely to have much more  density inhomogeneities  than what we have assumed here. 
In addition to above, the  initial cloud configuration also plays a decisive role where stars would form.  
For example, if the original parental cloud is not infinitely long or spherical, gravitational edge focusing results in enhanced
 concentrations of mass at one end  of the filament, where 
local collapse would proceed faster than  collapse of the entire filament \citep[e.g.,][]{burkert04}. 
We do not observe high concentration of stars or massive condensations at any end of the 
IRDC G192.76+00.10 region, thus the gravitational edge focusing effect is not preferably happening here.

In summary,  the caveats outlined above prevent us to make any definite statement on the YSO formation; however, if we 
consider the role of the helical magnetic field, and star formation by other dynamical processes are minimum, then we find a
remarkable reconciliation of the observed properties with the  
model predictions, suggesting that gravitational 
fragmentation is probably the dominant cause of core formation in 
the IRDC G192.76+00.10 filamentary  cloud.
Confirming and refining  
the scenario require  high angular resolution mid-infrared, dust polarimetric and 
millimeter line observations to identify  other possible low-mass YSOs, to constrain roles of
magnetic field, and influence of gravity, respectively in the star formation processes  of the region.
\subsection{Star formation efficiency and rate}
 In this section, we discuss the star formation efficiency (SFE) and star formation rate (SFR) of the  IRDC G192.76+00.10  region based 
on the YSOs identified in the present work. SFE and SFR are the fundamental physical parameters  that are essential for the understanding of 
evolution of star-forming regions and galaxies. 

SFE  is defined as the ratio of the total stellar mass to 
the total mass of stars and gas.  We estimate the total gaseous mass (M$_{gas}$) associated to the IRDC region $\sim$ 1100 $\msun$.
To estimate the M$_{gas}$, we used the $^{13}$CO  column density map and  integrated the column density  
 four times above local background, i.e.,  column density value  $>$ 1.1 $\times$ 10$^{16}$ cm$^{-2}$ over 11 pc$^2$ area.
 We then  converted
the total $^{13}$CO  column density to the total H$_2$ column density as discussed in Sect. 3.5, and  used the 
following equation to estimate the mass:
\be
M = \mu {\rm m_{H}} A_{\rm pix} \Sigma {\rm H_{2}} \, \label{eq:mass}
\en
where, $\mu$ is the mean molecular weight, ${\rm m_{H}}$ is the mass of the hydrogen atom, $\Sigma$  H$_2$ is the summed 
H$_2$ column density, and $A_{\rm pix}$ is the area of a pixel in cm$^{-2}$ at the distance of the region. 
The determination of the stellar mass in accreting stars is not trivial, as emission due to circumstellar disk surrounding the star and accreting material from the 
protoplanetary disk onto the central star affects the observed spectrum. 
Broad-band spectroscopic observation  would reveal  more accurate mass 
of the YSOs \citep[e.g.,][]{manara13};} however, if we assume that each source of the IRDC G192.76+00.10  region has a mass 
of 0.5 \msun  (close to the mean stellar mass of the region; see Sect. 3.4), consistent with the characteristic mass from the studies 
of \citet{cha03} and \citet{kro01}  type of initial mass function (IMF), then 
 the total mass of all the YSOs (M$_{YSOs}$) is $\sim$ 30\msun. Using the  M$_{gas}$ and  M$_{YSOs}$, 
we  estimated  the SFE  $\sim$ 0.03 or 3\%  in the IRDC G192.76+00.10 region.  
.

SFR describes the rate at which the gas in a cloud is turning into stars. Assuming 1 Myr as the duration of star formation (see Sect. 4.1) and 
using the derived  M$_{gas}$ and  M$_{YSOs}$, we estimated the 
SFR (=  M$_{gas}$ $\times$SFE/$t_{sf}$, where  $t_{sf}$ is 
the star formation time scale) of the  IRDC G192.76+00.10 region $\sim$ 30 $\msun$~ Myr$^{-1}$.  
The projected area of the IRDC G192.76+00.10  region over which we estimated the cloud mass is $\sim$ 11 pc$^{2}$. 
If we normalized the derived SFR  by the cloud area, this leads to SFR per unit surface area  $\sim$ 3 $\msun$~ Myr$^{-1}$  pc$^{-2}$.
The SFE and SFR per unit surface area of the 
IRDC G192.76+00.10 region  are comparable to the  SFE (i.e., 3-6\%) and SFR per unit surface area 
(i.e., 0.6-3.2 $\msun$~ Myr$^{-1}$  pc$^{-2}$) values 
reported by \citet{evans09} for nearby  molecular cloud complexes.

Since our YSOs sample  is  complete down to $\sim$ 0.15 $\msun$, thus we may be missing a population of deeply embedded YSOs of masses below  
this completeness level. If we assume that the cloud has already  formed stars down to the hydrogen burning limit (i.e., $\sim$ 0.08 \msun), this would 
not drastically alter the SFE of the complex, because most of the observations in our Galaxy are consistent with an IMF that declines below 0.1 
\msun \citep[e.g.,][]{lada05,olive07,and08,benoit15}.  Thus, we do not expect a
large population of Class II and Class I YSOs below our completeness limit. 
The shape and universality of the IMF at the sub-stellar regime 
is still under debate though, however, assuming \citet{cha03}  
type of mass function, we may probably 
be missing only about 20\% of the total number of YSOs.  
This may increase the  SFE of the region by only about 1\%,  assuming each YSO has a mass 0.5 \msun.

\subsection{Emerging young cluster}

The detection and identification of clumps and cores in star-forming complexes using long wavelength observations 
depend actually on the resolution and sensitivity  of surveys. For example, the 
 dust clump ``P1'' of the G28.34+0.06 region detected with the IRAM 30 m
telescope and JCMT at resolutions from 11\arcs to 15\arcs \citep{rath06,carey98} are resolved by 
the  Submillimeter Array (SMA) into
five cores at 1\arcsec.2 resolution \citep{zhang09}. So far, no high resolution observations have been conducted to identify clumps and cores of the  IRDC G192.76+00.10 region. In this work,
we have detected protostellar sources of mass down to 0.15 \msun,  so millimeter continuum observations of mass sensitivities  better than 0.45 \msun~ 
(assuming SFE of cores is $\sim$ 30\%; \citet[][]{alves07}) and spatial resolution better than {\it Spitzer} observations would 
be very helpful for the detection of resolved cores in  the region. Nonetheless, using 1.1 millimeter Bolocam Galactic Plane Survey (BGPS; beam $\sim$ 33\arcsec), \citet{dunham10} observed an elongated clump (their ID 50) at the center area  of IRDC G192.76+00.10. From their observation,
we estimated the  virial parameter  \citep[][]{bertoldi92} of the clump $\sim$ 0.3.  Virial parameter represents   
the dynamical stability of a core or clump. A clump is assumed to
be gravitationally bound if its virial parameter value is $\leq$ 1, unbound otherwise.
The virial paramete value of the BGPS clump  suggests the clump is susceptible to gravitational collapse, and  can lead to star formation.

The filament in which the BGPS clump is embedded, 
is in  super-critical (discussed in Sect. 4.2.1) stage, implying that the  filamentary cloud may be  contracting.  
Moreover, observations and simulations of filaments show  accretion flows along the filaments onto the
clumps and cores located at the bottom of their potential \citep{bal01,kirk13,tack14}, and some clumps and cores possibly accrete mass faster than others 
due to their higher  gravitational potential \citep[e.g.,][]{tack14}.
Observations suggest that star-forming clouds  dynamically evolve on time-scales of a few Myr. If it is to be believed that  cluster 
formation mainly occurs in filamentary clouds, then the lifetime of the typical filament should be a few Myr, since  clusters with age greater than 5 Myr 
are found to be seldom associated with molecular gas \citep[e.g.,][]{leisa89}.  
In Sect. 4.1, we have  estimated that the  age of the IRDC G192.76+00.10  region is around 
1 Myr. If we put in context all our results, they suggest that the YSOs of the IRDC G192.76+00.10 region are young, they are embedded in a gas reservoir of mass $\sim$ 1100 \msun, and the 
dark cloud
is forming stars at a high rate of $\sim$ 30 $\msun$~ Myr$^{-1}$ and is part of a large filament environment where individual clumps and cores can grow in mass by  
accreting matter from the parental filament to their potentials. These evidences suggest that if the dark cloud will evolve 
dynamically for a few Myr, then the possibility that it may emerge to a richer cluster exists.

\subsection {Overall picture of star formation in the complex}
 The 20 pc long filament of the S254-S258 complex consists of six evolved \hii regions. Among these, 
four \hii regions (S254, S255, S256, and S257) are located at the center of the complex at a mean separation of 2.6 pc (see Fig. 1).
The exciting stars of  S254, S255 and S257 are isolated sources,
without any co-spatial clustering of low-mass stars around them. 
The mean dynamical age of these evolved \hii regions is $\sim$ 2.5 Myr \citep[see][]{chava08}.
It seems that the evolved massive stars (exciting stars of evolved \hii regions) in 
this complex must have formed about 2.5 Myr ago. 

In this work, we characterized the protostellar population of the  IRDC G192.76+00.10 region, which is located
farther away from the evolved \hii regions of the complex. 
Although the low-mass 
YSOs in the IRDC G192.76+00.10 region are at different evolutionary stages, the majority ($\sim$ 70\%) of them 
are Class I and Flat-spectrum in nature.  The  typical lifetime of  Class I and Flat-spectrum objects is around 0.4
Myr \citep{evans09}. Therefore, it is very likely that the majority of the YSOs in the  IRDC G192.76+00.10 region
were formed  simultaneously or over a narrow  range of time, possibly less than 0.5 Myr of time.

The evolutionary status of the hig-mass OB stars (age $\sim$ 2.5 Myr) and  low-mass protostars of the IRDC G192.76+00.10 region (age $\sim$ 0.5 Myr) suggests  existence of 
multi-generation star formation in the complex. 
The evolved \hii regions (e.g., S255, and S257) appear to be projected
on the long filament.  If  all the optically visible 
evolved \hii regions  and the low-mass protostars of the IRDC G192.76+00.10 
region are part of the  same long ($\sim$ 20 pc) filament, it is then plausible to think that 
multiple generations of stars may be residing in the whole filament. If this is the case, then the whole 
filament probably has not fragmented  as a single entity. 
Hierarchical fragmentation has also been  observed in several filamentary clouds \citep[e.g.,][]{hacer11,taka13}.
For example, in Orion filamentary cloud, \citet{taka13} found different fragmentation scales for large-scale clumps (9-10 pc), 
small-scale clumps ($\sim$ 2.5 pc), and dense cores (0.15-0.55 pc), and interpreted this result as the  hierarchical 
fragmentation of Orion A filament. Thus, the possibility that the long filament of the S254-S258 complex has undergone  hierarchical fragmentation can not be ruled out.
Hierarchical fragmentation in clouds could be of various origins. For example, it could be due to combined effect of  turbulence, cloud temperature, 
magnetic field, and ionization from the massive stars as discussed in \citet[][]{taka13} or as discussed in \citet{semadeni09}, where low-mass and 
high-mass star forming regions  can arise from hierarchical gravitational collapse of large cloud, where the former regions arise from collapse of small-scale fragments, 
while the latter may appear due to large-scale  collapse.

\subsection{Concluding remarks}
 In this paper, we have presented a  dark cloud (mainly consists of low-mass stars) based on MIPS, WISE and JHK
photometry.  It harbors 62 YSOs, 49 
of which contain 24 $\mu$m sources, indicating that the cluster  is  young.
Interestingly, the dark cloud  is found in the filament of 
an OB complex, with  members distributed in an aligned fashion 
along the long axis of the filament and at 
a NN separation $\sim$ 0.2 pc.  We find that the protostars 
to Class II ratio in the dark cloud is high ($\sim$ 2). When compared to other star-forming regions, it indicates that 
the region is young, possibly younger than  1 Myr. 
 From our results, we  also find 
evidences favoring gravitational fragmentation as the dominant cause of YSOs formation in this dark cloud, but a   
conclusive answer would require a further large-scale study of the filament concerning 
magnetic field, turbulence, and gravity.
The  lack of deep infrared observations 
prevent us making a definite conclusion on the properties of the cloud; however, from the present observations, 
we argue that the cloud is currently forming a cluster at an efficiency of $\sim$
3\% and a rate of  $\sim$ 30 $\msun$ Myr$^{-1}$. The cluster is embedded in a supercritical filamentary 
 dark cloud of mass $\sim$ 1100 \msun~ and is part of a large filament geometry. We hypothesized that 
it may become a richer cluster with time. 
Due to the lack of massive protostars,  the possibility of radiative heating 
and  photo-evaporation of the dense gas in the studied filament region is low, 
hence the dark cloud provides an unique opportunity with high sensitive observations to gain a deeper 
understanding of cluster formation and evolution within the filamentary environment.

\section*{Acknowledgments}
 We thank the anonymous referee for a critical reading of
the paper and several useful constructive comments and suggestions, which greatly improved
the scientific content of the paper.
We thank Dr. L. Chavarria for allowing us to use their CO column density map.
MRS thanks the Tata Institute of Fundamental Research (TIFR) for the kind 
hospitality during his visits to the institute, where a part of the work 
reported was carried out. MRS  also acknowledges the financial
support provided by the French Space Agency (CNES) for his postdoctoral fellowship.
DKO and IZ acknowledge support 
from DST-RFBR project (P-142; 13-02-92627). The I.Z. research was partly supported
by the grant within the agreement No. 02.B.49.21.0003 between 
the Ministry of education and science of the Russian Federation and Lobachevsky State University of Nizhni Novgorod. 

\bibliography{myref}

\begin{thebibliography}{138}
\expandafter\ifx\csname natexlab\endcsname\relax\def\natexlab#1{#1}\fi

\bibitem[{{Alcal{\'a}} {et~al.}(2008){Alcal{\'a}}, {Spezzi}, {Chapman},
  {Evans}, {Huard}, {J{\o}rgensen}, {Mer{\'{\i}}n}, {Stapelfeldt}, {Covino},
  {Frasca}, {Gandolfi}, \& {Oliveira}}]{alc08}
{Alcal{\'a}}, J.~M., {Spezzi}, L., {Chapman}, N., {et~al.} 2008, \apj, 676, 427

\bibitem[{{Alves} {et~al.}(2007){Alves}, {Lombardi}, \& {Lada}}]{alves07}
{Alves}, J., {Lombardi}, M., \& {Lada}, C.~J. 2007, \aap, 462, L17

\bibitem[{{Andersen} {et~al.}(2008){Andersen}, {Meyer}, {Greissl}, \&
  {Aversa}}]{and08}
{Andersen}, M., {Meyer}, M.~R., {Greissl}, J., \& {Aversa}, A. 2008, \apjl,
  683, L183

\bibitem[{{Andr{\'e}}(2013)}]{andre13}
{Andr{\'e}}, P. 2013, arXiv1309.7762A

\bibitem[{{Andr{\'e}} {et~al.}(2010){Andr{\'e}}, {Men'shchikov}, {Bontemps},
  {K{\"o}nyves}, {Motte}, {Schneider}, {Didelon}, {Minier}, {Saraceno},
  {Ward-Thompson}, {di Francesco}, {White}, {Molinari}, {Testi}, {Abergel},
  {Griffin}, {Henning}, {Royer}, {Mer{\'{\i}}n}, {Vavrek}, {Attard},
  {Arzoumanian}, {Wilson}, {Ade}, {Aussel}, {Baluteau}, {Benedettini},
  {Bernard}, {Blommaert}, {Cambr{\'e}sy}, {Cox}, {di Giorgio}, {Hargrave},
  {Hennemann}, {Huang}, {Kirk}, {Krause}, {Launhardt}, {Leeks}, {Le Pennec},
  {Li}, {Martin}, {Maury}, {Olofsson}, {Omont}, {Peretto}, {Pezzuto}, {Prusti},
  {Roussel}, {Russeil}, {Sauvage}, {Sibthorpe}, {Sicilia-Aguilar}, {Spinoglio},
  {Waelkens}, {Woodcraft}, \& {Zavagno}}]{andre10}
{Andr{\'e}}, P., {Men'shchikov}, A., {Bontemps}, S., {et~al.} 2010, \aap, 518,
  L102

\bibitem[{{Andr{\'e}} {et~al.}(2011){Andr{\'e}}, {Men'shchikov}, {K{\"o}nyves},
  \& {Arzoumanian}}]{andre11}
{Andr{\'e}}, P., {Men'shchikov}, A., {K{\"o}nyves}, V., \& {Arzoumanian}, D.
  2011, in IAU Symposium, Vol. 270, Computational Star Formation, ed.
  J.~{Alves}, B.~G. {Elmegreen}, J.~M. {Girart}, \& V.~{Trimble}, 255--262

\bibitem[{{Andr{\'e}} {et~al.}(2000){Andr{\'e}}, {Ward-Thompson}, \&
  {Barsony}}]{and00}
{Andr{\'e}}, P., {Ward-Thompson}, D., \& {Barsony}, M. 2000, Protostars and
  Planets IV, 59

\bibitem[{{Arzoumanian} {et~al.}(2011){Arzoumanian}, {Andr{\'e}}, {Didelon},
  {K{\"o}nyves}, {Schneider}, {Men'shchikov}, {Sousbie}, {Zavagno}, {Bontemps},
  {di Francesco}, {Griffin}, {Hennemann}, {Hill}, {Kirk}, {Martin}, {Minier},
  {Molinari}, {Motte}, {Peretto}, {Pezzuto}, {Spinoglio}, {Ward-Thompson},
  {White}, \& {Wilson}}]{arzoum11}
{Arzoumanian}, D., {Andr{\'e}}, P., {Didelon}, P., {et~al.} 2011, \aap, 529, L6

\bibitem[{{Ballesteros-Paredes}(2004)}]{Ball04}
{Ballesteros-Paredes}, J. 2004, \apss, 292, 193

\bibitem[{{Balsara} {et~al.}(2001){Balsara}, {Ward-Thompson}, \&
  {Crutcher}}]{bal01}
{Balsara}, D., {Ward-Thompson}, D., \& {Crutcher}, R.~M. 2001, \mnras, 327, 715

\bibitem[{{Banerjee} {et~al.}(2009){Banerjee}, {V{\'a}zquez-Semadeni},
  {Hennebelle}, \& {Klessen}}]{banerjee09}
{Banerjee}, R., {V{\'a}zquez-Semadeni}, E., {Hennebelle}, P., \& {Klessen},
  R.~S. 2009, \mnras, 398, 1082

\bibitem[{{Baraffe} {et~al.}(2003){Baraffe}, {Chabrier}, {Barman}, {Allard}, \&
  {Hauschildt}}]{bar03}
{Baraffe}, I., {Chabrier}, G., {Barman}, T.~S., {Allard}, F., \& {Hauschildt},
  P.~H. 2003, \aap, 402, 701

\bibitem[{{Bate}(2009)}]{bate09}
{Bate}, M.~R. 2009, \mnras, 397, 232

\bibitem[{{Battersby} {et~al.}(2011){Battersby}, {Bally}, {Ginsburg},
  {Bernard}, {Brunt}, {Fuller}, {Martin}, {Molinari}, {Mottram}, {Peretto},
  {Testi}, \& {Thompson}}]{bata11}
{Battersby}, C., {Bally}, J., {Ginsburg}, A., {et~al.} 2011, \aap, 535, A128

\bibitem[{{Bertoldi} \& {McKee}(1992)}]{bertoldi92}
{Bertoldi}, F. \& {McKee}, C.~F. 1992, \apj, 395, 140

\bibitem[{{Beuther} \& {Sridharan}(2007)}]{beuth07}
{Beuther}, H. \& {Sridharan}, T.~K. 2007, \apj, 668, 348

\bibitem[{{Bieging} {et~al.}(2009){Bieging}, {Peters}, {Vila Vilaro},
  {Schlottman}, \& {Kulesa}}]{beig09}
{Bieging}, J.~H., {Peters}, W.~L., {Vila Vilaro}, B., {Schlottman}, K., \&
  {Kulesa}, C. 2009, \aj, 138, 975

\bibitem[{{Bohlin} {et~al.}(1978){Bohlin}, {Savage}, \& {Drake}}]{bohlin78}
{Bohlin}, R.~C., {Savage}, B.~D., \& {Drake}, J.~F. 1978, \apj, 224, 132

\bibitem[{{Brand} {et~al.}(2011){Brand}, {Massi}, {Zavagno}, {Deharveng}, \&
  {Lefloch}}]{brand11}
{Brand}, J., {Massi}, F., {Zavagno}, A., {Deharveng}, L., \& {Lefloch}, B.
  2011, \aap, 527, A62

\bibitem[{{Burkert} \& {Hartmann}(2004)}]{burkert04}
{Burkert}, A. \& {Hartmann}, L. 2004, \apj, 616, 288

\bibitem[{{Carey} {et~al.}(1998){Carey}, {Clark}, {Egan}, {Price}, {Shipman},
  \& {Kuchar}}]{carey98}
{Carey}, S.~J., {Clark}, F.~O., {Egan}, M.~P., {et~al.} 1998, \apj, 508, 721

\bibitem[{{Carey} {et~al.}(2009){Carey}, {Noriega-Crespo}, {Mizuno}, {Shenoy},
  {Paladini}, {Kraemer}, {Price}, {Flagey}, {Ryan}, {Ingalls}, {Kuchar},
  {Pinheiro Gon{\c c}alves}, {Indebetouw}, {Billot}, {Marleau}, {Padgett},
  {Rebull}, {Bressert}, {Ali}, {Molinari}, {Martin}, {Berriman}, {Boulanger},
  {Latter}, {Miville-Deschenes}, {Shipman}, \& {Testi}}]{carey09}
{Carey}, S.~J., {Noriega-Crespo}, A., {Mizuno}, D.~R., {et~al.} 2009, \pasp,
  121, 76

\bibitem[{{Carpenter} {et~al.}(1995){Carpenter}, {Snell}, \&
  {Schloerb}}]{carpen95}
{Carpenter}, J.~M., {Snell}, R.~L., \& {Schloerb}, F.~P. 1995, \apj, 450, 201

\bibitem[{{Casali} {et~al.}(2007){Casali}, {Adamson}, {Alves de Oliveira},
  {Almaini}, {Burch}, {Chuter}, {Elliot}, {Folger}, {Foucaud}, {Hambly},
  {Hastie}, {Henry}, {Hirst}, {Irwin}, {Ives}, {Lawrence}, {Laidlaw}, {Lee},
  {Lewis}, {Lunney}, {McLay}, {Montgomery}, {Pickup}, {Read}, {Rees}, {Robson},
  {Sekiguchi}, {Vick}, {Warren}, \& {Woodward}}]{casali07}
{Casali}, M., {Adamson}, A., {Alves de Oliveira}, C., {et~al.} 2007, \aap, 467,
  777

\bibitem[{{Chabrier}(2003)}]{cha03}
{Chabrier}, G. 2003, \pasp, 115, 763

\bibitem[{{Chavarr{\'{\i}}a} {et~al.}(2008){Chavarr{\'{\i}}a}, {Allen}, {Hora},
  {Brunt}, \& {Fazio}}]{chava08}
{Chavarr{\'{\i}}a}, L.~A., {Allen}, L.~E., {Hora}, J.~L., {Brunt}, C.~M., \&
  {Fazio}, G.~G. 2008, \apj, 682, 445

\bibitem[{{Chen} {et~al.}(2011){Chen}, {Pandey}, {Sharma}, {Chen}, {Chen},
  {Sperauskas}, {Ogura}, {Chuang}, \& {Boyle}}]{chen11}
{Chen}, W.~P., {Pandey}, A.~K., {Sharma}, S., {et~al.} 2011, \aj, 142, 71

\bibitem[{{Chu} \& {Gruendl}(2008)}]{chu08}
{Chu}, Y.-H. \& {Gruendl}, R.~A. 2008, in Astronomical Society of the Pacific
  Conference Series, Vol. 387, Massive Star Formation: Observations Confront
  Theory, ed. H.~{Beuther}, H.~{Linz}, \& T.~{Henning}, 415

\bibitem[{{Condon} {et~al.}(1998){Condon}, {Cotton}, {Greisen}, {Yin},
  {Perley}, {Taylor}, \& {Broderick}}]{condon98}
{Condon}, J.~J., {Cotton}, W.~D., {Greisen}, E.~W., {et~al.} 1998, \aj, 115,
  1693

\bibitem[{{Connelley} \& {Greene}(2010)}]{conn10}
{Connelley}, M.~S. \& {Greene}, T.~P. 2010, \aj, 140, 1214

\bibitem[{{Crapsi} {et~al.}(2008{\natexlab{a}}){Crapsi}, {van Dishoeck},
  {Hogerheijde}, {Pontoppidan}, \& {Dullemond}}]{crapsi08}
{Crapsi}, A., {van Dishoeck}, E.~F., {Hogerheijde}, M.~R., {Pontoppidan},
  K.~M., \& {Dullemond}, C.~P. 2008{\natexlab{a}}, \aap, 486, 245

\bibitem[{{Crapsi} {et~al.}(2008{\natexlab{b}}){Crapsi}, {van Dishoeck},
  {Hogerheijde}, {Pontoppidan}, \& {Dullemond}}]{carspi08}
{Crapsi}, A., {van Dishoeck}, E.~F., {Hogerheijde}, M.~R., {Pontoppidan},
  K.~M., \& {Dullemond}, C.~P. 2008{\natexlab{b}}, \aap, 486, 245

\bibitem[{{Cutri} \& {et al.}(2012)}]{cutri12}
{Cutri}, R.~M. \& {et al.} 2012, VizieR Online Data Catalog, 2311, 0

\bibitem[{{Cutri} {et~al.}(2003){Cutri}, {Skrutskie}, {van Dyk}, {Beichman},
  {Carpenter}, {Chester}, {Cambresy}, {Evans}, {Fowler}, {Gizis}, {Howard},
  {Huchra}, {Jarrett}, {Kopan}, {Kirkpatrick}, {Light}, {Marsh}, {McCallon},
  {Schneider}, {Stiening}, {Sykes}, {Weinberg}, {Wheaton}, {Wheelock}, \&
  {Zacarias}}]{cutri03}
{Cutri}, R.~M., {Skrutskie}, M.~F., {van Dyk}, S., {et~al.} 2003, VizieR Online
  Data Catalog, 2246, 0

\bibitem[{{Dahm}(2008)}]{dahm08}
{Dahm}, S.~E. 2008, \aj, 136, 521

\bibitem[{{Dale} \& {Bonnell}(2011)}]{dale11}
{Dale}, J.~E. \& {Bonnell}, I. 2011, \mnras, 414, 321

\bibitem[{{De Marchi} {et~al.}(2011){De Marchi}, {Panagia}, \&
  {Sabbi}}]{marchi11}
{De Marchi}, G., {Panagia}, N., \& {Sabbi}, E. 2011, \apj, 740, 10

\bibitem[{{Deharveng} {et~al.}(2010){Deharveng}, {Schuller}, {Anderson},
  {Zavagno}, {Wyrowski}, {Menten}, {Bronfman}, {Testi}, {Walmsley}, \&
  {Wienen}}]{deha10}
{Deharveng}, L., {Schuller}, F., {Anderson}, L.~D., {et~al.} 2010, \aap, 523,
  A6

\bibitem[{{Deharveng} {et~al.}(2012){Deharveng}, {Zavagno}, {Anderson},
  {Motte}, {Abergel}, {Andr{\'e}}, {Bontemps}, {Leleu}, {Roussel}, \&
  {Russeil}}]{dehar12}
{Deharveng}, L., {Zavagno}, A., {Anderson}, L.~D., {et~al.} 2012, \aap, 546,
  A74

\bibitem[{{Duarte-Cabral} {et~al.}(2012){Duarte-Cabral}, {Chrysostomou},
  {Peretto}, {Fuller}, {Matthews}, {Schieven}, \& {Davis}}]{ana12}
{Duarte-Cabral}, A., {Chrysostomou}, A., {Peretto}, N., {et~al.} 2012, \aap,
  543, A140

\bibitem[{{Dunham} {et~al.}(2010){Dunham}, {Rosolowsky}, {Evans}, {Cyganowski},
  {Aguirre}, {Bally}, {Battersby}, {Bradley}, {Dowell}, {Drosback}, {Ginsburg},
  {Glenn}, {Harvey}, {Merello}, {Schlingman}, {Shirley}, {Stringfellow},
  {Walawender}, \& {Williams}}]{dunham10}
{Dunham}, M.~K., {Rosolowsky}, E., {Evans}, II, N.~J., {et~al.} 2010, \apj,
  717, 1157

\bibitem[{{Elmegreen} {et~al.}(2000){Elmegreen}, {Efremov}, {Pudritz}, \&
  {Zinnecker}}]{elmegreen00}
{Elmegreen}, B.~G., {Efremov}, Y., {Pudritz}, R.~E., \& {Zinnecker}, H. 2000,
  Protostars and Planets IV, 179

\bibitem[{{Elmegreen} \& {Lada}(1977)}]{elm77}
{Elmegreen}, B.~G. \& {Lada}, C.~J. 1977, \apj, 214, 725

\bibitem[{{Engelbracht} {et~al.}(2007){Engelbracht}, {Blaylock}, {Su}, {Rho},
  {Rieke}, {Muzerolle}, {Padgett}, {Hines}, {Gordon}, {Fadda},
  {Noriega-Crespo}, {Kelly}, {Latter}, {Hinz}, {Misselt}, {Morrison},
  {Stansberry}, {Shupe}, {Stolovy}, {Wheaton}, {Young}, {Neugebauer},
  {Wachter}, {P{\'e}rez-Gonz{\'a}lez}, {Frayer}, \& {Marleau}}]{enge07}
{Engelbracht}, C.~W., {Blaylock}, M., {Su}, K.~Y.~L., {et~al.} 2007, \pasp,
  119, 994

\bibitem[{{Evans} {et~al.}(2003){Evans}, {Allen}, {Blake}, {Boogert}, {Bourke},
  {Harvey}, {Kessler}, {Koerner}, {Lee}, {Mundy}, {Myers}, {Padgett},
  {Pontoppidan}, {Sargent}, {Stapelfeldt}, {van Dishoeck}, {Young}, \&
  {Young}}]{eva03}
{Evans}, II, N.~J., {Allen}, L.~E., {Blake}, G.~A., {et~al.} 2003, \pasp, 115,
  965

\bibitem[{{Evans} {et~al.}(2009){Evans}, {Dunham}, {J{\o}rgensen}, {Enoch},
  {Mer{\'{\i}}n}, {van Dishoeck}, {Alcal{\'a}}, {Myers}, {Stapelfeldt},
  {Huard}, {Allen}, {Harvey}, {van Kempen}, {Blake}, {Koerner}, {Mundy},
  {Padgett}, \& {Sargent}}]{evans09}
{Evans}, II, N.~J., {Dunham}, M.~M., {J{\o}rgensen}, J.~K., {et~al.} 2009,
  \apjs, 181, 321

\bibitem[{{Fiege} \& {Pudritz}(2000)}]{fiege00}
{Fiege}, J.~D. \& {Pudritz}, R.~E. 2000, \mnras, 311, 105

\bibitem[{{Flaherty} {et~al.}(2007){Flaherty}, {Pipher}, {Megeath}, {Winston},
  {Gutermuth}, {Muzerolle}, {Allen}, \& {Fazio}}]{fla07}
{Flaherty}, K.~M., {Pipher}, J.~L., {Megeath}, S.~T., {et~al.} 2007, \apj, 663,
  1069

\bibitem[{{Foster} {et~al.}(2015){Foster}, {Cottaar}, {Covey}, {Arce}, {Meyer},
  {Nidever}, {Stassun}, {Tan}, {Chojnowski}, {da Rio}, {Flaherty}, {Rebull},
  {Frinchaboy}, {Majewski}, {Skrutskie}, {Wilson}, \& {Zasowski}}]{foster15}
{Foster}, J.~B., {Cottaar}, M., {Covey}, K.~R., {et~al.} 2015, \apj, 799, 136

\bibitem[{{Genzel} \& {Stutzki}(1989)}]{genzel89}
{Genzel}, R. \& {Stutzki}, J. 1989, \araa, 27, 41

\bibitem[{{Getman} {et~al.}(2012){Getman}, {Feigelson}, {Sicilia-Aguilar},
  {Broos}, {Kuhn}, \& {Garmire}}]{getman12}
{Getman}, K.~V., {Feigelson}, E.~D., {Sicilia-Aguilar}, A., {et~al.} 2012,
  \mnras, 426, 2917

\bibitem[{{Gouliermis} {et~al.}(2014){Gouliermis}, {Hony}, \&
  {Klessen}}]{gou14}
{Gouliermis}, D.~A., {Hony}, S., \& {Klessen}, R.~S. 2014, \mnras, 439, 3775

\bibitem[{{Guieu} {et~al.}(2010){Guieu}, {Rebull}, {Stauffer}, {Vrba},
  {Noriega-Crespo}, {Spuck}, {Roelofsen Moody}, {Sepulveda}, {Weehler},
  {Maranto}, {Cole}, {Flagey}, {Laher}, {Penprase}, {Ramirez}, \&
  {Stolovy}}]{guieu10}
{Guieu}, S., {Rebull}, L.~M., {Stauffer}, J.~R., {et~al.} 2010, \apj, 720, 46

\bibitem[{{Hacar} \& {Tafalla}(2011)}]{hacer11}
{Hacar}, A. \& {Tafalla}, M. 2011, \aap, 533, A34

\bibitem[{{Hambly} {et~al.}(2008){Hambly}, {Collins}, {Cross}, {Mann}, {Read},
  {Sutorius}, {Bond}, {Bryant}, {Emerson}, {Lawrence}, {Rimoldini}, {Stewart},
  {Williams}, {Adamson}, {Hirst}, {Dye}, \& {Warren}}]{hambly08}
{Hambly}, N.~C., {Collins}, R.~S., {Cross}, N.~J.~G., {et~al.} 2008, \mnras,
  384, 637

\bibitem[{{Hartmann}(2002)}]{hartmann02}
{Hartmann}, L. 2002, \apj, 578, 914

\bibitem[{{Heitsch} {et~al.}(2009){Heitsch}, {Ballesteros-Paredes}, \&
  {Hartmann}}]{hei09}
{Heitsch}, F., {Ballesteros-Paredes}, J., \& {Hartmann}, L. 2009, \apj, 704,
  1735

\bibitem[{{Hennebelle} {et~al.}(2008){Hennebelle}, {Banerjee},
  {V{\'a}zquez-Semadeni}, {Klessen}, \& {Audit}}]{hennebelle08}
{Hennebelle}, P., {Banerjee}, R., {V{\'a}zquez-Semadeni}, E., {Klessen}, R.~S.,
  \& {Audit}, E. 2008, \aap, 486, L43

\bibitem[{{Hennemann} {et~al.}(2012){Hennemann}, {Motte}, {Schneider},
  {Didelon}, {Hill}, {Arzoumanian}, {Bontemps}, {Csengeri}, {Andr{\'e}},
  {Konyves}, {Louvet}, {Marston}, {Men'shchikov}, {Minier}, {Nguyen Luong},
  {Palmeirim}, {Peretto}, {Sauvage}, {Zavagno}, {Anderson}, {Bernard}, {Di
  Francesco}, {Elia}, {Li}, {Martin}, {Molinari}, {Pezzuto}, {Russeil}, {Rygl},
  {Schisano}, {Spinoglio}, {Sousbie}, {Ward-Thompson}, \& {White}}]{henn12}
{Hennemann}, M., {Motte}, F., {Schneider}, N., {et~al.} 2012, \aap, 543, L3

\bibitem[{{Henning} {et~al.}(2010){Henning}, {Linz}, {Krause}, {Ragan},
  {Beuther}, {Launhardt}, {Nielbock}, \& {Vasyunina}}]{henn10}
{Henning}, T., {Linz}, H., {Krause}, O., {et~al.} 2010, \aap, 518, L95

\bibitem[{{Indebetouw} {et~al.}(2007){Indebetouw}, {Robitaille}, {Whitney},
  {Churchwell}, {Babler}, {Meade}, {Watson}, \& {Wolfire}}]{inde07}
{Indebetouw}, R., {Robitaille}, T.~P., {Whitney}, B.~A., {et~al.} 2007, \apj,
  666, 321

\bibitem[{{Jackson} {et~al.}(2010){Jackson}, {Finn}, {Chambers}, {Rathborne},
  \& {Simon}}]{jackson10}
{Jackson}, J.~M., {Finn}, S.~C., {Chambers}, E.~T., {Rathborne}, J.~M., \&
  {Simon}, R. 2010, \apjl, 719, L185

\bibitem[{{J{\o}rgensen} {et~al.}(2006){J{\o}rgensen}, {Harvey}, {Evans},
  {Huard}, {Allen}, {Porras}, {Blake}, {Bourke}, {Chapman}, {Cieza}, {Koerner},
  {Lai}, {Mundy}, {Myers}, {Padgett}, {Rebull}, {Sargent}, {Spiesman},
  {Stapelfeldt}, {van Dishoeck}, {Wahhaj}, \& {Young}}]{jorgen06}
{J{\o}rgensen}, J.~K., {Harvey}, P.~M., {Evans}, II, N.~J., {et~al.} 2006,
  \apj, 645, 1246

\bibitem[{{Jose} {et~al.}(2012){Jose}, {Pandey}, {Ogura}, {Samal}, {Ojha},
  {Bhatt}, {Chauhan}, {Eswaraiah}, {Mito}, {Kobayashi}, \& {Yadav}}]{jose12}
{Jose}, J., {Pandey}, A.~K., {Ogura}, K., {et~al.} 2012, \mnras, 424, 2486

\bibitem[{{Jose} {et~al.}(2013){Jose}, {Pandey}, {Samal}, {Ojha}, {Ogura},
  {Kim}, {Kobayashi}, {Goyal}, {Chauhan}, \& {Eswaraiah}}]{jose13}
{Jose}, J., {Pandey}, A.~K., {Samal}, M.~R., {et~al.} 2013, \mnras, 432, 3445

\bibitem[{{Kirk} {et~al.}(2013){Kirk}, {Myers}, {Bourke}, {Gutermuth},
  {Hedden}, \& {Wilson}}]{kirk13}
{Kirk}, H., {Myers}, P.~C., {Bourke}, T.~L., {et~al.} 2013, \apj, 766, 115

\bibitem[{{Klessen} {et~al.}(2004){Klessen}, {Ballesteros-Paredes}, {Li}, \&
  {Mac Low}}]{klessen04}
{Klessen}, R.~S., {Ballesteros-Paredes}, J., {Li}, Y., \& {Mac Low}, M.-M.
  2004, in Astronomical Society of the Pacific Conference Series, Vol. 322, The
  Formation and Evolution of Massive Young Star Clusters, ed. H.~J.~G.~L.~M.
  {Lamers}, L.~J. {Smith}, \& A.~{Nota}, 299--308

\bibitem[{{Klessen} \& {Burkert}(2000)}]{klessen00}
{Klessen}, R.~S. \& {Burkert}, A. 2000, \apjs, 128, 287

\bibitem[{{Klessen} \& {Hennebelle}(2010)}]{kle10}
{Klessen}, R.~S. \& {Hennebelle}, P. 2010, \aap, 520, A17

\bibitem[{{Koenig} {et~al.}(2012){Koenig}, {Leisawitz}, {Benford}, {Rebull},
  {Padgett}, \& {Assef}}]{koenig12}
{Koenig}, X.~P., {Leisawitz}, D.~T., {Benford}, D.~J., {et~al.} 2012, \apj,
  744, 130

\bibitem[{{Kroupa}(2001)}]{kro01}
{Kroupa}, P. 2001, \mnras, 322, 231

\bibitem[{{Kumar} {et~al.}(2007){Kumar}, {Davis}, {Grave}, {Ferreira}, \&
  {Froebrich}}]{kumar07}
{Kumar}, M.~S.~N., {Davis}, C.~J., {Grave}, J.~M.~C., {Ferreira}, B., \&
  {Froebrich}, D. 2007, \mnras, 374, 54

\bibitem[{{Lada}(1987)}]{lada87}
{Lada}, C.~J. 1987, in IAU Symposium, Vol. 115, Star Forming Regions, ed.
  M.~{Peimbert} \& J.~{Jugaku}, 1--17

\bibitem[{{Lada}(2005)}]{lada05}
{Lada}, C.~J. 2005, in Astrophysics and Space Science Library, Vol. 327, The
  Initial Mass Function 50 Years Later, ed. E.~{Corbelli}, F.~{Palla}, \&
  H.~{Zinnecker}, 109

\bibitem[{{Larson}(1985)}]{larson85}
{Larson}, R.~B. 1985, \mnras, 214, 379

\bibitem[{{Leisawitz} {et~al.}(1989){Leisawitz}, {Bash}, \&
  {Thaddeus}}]{leisa89}
{Leisawitz}, D., {Bash}, F.~N., \& {Thaddeus}, P. 1989, \apjs, 70, 731

\bibitem[{{Longmore} {et~al.}(2012){Longmore}, {Rathborne}, {Bastian}, {Alves},
  {Ascenso}, {Bally}, {Testi}, {Longmore}, {Battersby}, {Bressert}, {Purcell},
  {Walsh}, {Jackson}, {Foster}, {Molinari}, {Meingast}, {Amorim}, {Lima},
  {Marques}, {Moitinho}, {Pinhao}, {Rebordao}, \& {Santos}}]{long12}
{Longmore}, S.~N., {Rathborne}, J., {Bastian}, N., {et~al.} 2012, \apj, 746,
  117

\bibitem[{{Lonsdale} {et~al.}(2003){Lonsdale}, {Smith}, {Rowan-Robinson},
  {Surace}, {Shupe}, {Xu}, {Oliver}, {Padgett}, {Fang}, {Conrow},
  {Franceschini}, {Gautier}, {Griffin}, {Hacking}, {Masci}, {Morrison},
  {O'Linger}, {Owen}, {P{\'e}rez-Fournon}, {Pierre}, {Puetter}, {Stacey},
  {Castro}, {Polletta}, {Farrah}, {Jarrett}, {Frayer}, {Siana}, {Babbedge},
  {Dye}, {Fox}, {Gonzalez-Solares}, {Salaman}, {Berta}, {Condon}, {Dole}, \&
  {Serjeant}}]{londse03}
{Lonsdale}, C.~J., {Smith}, H.~E., {Rowan-Robinson}, M., {et~al.} 2003, \pasp,
  115, 897

\bibitem[{{MacLaren} {et~al.}(1988){MacLaren}, {Richardson}, \&
  {Wolfendale}}]{mac88}
{MacLaren}, I., {Richardson}, K.~M., \& {Wolfendale}, A.~W. 1988, \apj, 333,
  821

\bibitem[{{Madsen} {et~al.}(2002){Madsen}, {Dravins}, \&
  {Lindegren}}]{madsen02}
{Madsen}, S., {Dravins}, D., \& {Lindegren}, L. 2002, \aap, 381, 446

\bibitem[{{Mallick} {et~al.}(2013){Mallick}, {Kumar}, {Ojha}, {Bachiller},
  {Samal}, \& {Pirogov}}]{mallick13}
{Mallick}, K.~K., {Kumar}, M.~S.~N., {Ojha}, D.~K., {et~al.} 2013, \apj, 779,
  113

\bibitem[{{Manara} {et~al.}(2013){Manara}, {Beccari}, {Da Rio}, {De Marchi},
  {Natta}, {Ricci}, {Robberto}, \& {Testi}}]{manara13}
{Manara}, C.~F., {Beccari}, G., {Da Rio}, N., {et~al.} 2013, \aap, 558, A114

\bibitem[{{Massi} {et~al.}(2015){Massi}, {Giannetti}, {Di Carlo}, {Brand},
  {Beltr{\'a}n}, \& {Marconi}}]{massi15}
{Massi}, F., {Giannetti}, A., {Di Carlo}, E., {et~al.} 2015, \aap, 573, A95

\bibitem[{{Matzner}(2002)}]{matzner02}
{Matzner}, C.~D. 2002, \apj, 566, 302

\bibitem[{{Meyer} {et~al.}(1997){Meyer}, {Calvet}, \& {Hillenbrand}}]{mey97}
{Meyer}, M.~R., {Calvet}, N., \& {Hillenbrand}, L.~A. 1997, \aj, 114, 288

\bibitem[{{Miettinen} \& {Offner}(2013)}]{miet13}
{Miettinen}, O. \& {Offner}, S.~S.~R. 2013, \aap, 555, A41

\bibitem[{{Molinari} {et~al.}(2010){Molinari}, {Swinyard}, {Bally}, {Barlow},
  {Bernard}, {Martin}, {Moore}, {Noriega-Crespo}, {Plume}, {Testi}, {Zavagno},
  {Abergel}, {Ali}, {Anderson}, {Andr{\'e}}, {Baluteau}, {Battersby},
  {Beltr{\'a}n}, {Benedettini}, {Billot}, {Blommaert}, {Bontemps}, {Boulanger},
  {Brand}, {Brunt}, {Burton}, {Calzoletti}, {Carey}, {Caselli}, {Cesaroni},
  {Cernicharo}, {Chakrabarti}, {Chrysostomou}, {Cohen}, {Compiegne}, {de
  Bernardis}, {de Gasperis}, {di Giorgio}, {Elia}, {Faustini}, {Flagey},
  {Fukui}, {Fuller}, {Ganga}, {Garcia-Lario}, {Glenn}, {Goldsmith}, {Griffin},
  {Hoare}, {Huang}, {Ikhenaode}, {Joblin}, {Joncas}, {Juvela}, {Kirk},
  {Lagache}, {Li}, {Lim}, {Lord}, {Marengo}, {Marshall}, {Masi}, {Massi},
  {Matsuura}, {Minier}, {Miville-Desch{\^e}nes}, {Montier}, {Morgan}, {Motte},
  {Mottram}, {M{\"u}ller}, {Natoli}, {Neves}, {Olmi}, {Paladini}, {Paradis},
  {Parsons}, {Peretto}, {Pestalozzi}, {Pezzuto}, {Piacentini}, {Piazzo},
  {Polychroni}, {Pomar{\`e}s}, {Popescu}, {Reach}, {Ristorcelli}, {Robitaille},
  {Robitaille}, {Rod{\'o}n}, {Roy}, {Royer}, {Russeil}, {Saraceno}, {Sauvage},
  {Schilke}, {Schisano}, {Schneider}, {Schuller}, {Schulz}, {Sibthorpe},
  {Smith}, {Smith}, {Spinoglio}, {Stamatellos}, {Strafella}, {Stringfellow},
  {Sturm}, {Taylor}, {Thompson}, {Traficante}, {Tuffs}, {Umana}, {Valenziano},
  {Vavrek}, {Veneziani}, {Viti}, {Waelkens}, {Ward-Thompson}, {White},
  {Wilcock}, {Wyrowski}, {Yorke}, \& {Zhang}}]{molinari10}
{Molinari}, S., {Swinyard}, B., {Bally}, J., {et~al.} 2010, \aap, 518, L100

\bibitem[{{Mu{\~n}oz} {et~al.}(2007){Mu{\~n}oz}, {Mardones}, {Garay},
  {Rebolledo}, {Brooks}, \& {Bontemps}}]{munoz07}
{Mu{\~n}oz}, D.~J., {Mardones}, D., {Garay}, G., {et~al.} 2007, \apj, 668, 906

\bibitem[{{Mucciarelli} {et~al.}(2011){Mucciarelli}, {Preibisch}, \&
  {Zinnecker}}]{mucci11}
{Mucciarelli}, P., {Preibisch}, T., \& {Zinnecker}, H. 2011, \aap, 533, A121

\bibitem[{{Myers}(2009)}]{myers09}
{Myers}, P.~C. 2009, \apj, 700, 1609

\bibitem[{{Myers} {et~al.}(1998){Myers}, {Adams}, {Chen}, \&
  {Schaff}}]{myers98}
{Myers}, P.~C., {Adams}, F.~C., {Chen}, H., \& {Schaff}, E. 1998, \apj, 492,
  703

\bibitem[{{Nagasawa}(1987)}]{naga87}
{Nagasawa}, M. 1987, Progress of Theoretical Physics, 77, 635

\bibitem[{{Neichel} {et~al.}(2015){Neichel}, {Samal}, {Plana}, {Zavagno},
  {Bernard}, \& {Fusco}}]{benoit15}
{Neichel}, B., {Samal}, M.~R., {Plana}, H., {et~al.} 2015, arXiv:1502.02102

\bibitem[{{Ojha} {et~al.}(2011){Ojha}, {Samal}, {Pandey}, {Bhatt}, {Ghosh},
  {Sharma}, {Tamura}, {Mohan}, \& {Zinchenko}}]{ojha11}
{Ojha}, D.~K., {Samal}, M.~R., {Pandey}, A.~K., {et~al.} 2011, \apj, 738, 156

\bibitem[{{Oliveira} {et~al.}(2009){Oliveira}, {Jeffries}, \& {van
  Loon}}]{olive07}
{Oliveira}, J.~M., {Jeffries}, R.~D., \& {van Loon}, J.~T. 2009, \mnras, 392,
  1034

\bibitem[{{Ostriker} {et~al.}(2001){Ostriker}, {Stone}, \&
  {Gammie}}]{ostriker01}
{Ostriker}, E.~C., {Stone}, J.~M., \& {Gammie}, C.~F. 2001, \apj, 546, 980

\bibitem[{{Ostriker}(1964)}]{ostriker64}
{Ostriker}, J. 1964, \apj, 140, 1056

\bibitem[{{Padoan} {et~al.}(2001){Padoan}, {Juvela}, {Goodman}, \&
  {Nordlund}}]{padoan01}
{Padoan}, P., {Juvela}, M., {Goodman}, A.~A., \& {Nordlund}, {\AA}. 2001, \apj,
  553, 227

\bibitem[{{Pandey} {et~al.}(2013){Pandey}, {Eswaraiah}, {Sharma}, {Samal},
  {Chauhan}, {Chen}, {Jose}, {Ojha}, {Kesh Yadav}, \& {Chandola}}]{pandey13}
{Pandey}, A.~K., {Eswaraiah}, C., {Sharma}, S., {et~al.} 2013, \apj, 764, 172

\bibitem[{{Panwar} {et~al.}(2014){Panwar}, {Chen}, {Pandey}, {Samal}, {Ogura},
  {Ojha}, {Jose}, \& {Bhatt}}]{neelam14}
{Panwar}, N., {Chen}, W.~P., {Pandey}, A.~K., {et~al.} 2014, \mnras, 443, 1614

\bibitem[{{Peretto} \& {Fuller}(2009)}]{pere09}
{Peretto}, N. \& {Fuller}, G.~A. 2009, \aap, 505, 405

\bibitem[{{Pon} {et~al.}(2011){Pon}, {Johnstone}, \& {Heitsch}}]{pon11}
{Pon}, A., {Johnstone}, D., \& {Heitsch}, F. 2011, \apj, 740, 88

\bibitem[{{Pontoppidan} {et~al.}(2005){Pontoppidan}, {Dullemond}, {van
  Dishoeck}, {Blake}, {Boogert}, {Evans}, {Kessler-Silacci}, \&
  {Lahuis}}]{ponto05}
{Pontoppidan}, K.~M., {Dullemond}, C.~P., {van Dishoeck}, E.~F., {et~al.} 2005,
  \apj, 622, 463

\bibitem[{{Povich} {et~al.}(2009){Povich}, {Churchwell}, {Bieging}, {Kang},
  {Whitney}, {Brogan}, {Kulesa}, {Cohen}, {Babler}, {Indebetouw}, {Meade}, \&
  {Robitaille}}]{povich09}
{Povich}, M.~S., {Churchwell}, E., {Bieging}, J.~H., {et~al.} 2009, \apj, 696,
  1278

\bibitem[{{Price} {et~al.}(2001){Price}, {Egan}, {Carey}, {Mizuno}, \&
  {Kuchar}}]{price01}
{Price}, S.~D., {Egan}, M.~P., {Carey}, S.~J., {Mizuno}, D.~R., \& {Kuchar},
  T.~A. 2001, \aj, 121, 2819

\bibitem[{{Ragan} {et~al.}(2012){Ragan}, {Henning}, {Krause}, {Pitann},
  {Beuther}, {Linz}, {Tackenberg}, {Balog}, {Hennemann}, {Launhardt}, {Lippok},
  {Nielbock}, {Schmiedeke}, {Schuller}, {Steinacker}, {Stutz}, \&
  {Vasyunina}}]{ragan12}
{Ragan}, S., {Henning}, T., {Krause}, O., {et~al.} 2012, \aap, 547, A49

\bibitem[{{Ragan} {et~al.}(2015){Ragan}, {Henning}, {Beuther}, {Linz}, \&
  {Zahorecz}}]{ragan15}
{Ragan}, S.~E., {Henning}, T., {Beuther}, H., {Linz}, H., \& {Zahorecz}, S.
  2015, \aap, 573, A119

\bibitem[{{Rathborne} {et~al.}(2006){Rathborne}, {Jackson}, \&
  {Simon}}]{rath06}
{Rathborne}, J.~M., {Jackson}, J.~M., \& {Simon}, R. 2006, \apj, 641, 389

\bibitem[{{Reach}(2007)}]{reach07}
{Reach}, W.~T. 2007, in IAU Symposium, Vol. 237, IAU Symposium, ed. B.~G.
  {Elmegreen} \& J.~{Palous}, 188--191

\bibitem[{{Rebull} {et~al.}(2011){Rebull}, {Guieu}, {Stauffer}, {Hillenbrand},
  {Noriega-Crespo}, {Stapelfeldt}, {Carey}, {Carpenter}, {Cole}, {Padgett},
  {Strom}, \& {Wolff}}]{rebul11}
{Rebull}, L.~M., {Guieu}, S., {Stauffer}, J.~R., {et~al.} 2011, \apjs, 193, 25

\bibitem[{{Rebull} {et~al.}(2010){Rebull}, {Padgett}, {McCabe}, {Hillenbrand},
  {Stapelfeldt}, {Noriega-Crespo}, {Carey}, {Brooke}, {Huard}, {Terebey},
  {Audard}, {Monin}, {Fukagawa}, {G{\"u}del}, {Knapp}, {Menard}, {Allen},
  {Angione}, {Baldovin-Saavedra}, {Bouvier}, {Briggs}, {Dougados}, {Evans},
  {Flagey}, {Guieu}, {Grosso}, {Glauser}, {Harvey}, {Hines}, {Latter},
  {Skinner}, {Strom}, {Tromp}, \& {Wolf}}]{rebul10}
{Rebull}, L.~M., {Padgett}, D.~L., {McCabe}, C.-E., {et~al.} 2010, \apjs, 186,
  259

\bibitem[{{Rebull} {et~al.}(2007){Rebull}, {Stapelfeldt}, {Evans},
  {J{\o}rgensen}, {Harvey}, {Brooke}, {Bourke}, {Padgett}, {Chapman}, {Lai},
  {Spiesman}, {Noriega-Crespo}, {Mer{\'{\i}}n}, {Huard}, {Allen}, {Blake},
  {Jarrett}, {Koerner}, {Mundy}, {Myers}, {Sargent}, {van Dishoeck}, {Wahhaj},
  \& {Young}}]{rebul07}
{Rebull}, L.~M., {Stapelfeldt}, K.~R., {Evans}, II, N.~J., {et~al.} 2007,
  \apjs, 171, 447

\bibitem[{{Robitaille} {et~al.}(2008){Robitaille}, {Meade}, {Babler},
  {Whitney}, {Johnston}, {Indebetouw}, {Cohen}, {Povich}, {Sewilo}, {Benjamin},
  \& {Churchwell}}]{robi08a}
{Robitaille}, T.~P., {Meade}, M.~R., {Babler}, B.~L., {et~al.} 2008, \aj, 136,
  2413

\bibitem[{{Robitaille} {et~al.}(2007){Robitaille}, {Whitney}, {Indebetouw}, \&
  {Wood}}]{robi07}
{Robitaille}, T.~P., {Whitney}, B.~A., {Indebetouw}, R., \& {Wood}, K. 2007,
  \apjs, 169, 328

\bibitem[{{Robitaille} {et~al.}(2006){Robitaille}, {Whitney}, {Indebetouw},
  {Wood}, \& {Denzmore}}]{robi06}
{Robitaille}, T.~P., {Whitney}, B.~A., {Indebetouw}, R., {Wood}, K., \&
  {Denzmore}, P. 2006, \apjs, 167, 256

\bibitem[{{Russeil} {et~al.}(2013){Russeil}, {Schneider}, {Anderson},
  {Zavagno}, {Molinari}, {Persi}, {Bontemps}, {Motte}, {Ossenkopf},
  {Andr{\'e}}, {Arzoumanian}, {Bernard}, {Deharveng}, {Didelon}, {Di
  Francesco}, {Elia}, {Hennemann}, {Hill}, {K{\"o}nyves}, {Li}, {Martin},
  {Nguyen Luong}, {Peretto}, {Pezzuto}, {Polychroni}, {Roussel}, {Rygl},
  {Spinoglio}, {Testi}, {Tig{\'e}}, {Vavrek}, {Ward-Thompson}, \&
  {White}}]{russeil13}
{Russeil}, D., {Schneider}, N., {Anderson}, L.~D., {et~al.} 2013, \aap, 554,
  A42

\bibitem[{{Samal} {et~al.}(2012){Samal}, {Pandey}, {Ojha}, {Chauhan}, {Jose},
  \& {Pandey}}]{samal12}
{Samal}, M.~R., {Pandey}, A.~K., {Ojha}, D.~K., {et~al.} 2012, \apj, 755, 20

\bibitem[{{Samal} {et~al.}(2007){Samal}, {Pandey}, {Ojha}, {Ghosh}, {Kulkarni},
  \& {Bhatt}}]{samal07}
{Samal}, M.~R., {Pandey}, A.~K., {Ojha}, D.~K., {et~al.} 2007, \apj, 671, 555

\bibitem[{{Samal} {et~al.}(2010){Samal}, {Pandey}, {Ojha}, {Ghosh}, {Kulkarni},
  {Kusakabe}, {Tamura}, {Bhatt}, {Thompson}, \& {Sagar}}]{samal10}
{Samal}, M.~R., {Pandey}, A.~K., {Ojha}, D.~K., {et~al.} 2010, \apj, 714, 1015

\bibitem[{{Samal} {et~al.}(2014){Samal}, {Zavagno}, {Deharveng}, {Molinari},
  {Ojha}, {Paradis}, {Tig{\'e}}, {Pandey}, \& {Russeil}}]{samal14}
{Samal}, M.~R., {Zavagno}, A., {Deharveng}, L., {et~al.} 2014, \aap, 566, A122

\bibitem[{{Schneider} {et~al.}(2010){Schneider}, {Csengeri}, {Bontemps},
  {Motte}, {Simon}, {Hennebelle}, {Federrath}, \& {Klessen}}]{schneider10}
{Schneider}, N., {Csengeri}, T., {Bontemps}, S., {et~al.} 2010, \aap, 520, A49

\bibitem[{{Sciortino}(2007)}]{sciortino07}
{Sciortino}, S. 2007, \memsai, 78, 616

\bibitem[{{Simon} {et~al.}(2006){Simon}, {Rathborne}, {Shah}, {Jackson}, \&
  {Chambers}}]{simon06}
{Simon}, R., {Rathborne}, J.~M., {Shah}, R.~Y., {Jackson}, J.~M., \&
  {Chambers}, E.~T. 2006, \apj, 653, 1325

\bibitem[{{Smith} {et~al.}(2014){Smith}, {Glover}, \& {Klessen}}]{smith14}
{Smith}, R.~J., {Glover}, S.~C.~O., \& {Klessen}, R.~S. 2014, \mnras, 445, 2900

\bibitem[{{Spezzi} {et~al.}(2008){Spezzi}, {Alcal{\'a}}, {Covino}, {Frasca},
  {Gandolfi}, {Oliveira}, {Chapman}, {Evans}, {Huard}, {J{\o}rgensen},
  {Mer{\'{\i}}n}, \& {Stapelfeldt}}]{spe08}
{Spezzi}, L., {Alcal{\'a}}, J.~M., {Covino}, E., {et~al.} 2008, \apj, 680, 1295

\bibitem[{{Spezzi} {et~al.}(2013){Spezzi}, {Cox}, {Prusti}, {Mer{\'{\i}}n},
  {Ribas}, {Alves de Oliveira}, {Winston}, {K{\'o}sp{\'a}l}, {Royer}, {Vavrek},
  {Andr{\'e}}, {Pilbratt}, {Testi}, {Bressert}, {Ricci}, {Men'shchikov}, \&
  {K{\"o}nyves}}]{spe13}
{Spezzi}, L., {Cox}, N.~L.~J., {Prusti}, T., {et~al.} 2013, \aap, 555, A71

\bibitem[{{Tackenberg} {et~al.}(2014){Tackenberg}, {Beuther}, {Henning},
  {Linz}, {Sakai}, {Ragan}, {Krause}, {Nielbock}, {Hennemann}, {Pitann}, \&
  {Schmiedeke}}]{tack14}
{Tackenberg}, J., {Beuther}, H., {Henning}, T., {et~al.} 2014, \aap, 565, A101

\bibitem[{{Takahashi} {et~al.}(2013){Takahashi}, {Ho}, {Teixeira}, {Zapata}, \&
  {Su}}]{taka13}
{Takahashi}, S., {Ho}, P.~T.~P., {Teixeira}, P.~S., {Zapata}, L.~A., \& {Su},
  Y.-N. 2013, \apj, 763, 57

\bibitem[{{Tasker} \& {Tan}(2009)}]{tasker09}
{Tasker}, E.~J. \& {Tan}, J.~C. 2009, \apj, 700, 358

\bibitem[{{Tomisaka}(1995)}]{tomi95}
{Tomisaka}, K. 1995, \apj, 438, 226

\bibitem[{{V{\'a}zquez-Semadeni} {et~al.}(2009){V{\'a}zquez-Semadeni},
  {G{\'o}mez}, {Jappsen}, {Ballesteros-Paredes}, \& {Klessen}}]{semadeni09}
{V{\'a}zquez-Semadeni}, E., {G{\'o}mez}, G.~C., {Jappsen}, A.-K.,
  {Ballesteros-Paredes}, J., \& {Klessen}, R.~S. 2009, \apj, 707, 1023

\bibitem[{{Wang} {et~al.}(2011){Wang}, {Zhang}, {Wu}, \& {Zhang}}]{wang11}
{Wang}, K., {Zhang}, Q., {Wu}, Y., \& {Zhang}, H. 2011, \apj, 735, 64

\bibitem[{{Wright} {et~al.}(2010){Wright}, {Eisenhardt}, {Mainzer}, {Ressler},
  {Cutri}, {Jarrett}, {Kirkpatrick}, {Padgett}, {McMillan}, {Skrutskie},
  {Stanford}, {Cohen}, {Walker}, {Mather}, {Leisawitz}, {Gautier}, {McLean},
  {Benford}, {Lonsdale}, {Blain}, {Mendez}, {Irace}, {Duval}, {Liu}, {Royer},
  {Heinrichsen}, {Howard}, {Shannon}, {Kendall}, {Walsh}, {Larsen}, {Cardon},
  {Schick}, {Schwalm}, {Abid}, {Fabinsky}, {Naes}, \& {Tsai}}]{wright10}
{Wright}, E.~L., {Eisenhardt}, P.~R.~M., {Mainzer}, A.~K., {et~al.} 2010, \aj,
  140, 1868

\bibitem[{{Yamamura} {et~al.}(2009){Yamamura}, {Tsuji}, {Tanab{\'e}}, \&
  {Nakajima}}]{Yamam09}
{Yamamura}, I., {Tsuji}, T., {Tanab{\'e}}, T., \& {Nakajima}, T. 2009, in
  Astronomical Society of the Pacific Conference Series, Vol. 418, AKARI, a
  Light to Illuminate the Misty Universe, ed. T.~{Onaka}, G.~J. {White},
  T.~{Nakagawa}, \& I.~{Yamamura}, 143

\bibitem[{{Yun} {et~al.}(2008){Yun}, {Djupvik}, {Delgado}, \& {Alfaro}}]{yun08}
{Yun}, J.~L., {Djupvik}, A.~A., {Delgado}, A.~J., \& {Alfaro}, E.~J. 2008,
  \aap, 483, 209

\bibitem[{{Zavagno} {et~al.}(2007){Zavagno}, {Pomar{\`e}s}, {Deharveng},
  {Hosokawa}, {Russeil}, \& {Caplan}}]{zava07}
{Zavagno}, A., {Pomar{\`e}s}, M., {Deharveng}, L., {et~al.} 2007, \aap, 472,
  835

\bibitem[{{Zhang} {et~al.}(2009){Zhang}, {Wang}, {Pillai}, \&
  {Rathborne}}]{zhang09}
{Zhang}, Q., {Wang}, Y., {Pillai}, T., \& {Rathborne}, J. 2009, \apj, 696, 268

\bibitem[{{Zinchenko} {et~al.}(2012){Zinchenko}, {Liu}, {Su}, {Kurtz}, {Ojha},
  {Samal}, \& {Ghosh}}]{zinch12}
{Zinchenko}, I., {Liu}, S.-Y., {Su}, Y.-N., {et~al.} 2012, \apj, 755, 177

\end{thebibliography}

\begin{table}
\scriptsize
\caption{Inferred Physical Parameters from SED Fits to YSOs}
\begin{tabular}{cccc} 
\hline\hline
 ID & \multicolumn{1}{c}{$M_{\rm disk}$ }
& \multicolumn{1}{c}{$\dot{M}_{\rm disk}$ }  \\
   & \multicolumn{1}{c}{( $M_\odot$)} 
& \multicolumn{1}{c}{(10$^{-6}$ $M_\odot$/yr)}   \\
 \hline
\hline
01 & 0.006      $\pm$   0.01 &0.10      $\pm$   0.22\\
02 & 0.007      $\pm$   0.02 &0.08      $\pm$   0.27\\
03 & 0.013      $\pm$   0.02 &0.64      $\pm$   1.49\\
05 & 0.006      $\pm$   0.01 &0.24      $\pm$   0.66\\
06 & 0.007      $\pm$   0.01 &0.31      $\pm$   0.78\\
07 & 0.009      $\pm$   0.02 &0.23      $\pm$   0.72\\
08 & 0.007      $\pm$   0.01 &0.09      $\pm$   0.60\\
09 & 0.012      $\pm$   0.02 &0.95      $\pm$   1.43\\
10 & 0.025      $\pm$   0.04 &3.10      $\pm$   3.38\\
11 & 0.003      $\pm$   0.01 &0.01      $\pm$   0.04\\
12 & 0.030      $\pm$   0.03 &5.79      $\pm$   12.35\\
14 & 0.006      $\pm$   0.02 &0.14      $\pm$   0.47\\
15 & 0.005      $\pm$   0.02 &0.47      $\pm$   2.14\\
16 & 0.043      $\pm$   0.04 &3.88      $\pm$   5.48\\
17 & 0.007      $\pm$   0.01 &0.21      $\pm$   0.47\\
20 & 0.006      $\pm$   0.01 &0.34      $\pm$   1.37\\
21 & 0.041      $\pm$   0.03 &4.63     $\pm$    3.58\\
22 & 0.005      $\pm$   0.01 &0.65      $\pm$   1.03\\
23 & 0.002      $\pm$   0.01 &0.02      $\pm$   0.09\\
24 & 0.002      $\pm$   0.01 &0.02      $\pm$   0.09\\
25 & 0.078      $\pm$   0.09 &0.48      $\pm$   0.79\\
26 & 0.014      $\pm$   0.03 &2.96      $\pm$   3.99\\
28 & 0.016      $\pm$   0.02 &1.46      $\pm$   2.38\\
31 & 0.005      $\pm$   0.01 &0.11      $\pm$   0.29\\
32 & 0.006      $\pm$   0.01 &0.40      $\pm$   0.97\\
33 & 0.015      $\pm$   0.02 &1.89      $\pm$   2.67\\
34 & 0.005      $\pm$   0.01 &0.14      $\pm$   0.26\\
36 & 0.015      $\pm$   0.02 &0.37      $\pm$   0.85\\
42 & 0.021      $\pm$   0.02 &1.42      $\pm$   2.37\\
44 & 0.015      $\pm$   0.02 &0.32      $\pm$   0.73\\
45 & 0.018      $\pm$   0.04 &0.48      $\pm$   1.09\\
47 & 0.029      $\pm$   0.03 &8.88      $\pm$   14.02\\
48 & 0.005      $\pm$   0.01 &0.18      $\pm$   0.50\\
49 & 0.001      $\pm$   0.02 &0.13      $\pm$   0.02\\
\hline
\end{tabular}
 \end{table}

\end{document}